\documentclass{IEEEtran}

\usepackage{cite}
\usepackage{amsmath,amssymb,amsfonts}
\usepackage{algorithmic}
\usepackage{graphicx}
\usepackage{epsfig}
\usepackage{textcomp}
\usepackage{epsfig}
\usepackage{epstopdf}
\usepackage{subfigure}

\usepackage{xcolor}
\usepackage{hyperref}
\hypersetup{
    colorlinks = true,
    linkbordercolor = {white},
    linkcolor= {black},
    citecolor = [rgb]{ 0.4275 ,   0.6627 ,   0.2706}
}
\usepackage{array}  
\usepackage{color}
\usepackage{graphicx}% Include figure files
\usepackage{epstopdf}
\usepackage{dcolumn}% Align table columns on decimal point
\usepackage{bm}% bold math
\usepackage{textcomp}
\usepackage{array}
\usepackage{tabularx,booktabs}
\usepackage{multirow}
\usepackage{gensymb}
\usepackage{fancyhdr}
\usepackage{float}
\usepackage{mathtools, cuted} %long equation break the columns
\usepackage{mathtools} %long equation break the columns

\def\e{\begin{equation}}
\def\f{\end{equation}}
\def\*{^{\displaystyle*}}
\def\=#1{\overline{\overline #1}}

\def\.{\cdot}

\def\##1{{\bf #1\mit}}
\def\_#1{{\bf #1\mit}}
\def\-#1{{\bf #1\mit}}

\def\Re{{\rm Re\mit}}

\def\time#1{\tilde{#1}}
\def\freq#1{{#1}}

\def\b{\delta}

\def\l#1{\label{eq:#1}}
\def\r#1{(\ref{eq:#1})}
\def\am{\left(\begin{array}{c}}
\def\amm{\left(\begin{array}{cc}}
\def\a{\end{array}\right)}

\def\amm{\overline{\overline{\alpha}}_{\rm mm}}

\newcommand{\ve}[1]{\mathbf{#1}}

\newcommand{\dya}[1]{\bar{\bar{#1}}}

\def\BibTeX{{\rm B\kern-.05em{\sc i\kern-.025em b}\kern-.08em
    T\kern-.1667em\lower.7ex\hbox{E}\kern-.125emX}}
\begin{document}

\title{Tutorial on electromagnetic nonreciprocity \\ and its origins}
\author{Viktar S. Asadchy, Mohammad Sajjad Mirmoosa, \IEEEmembership{Member, IEEE},  Ana D\'iaz-Rubio,  \IEEEmembership{Member, IEEE}, \\ Shanhui Fan, \IEEEmembership{Fellow, IEEE},  and Sergei~A.~Tretyakov, \IEEEmembership{Fellow, IEEE} 
\thanks{This work was supported  by Finnish Foundation for Technology Promotion, the Academy of Finland (projects 287894 and 309421), and by a U.S. Air Force Office of Scientific Research MURI project (Grant No. FA9550-18-1-0379). }
\thanks{V.~S. Asadchy  and S. Fan are with the Department of Electrical Engineering, Stanford University, Stanford, California 94305, USA  (e-mail: asadchy@stanford.edu)

V.~S. Asadchy, A. D\'iaz-Rubio, M. S. Mirmoosa, and S.~A.~Tretyakov are with the Department of Electronics and Nanoengineering, Aalto University, P.O. 15500, FI-00076 Aalto, Finland. 
}
}

%\author{S.~N.~Tcvetkova}
%\email{svetlana.tcvetkova@aalto.fi}
%\affiliation{Aalto University, Department of Electronics and Nanoengineering, P.O. 15500, FI-00076 Aalto, Finland}
%\author{S. Maci}
%\affiliation{University of Siena, Department of Information Engineering and Mathematics, Via Roma 56, 53100, Siena, Italy}
%\author{S.~A.~Tretyakov}
%\affiliation{Aalto University, Department of Electronics and Nanoengineering, P.O. 15500, FI-00076 Aalto, Finland}

\maketitle

\begin{abstract}
% This tutorial provides an intuitive and concrete description of the  phenomena of electromagnetic nonreciprocity that will be  useful   for readers with  engineering or physics backgrounds.  
% We describe the notions of reciprocity and nonreciprocity and explain the origins and nature of different nonreciprocal effects. Time-reversal and space-reversal symmetries in the most general case of external bias fields or external modulations are investigated and their relation with nonreciprocity is discussed. Classification of nonreciprocal phenomena in most general bianisotropic scatterers and media is presented and discussed.   Equipped with the knowledge of the reciprocity relations, we list the physical conditions which have to be satisfied to ensure reciprocity of a system. This list allows us to identify all possible routes towards creating nonreciprocal devices. Finally, we present examples of nonreciprocal devices which are based on the identified nonreciprocity mechanisms, discuss the current status of research in this field and provide an outlook for future research and development of nonreciprocal devices, including microwave and optical components as well as metamaterial and metasurface devices.  
%
This tutorial provides an intuitive and concrete description of the  phenomena of electromagnetic nonreciprocity that will be  useful   for readers with  engineering or physics backgrounds. The notion of time reversal and its different definitions are discussed with special emphasis to its relationship  with the reciprocity concept. Starting from the  Onsager reciprocal relations generally applicable to    many physical  processes, we present the derivation of the Lorentz theorem and discuss other implications of reciprocity for electromagnetic systems. Next, we identify all possible  routes  towards  engineering  nonreciprocal  devices and analyze in detail three of them: Based on external bias, based on nonlinear and time-variant systems. The principles of the operation of different nonreciprocal devices are explained. 
We address  the similarity and fundamental  difference  between  nonreciprocal effects and asymmetric transmission in reciprocal systems. 
In addition to the tutorial description of the topic, the manuscript  also contains   original findings. In particular, general classification of reciprocal and nonreciprocal phenomena in   linear  bianisotropic   media based on the space- and time-reversal symmetries is presented. This classification serves as a   powerful tool  for drawing analogies between seemingly distinct effects having the same physical origin and can be used for predicting novel electromagnetic phenomena. Furthermore, electromagnetic reciprocity theorem for time-varying systems is derived and its applicability is discussed. 
\end{abstract}


\tableofcontents

\section{\label{sec:1}Introduction }

The notion of  reciprocity and nonreciprocity is a fundamental scientific concept, important in many different branches of physics, chemistry, and engineering. In the general sense, electromagnetic nonreciprocity implies that the fields created by a source at the observation point are not the same compared to the case when the source and observation point  are interchanged.  
Nonreciprocal systems     are essential  for applications where one-way propagation is required~\cite[\textsection~13.3]{zvezdin_modern_1997},\cite[\textsection~9.4]{pozar_microwave_2012}: Radars using a single antenna for transmitting and receiving at the same time, suppression of destabilizing reflections in lasers, isolating signals from a power supply, etc. 
\begin{figure*}[tb!]
	\centering
	\subfigure[]{\includegraphics[height=0.2\linewidth]{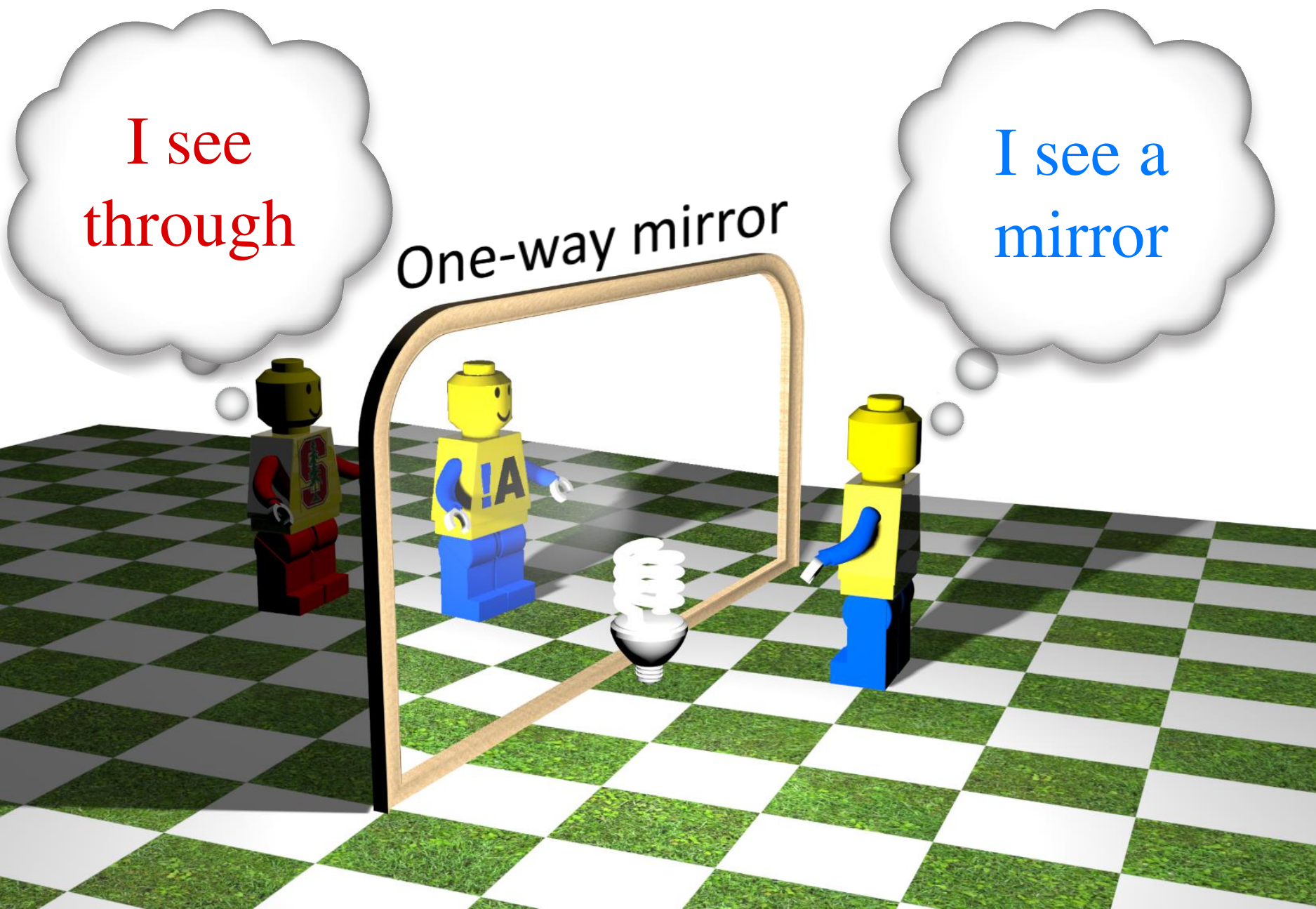} \label{fig1a}} 
	\subfigure[]{\includegraphics[height=0.15\linewidth]{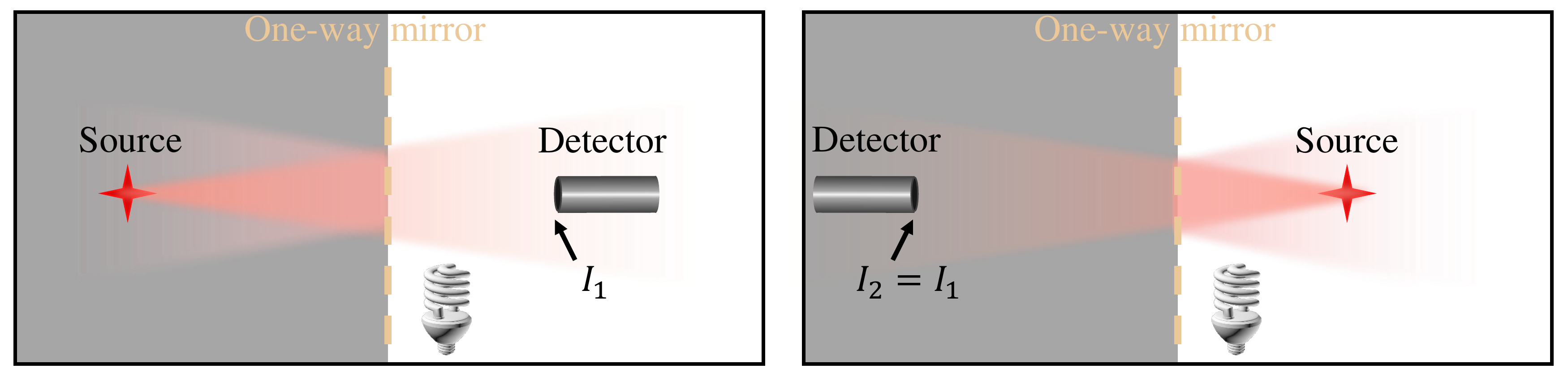}\label{fig1b}}\\
	\subfigure[]{\includegraphics[height=0.2\linewidth]{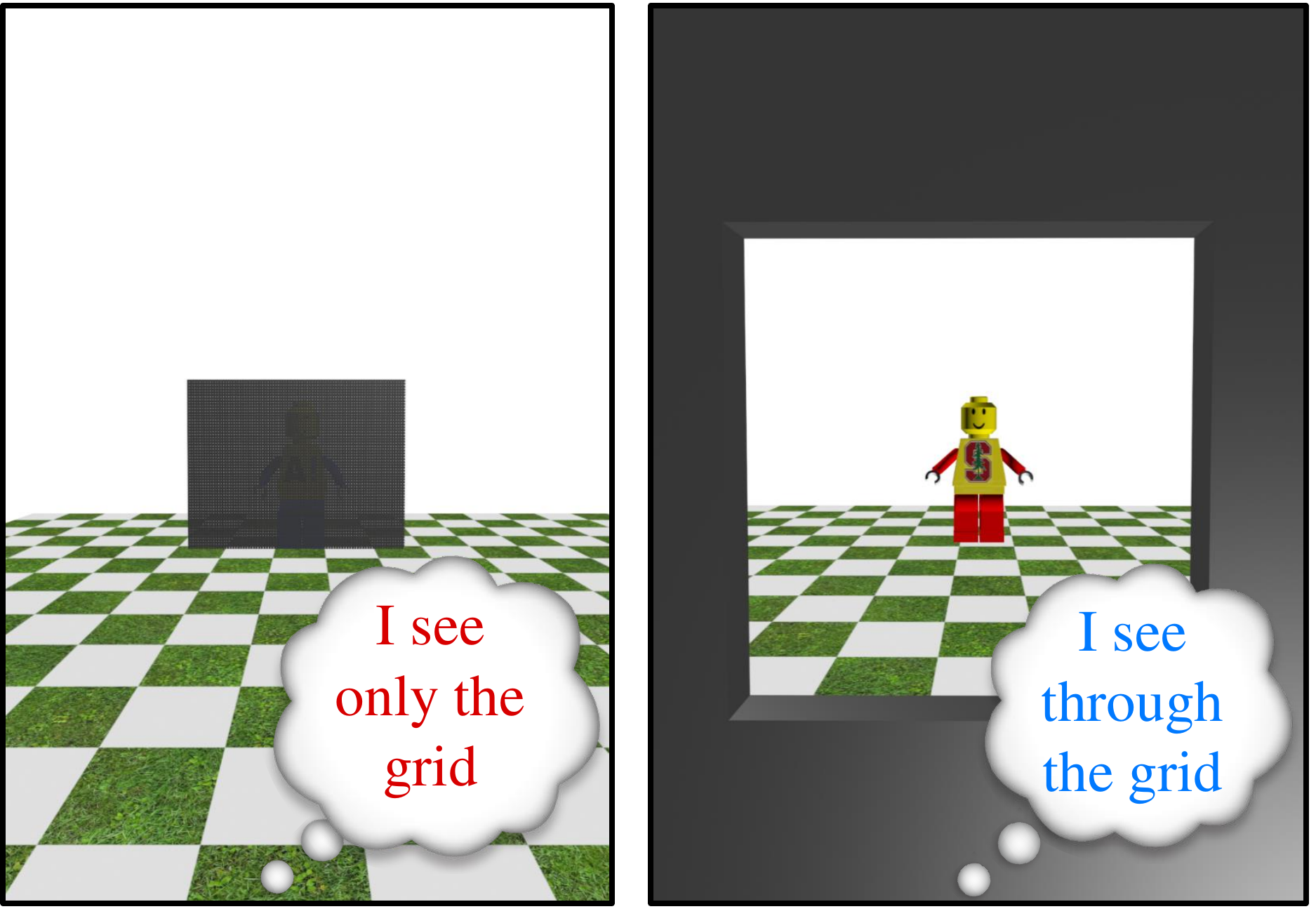} \label{fig1c}}
    \subfigure[]{\includegraphics[height=0.15\linewidth]{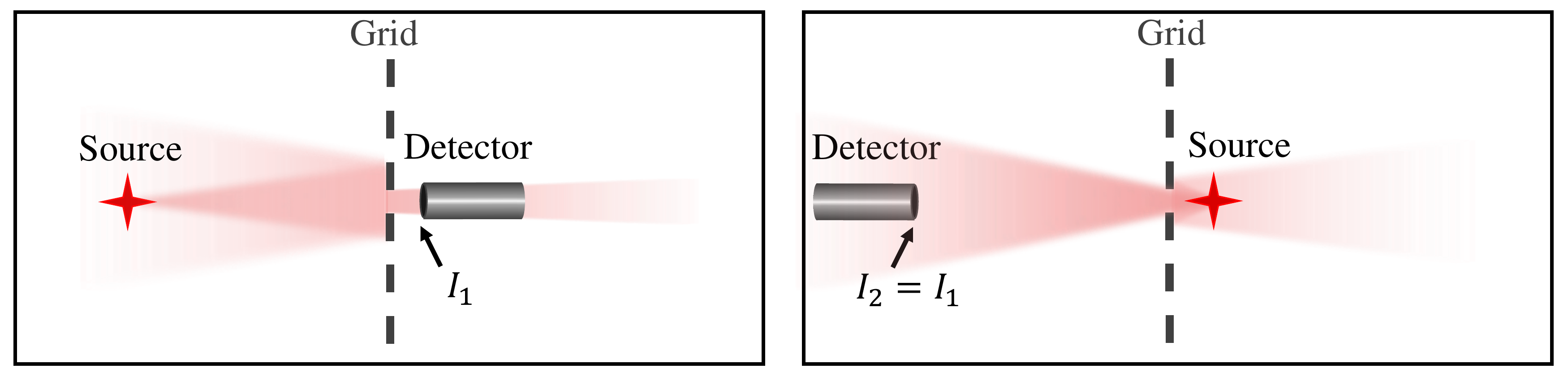} \label{fig1d}}
	\caption{ Examples from everyday life which resemble nonrecirpocal light propagation. In fact, reciprocity is preserved in both examples.  (a) Two observers are separated by a one-way mirror (silver coated glass). The blue observer is located on the bright side, illuminated by a bulb. The strong light from the bulb obscures the dimmed light coming from the red observer: The blue observer sees his own image. Meanwhile, the red observer clearly sees through the mirror. (b) A thought experiment proving reciprocity of the one-way mirror. Grey and white backgrounds denote dark and bright regions, respectively. Light intensities $I_1$ and $I_2$ measured by the detector at the two locations are equal (the detector measures only light from the source and not from the bulb). (c) Images seen by the two observers standing at the two sides of a metal grid fence at large (left figure) and very  short (right figure).  The blue observer sees the red observer, but not vice versa. (d) A thought experiment proving reciprocity of the grid fence.      }
	\label{fig1}
\end{figure*}
Although there is a rigorous definition of reciprocity breaking, the concept of nonreciprocity is not trivial and may be easily interpreted erroneously. Consider an example from our everyday life, commercialized   one-way  mirrors~\cite{bloch_transparent_1903} (used for surveillance or as reflective windows in buildings) which exhibit fictitious one-way wave propagation. In fact, such mirrors are completely reciprocal and their operation is based on the brightness contrast at the two sides of the mirror. The observer located on the bright side will receive dimmed light coming from the objects in the dark, but this image will be strongly obscured by the bright reflection of the observer himself (see Fig.~\ref{fig1a}). Nevertheless, reciprocity of this system can be proven by replacing observers with a point light source and a detector, as shown in Fig.~\ref{fig1b}.  The detected  signal \emph{from the source} (excluding  the signal from the external bulb) remains the same even after interchanging the positions of the source and detector. 
Another example of fictitious perception of nonreciprocity is a metal-grid fence: An observer standing at the near proximity of the fence sees another observer, standing at the opposite side far from the fence, more clear than contrariwise (see Fig.~\ref{fig1c}). Such effect occurs due to the contrast in the viewing angles of the two observers. Obviously, the fence is reciprocal, which can be verified by positioning a light source and a light detector in place of the observers (see Fig.~\ref{fig1d}). Interchanging the locations of the source and the detector will not modify the signal  \emph{from the source} measured  by the detector. In a very simplistic way, one could define reciprocity of an optical system as ``if I can see your eyes, then you can see mine'', which holds for this example (both observers see eyes of one another equally well). However, it is obvious that such definition is not general and fails  for the previous example system with so-called one-way mirrors.  

These examples evidently demonstrate that the rigorous definition of electromagnetic nonreciprocity might come to  drastic contradictions with the simplistic commonplace  sense. Importantly, although reciprocal systems, such as one-way mirrors, are much easier for deployment and use than the truly nonreciprocal counterparts, they cannot provide  truly nonreciprocal effects. 

The concept of reciprocity (nonreciprocity) has a long history. Probably, the earliest theoretical works about this concept  were developed by Stokes in 1849~\cite{stokes_perfect_1849} and Helmholtz in 1856~\cite{helmholtz_handbuch_1856} for light waves. At the same time,   the reciprocity  property was postulated for thermoelectric phenomena by 
Thomson (Lord Kelvin) in 1854~\cite{notitle_1854}. In 1860 Kirchhoff reformulated the reciprocity principle for  thermal radiation~\cite{kirchhoff_i._1860}. 
Rayleigh described this principle  for acoustics in his book published in 1878~\cite{rayleigh_treatise_1878}. Hendrik Lorentz came up with his famous electromagnetic reciprocity theorem in 1896~\cite{lorentz_theorem_1896}. The fundamental relation between the time-reversal invariance on microscopic dynamic equations and reciprocity in dissipative systems was realized by L. Onsager~\cite{onsager_reciprocal_1931,onsager_reciprocal_1931-1} and developed and extended by H.~Casimir \cite{casimir_onsagers_1945} to electromagnetic systems and to response functions of materials under influence of external bias fields. 

Conventional route for achieving electromagnetic nonreciprocity  (breaking reciprocity) is based on the magneto-optical effect~\cite{smit_ferrites:_1959,lax_microwave_1962,landau_electrodynamics_1984,
rodrigue_generation_1988,
gurevich_magnetization_1996,zvezdin_modern_1997,adam_ferrite_2002,
ishimaru_electromagnetic_2017} which implies asymmetric wave propagation through a medium (e.g., ferrimagnetic material) in the presence of a static magnetic field.  Although nonreciprocal systems at microwaves have been extensively developed in the middle of the twentieth century, efficient and compact nonreciprocal components operating at optical frequencies (natural materials exhibit weak magneto-optical effects in optics since both the cyclotron frequency of \textit{free} electrons and the Larmor frequency of  spin precession of \textit{bound} electrons are typically in the microwave range~\cite[p.~571]{ashcroft_solid_1976},\cite[\textsection~9.1]{pozar_microwave_2012},\cite[\textsection~79]{landau_electrodynamics_1984}) are yet to be found.    During the last decade, several   alternative routes towards realization of  nonreciprocal wave propagation in optics have been actively developed. Among them the most promising are based on nonlinear~\cite{tocci_thinfilm_1995,gallo_all-optical_1999,Gallo2,Zhou1,Lin1,Manipatruni1,Roy1,miroshnichenko_reversible_2010,trendafilov_hamiltonian_2010,Zhukovsky1,lepri_asymmetric_2011,Shadrivov1,fan_subwavelength_2011,Grigoriev1,fan_all-silicon_2012,Ding22,Anand1,Fan22,roy_cascaded_2013,chang_paritytime_2014,Xu22,nazari_optical_2014,Peng1,shi_limitations_2015,Yu22,mahmoud_all-passive_2015,sounas_time-reversal_2017,roy_critical_2017,Aleahmad1,bino_microresonator_2018,
rosario_hamann_nonreciprocity_2018,sounas_broadband_2018,sounas_nonreciprocity_2018} and active~\cite{kodera_artificial_2011,kodera_nonreciprocal_2011,sounas_electromagnetic_2013,kodera_unidirectional_2018} materials as well as materials whose properties are modulated in time by some external source~\cite{cullen_travelling-wave_1958,slater_interaction_1958,fettweis_steady-state_1959,kamal_parametric_1960,currie_coupled-cavity_1960,simon_action_1960,macdonald_exact_1961,oliner_wave_1961,anderson_reciprocity_1965,holberg_parametric_1966,yu_complete_2009,yu_optical_2009,kang_reconfigurable_2011,lira_electrically_2012,fang_photonic_2012,fang_realizing_2012,wang_optical_2013,horsley_optical_2013,sounas_giant_2013,fang_controlling_2013,qin_nonreciprocal_2014,estep_magnetic-free_2014,tzuang_non-reciprocal_2014,lin_light_2014,shaltout_time-varying_2015,hadad2015space,hadad_breaking_2016,correas-serrano_nonreciprocal_2016,taravati_mixer-duplexer-antenna_2017}. Comprehensive classification of nonreciprocal systems and their theoretical description   can be found in recent  review papers on this topic~\cite{potton_reciprocity_2004,jalas_what_2013,Sounas_time_modulation_2017,
caloz_electromagnetic_2018,fan_nonreciprocal_2018,caloz_spacetime_2019,
caloz_spacetime_2019-1,guo_nonreciprocal_2019,zhang_breaking_2019,nagulu_non-reciprocal_2020,williamson_breaking_2020}.

In this tutorial, we present the reciprocity and nonreciprocity notions in the educational manner, accessible for the readers without solid background in this field. We explain the origins  of different nonreciprocal effects and demonstrate how they are related to time-reversal and space-reversal symmetries of the system. The paper also provides classification of non-reciprocal phenomena in most general bianisotropic scatterers and media. 
It is important to note that throughout the  paper time harmonic oscillations in the form ${\rm e}^{+j \omega t}$ are assumed according to the conventional electrical engineering notations~\cite{cheng_field_1983,pozar_microwave_2012}.

% Known technology: ferrimagnetic/magneto-optical materials (in circuits, also active components)
% Emerging \emph{meta-technology}: active, nonlinear, or time-space modulated metamaterials and metasurfaces. Compact realizations, new functionalities
% \begin{figure*}[tb]
% \epsfig{file=Tasym_0_to_t.eps, width=0.8\linewidth}\\
% \epsfig{file=Tasym_t_to_0.eps, width=0.8\linewidth}
% \caption{Nonreciprocity concept....This picture needs update.}
% \label{ill}
% \end{figure*}
 
% 14

\section{Time reversal in electrodynamics } \label{TR}
\subsection{Time reversal definition and Loschmidt's paradox}\label{Loschmidt}
The concept of electromagnetic reciprocity  is closely related to that of the time-reversal symmetry of Maxwell's equations. The same symmetry property of field equations leads also to time-reversability  of physical processes in lossless systems. Nevertheless, the concepts of reciprocity and time reversal of physical processes have fundamental differences   and, therefore, should be distinguished. For the subsequent description, we start our discussion with the concept of time reversal. 

One of the fundamental questions in physics is ``Can the direction of time flow be determined?''~\cite{sachs_can_1963}. Our everyday experience suggests that it can be. When we mix up two substances, we know that we cannot reverse the process of mixing and separate the substances from the mixture. When we see  a video of some macroscopic physical process, in most cases, we can definitely say whether it is played  forward or backward. Nevertheless, the question about the time flow is {\it not} related to our daily experience, and it required several generations of physicists   to obtain the answer. This question is directly linked to another one:   ``Are all physical laws   invariant under time reversal?'' Conventionally, a physical process is called time-reversal invariant or reversible if its evolution in the backward direction (think about the reversed video of the original process) is also a {\it realistic} process~\cite[\textsection~2.3]{sachs_physics_1987}. That is, this evolution also satisfies the dynamic equations. For example, the played-backward process of a ball scattering on the billiard table looks as realistic as the direct process (assuming elastic ball collisions).

\begin{figure}[bt]
	\centering
	\includegraphics[width=0.94\linewidth]{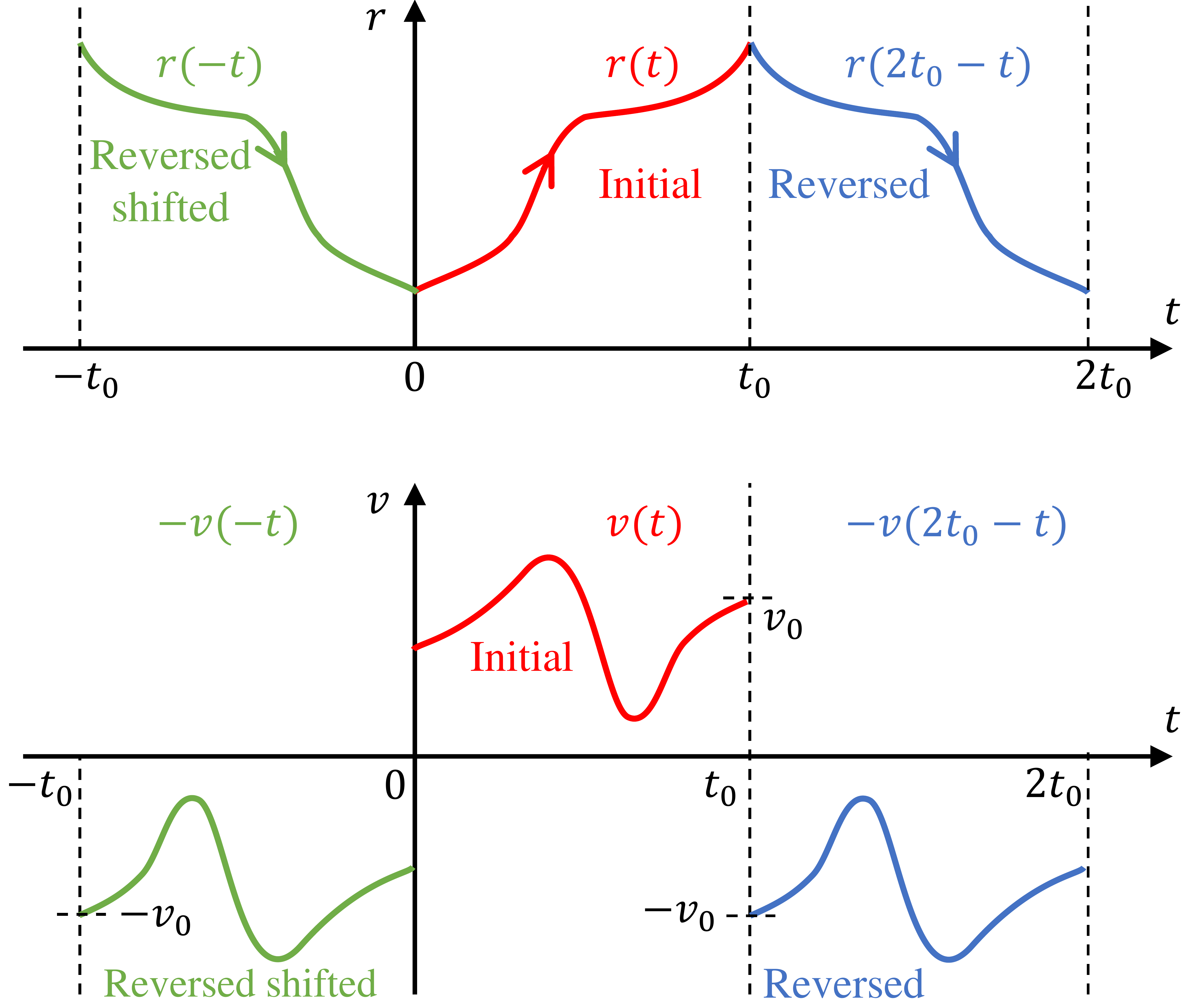} 
	\caption{Time reversal of the coordinate function of an arbitrary object (a billiard ball in this example). Although in the reversed process the time arrow direction does not change (${\rm d}t>0$), the displacement ${\rm d} r(t)$ and the derivative of the coordinate (velocity) in the reversed system flip the signs.   }
	\label{fig2}
\end{figure}

An equivalent definition of a time-reversal invariant process was formulated by H.~Casimir~\cite{casimir_reciprocity_1963} (see also~\cite{sachs_time_1972}):
``If a system of particles and fields moves
in a certain way during the period $0 < t <t_0$ and if at
the moment $t_0$ we would invert all velocities, currents,
magnetic fields and so on, then the system retraces its
steps: At the time $2t_0$ particle coordinates and fields are
what they were at $t =0$, at a time $2 t_0 - t$ the situation is
what it was at~$t$ ''. 
Figure~\ref{fig2} illustrates this definition by plotting a coordinate~$r$ and velocity projection $v$ of some particle as  functions of time~$t$. The initial particle motion is shown in red. At moment $t_0$, the speed of the particle flips its direction, while its coordinate   remains the same. This arrangement is equivalent to playing the video of the motion backward. The blue curve shows the reversed particle motion. At moment $2t_0$, the particle returns to its initial position as at $t=0$ with the same speed but in the {\it opposite} direction. 
The green curve in the plots of Fig.~\ref{fig2} depicts the curve for the reversed motion shifted  along the time axis by $-2t_0$. The blue and green curves are physically identical and correspond to the same motion due to the uniformity of time (the choice of time origin $t = 0$ is a matter of convenience). Comparing the initial motion, red curve, with the reversed shifted motion, green curve, one can see that the curves are identical mirrors of one another with respect to $t=0$ (the velocity curve has an additional  flip of the direction). This is why the physical notion of time reversal is conventionally written   mathematically as $t \rightarrow -t$~\cite{casimir_reciprocity_1963,sachs_time_1972,roberts_chaos_1992,silveirinha_hidden_2019},\cite[\textsection~1.9.2]{barron_molecular_2009},\cite[Ch.~8]{post_formal_1962},\cite[\textsection~4.1.2]{lindell_methods_1992},\cite[p.~270]{Jackson1999},\cite{caloz_electromagnetic_2018}. If the equations describing the process do not change under  substitution $t \rightarrow -t$, the process is called time-reversal invariant. As one can see from Fig.~\ref{fig2}, the coordinate of time-reversed process is  related to that of the original process as $T\{ r(t)\}= r(-t)$, while the corresponding velocities are related by $T\{ v(t)\}= -v(-t)$\footnote{Note that relation $T\{ r(t)\}= r(-t)$ does not impose any time symmetry on the original function $r(t)$ and should not be confused with $  r(t)= r(-t)$.}.  
Furthermore, time reversal operation does not change the sign of differential~${\rm d}t$ (the time arrow direction remains the same), while it flips the sign of the infinitesimal displacement ${\rm d} r(t)$. One can also note from the bottom panel of Fig.~\ref{fig2} that the infinitesimal increase of the velocity ${\rm d} v (t)$ does not change sign under time reversal (compare the slopes of the red curve at $t_0$ and green curve at $-t_0$). Therefore, the acceleration of an object $a = {\rm d} v /{\rm d} t$  is an even function with respect to the time-reversal operation. 

Although in this tutorial, as well as in the majority of books in the literature,  the  two definitions  of time reversal given above are likened,  it was pointed out in~\cite{north_two_2008} that they can yield different results for some special objects in spacetime. The first definition of time reversal based on  inversion of all velocities by H.~Casimir is sometimes referred to as ``active''~\cite{north_two_2008,post_logic_1979}, since the transformation is applied directly on the objects motion.  The second definition given by simple flipping of time $t \rightarrow -t$  also corresponds to the ``active'' scenario. One can also think of a ``passive'' time reversal for which the transformation does not act on the objects but rather on the time axis.  For the sake of exposition completeness, it should be mentioned that there exist alternative definitions of time reversal, such  as~\cite[Ch.~3]{horwich_asymmetries_1989}~\cite[Ch.~1]{albert_time_2000}, which are not generally accepted in the physics community~\cite{malament_time_2004}.

All  the physical laws, with the only exception of those corresponding to  weak interactions, are governed by the equations which are symmetric under time reversal. However, our daily experience tells us that most physical processes, governed by the very same physical laws, are irreversible in time. Moreover, our experience is also supported by the second law of thermodynamics expressed as ${\rm d} S/ {\rm d} t \geq 0 $ which says that any closed physical system cannot evolve from a  disordered to a more ordered state (e.g., reversing a process of dye mixing would violate this law). This contradiction of  irreversible processes which are governed by time-reversible physical laws was pointed out by   J.~Loschmidt in 1877~\cite{loschmidt_uber_1876}  and was named subsequently as Loschmidt's paradox. An elegant explanation of the paradox can be found in~\cite[Ch.~2]{sachs_physics_1987}. It is based on the fact that a physical process is not only determined by the physical laws, but it also depends on the {\it initial conditions}. 

\begin{figure}[bt]
\centering
\subfigure[]{\includegraphics[width=0.8\columnwidth]{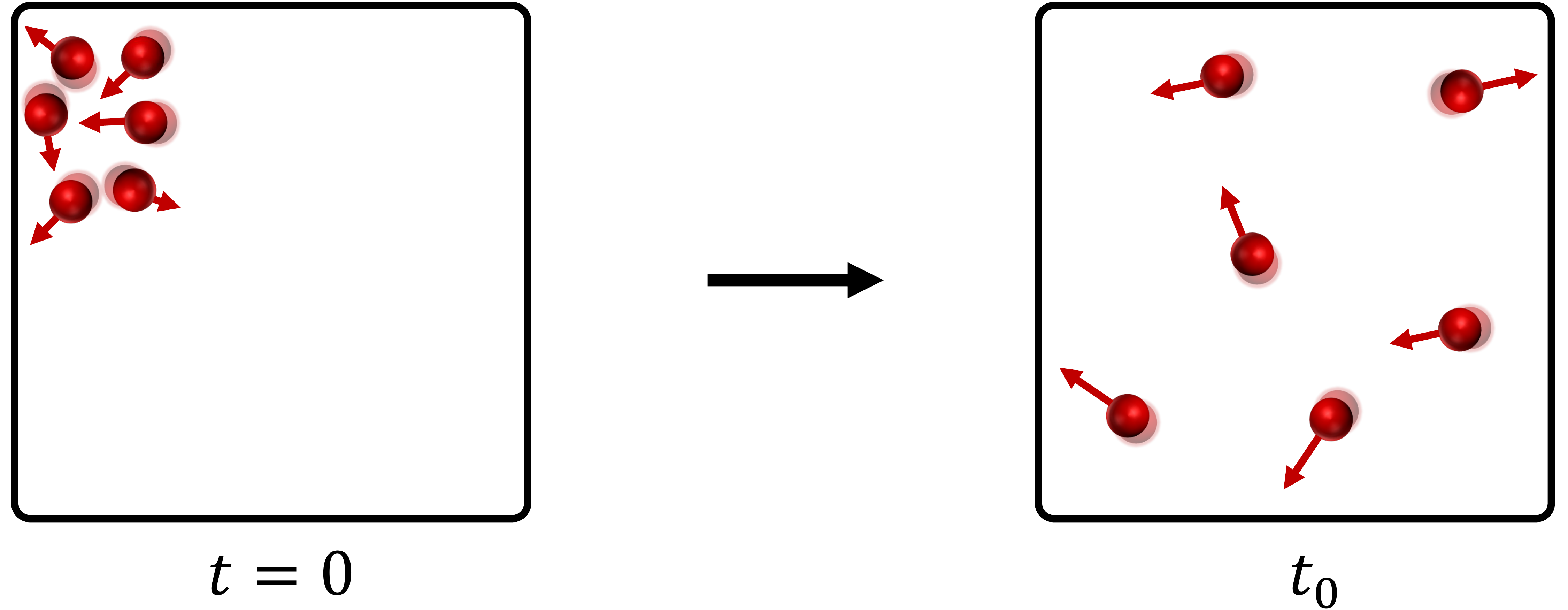} \label{mola}} 
\subfigure[]{\includegraphics[width=0.8\columnwidth]{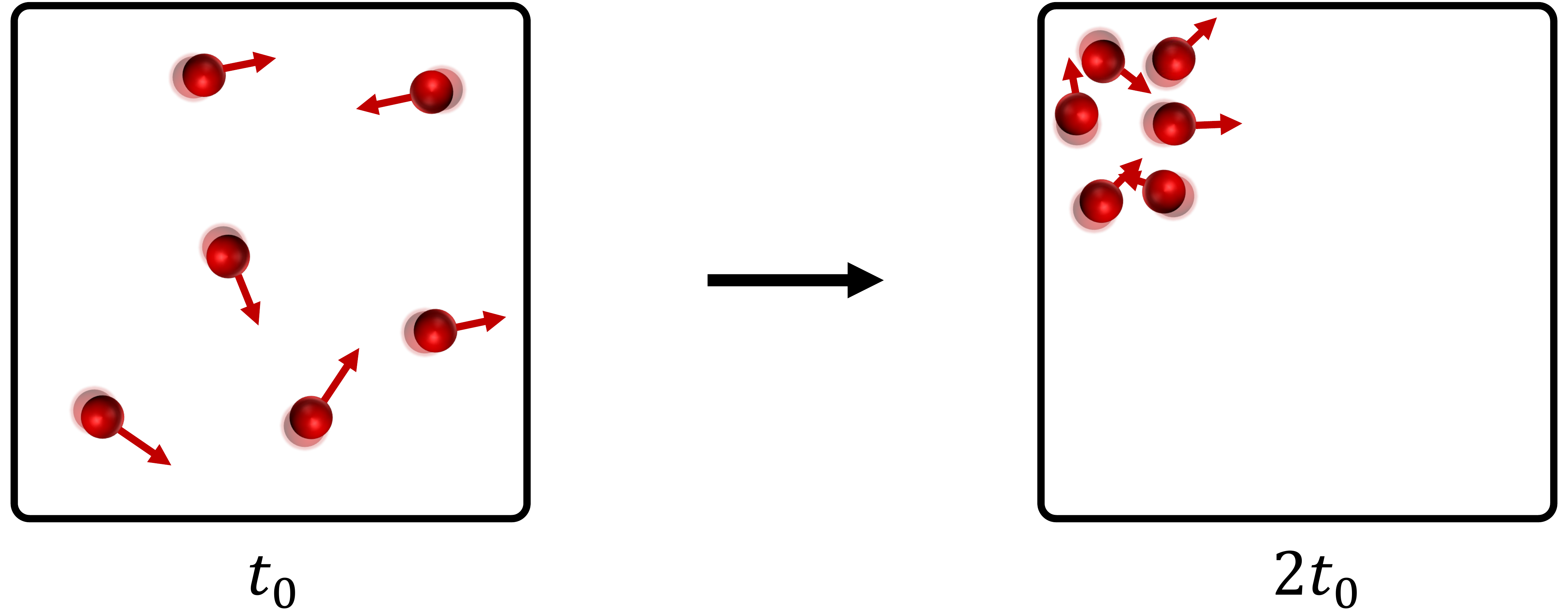} \label{molb}}
\subfigure[]{\includegraphics[width=0.8\columnwidth]{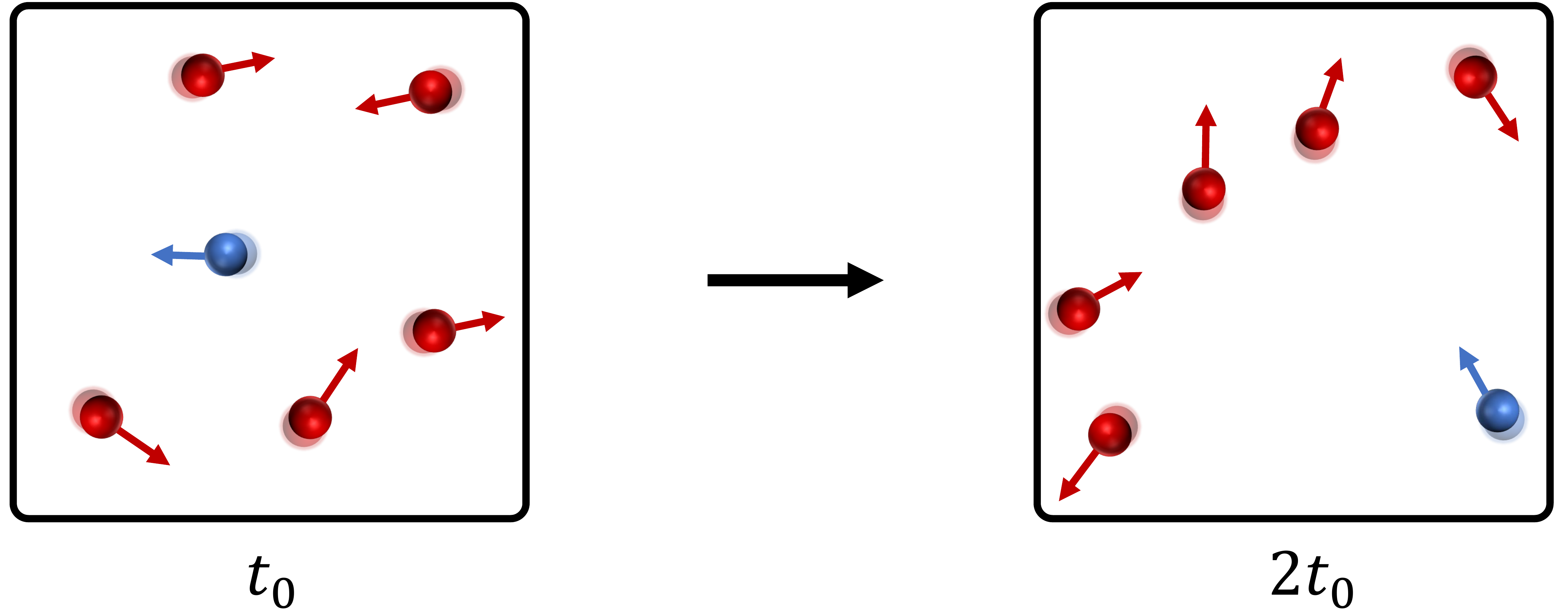} \label{molc}} 
\caption{ Gas molecules in a closed box. (a) Evolution of the molecules distribution from an ordered state to a disordered state. The arrows indicate the velocity directions. (b) Time reversal of the process in (a) with correct microscopic initial conditions (positions of the molecules are unchanged, while velocities are reversed). (c) Time reversal of the process in (a) with macroscopic initial conditions. Microscopic conditions are not fully satisfied, the blue molecule indicate the wrong initial state. }
\label{molecules}
\end{figure}

Figure~\ref{molecules} illustrates resolving the paradox. Consider a closed box with gas and assume that at $t=0$ all its molecules occupied only a small region in the corner of the box, as is shown in Fig.~\ref{mola}. The red arrows indicate the velocities of the molecules. After some time $t_0$, the molecules will spread somewhat uniform inside the box. 
%If the initial conditions of the reversed process  are not  {\it exactly} equal to those of the 
Time reversal of the process in Fig.~\ref{mola} requires fulfilment of the correct initial conditions. Let us assume that we are able to position all the molecules anywhere in the box and launch them at some moment $t_0$ with desired initial velocities. If we choose the position of the molecules as in the right illustration of  Fig.~\ref{mola}  and launch them with the same velocities but in the opposite directions, then after time period $t_0$ the molecules would come back exactly to the  initial ordered state (see Fig.~\ref{molb}). In other words, the process reversal would be possible if we were able to ensure the  {\it microscopic} initial conditions (position and velocity) for each molecule. In practice, assigning correct initial conditions even for a few molecules is a complicated engineering problem, therefore, for systems with large numbers of molecules the only parameters that we control are statistical  ones, such as the mean speed and the mean free path. These statistical parameters are related to macroscopic pressure and temperature. If we try to reverse the process in Fig.~\ref{mola} by satisfying only the {\it macroscopic} initial conditions (for example,   by wrongly assigning velocity of one of the molecules, as is shown in Fig.~\ref{molc}), the gas would not come back to the initial state and,  moreover, the new state would be drastically different from the initial one. The probability that by satisfying the macroscopic initial conditions we  also ensure by chance  the microscopic ones is proportional to $n^{-N}$, where $N$ is the number of molecules in the closed box and $n$ is the number of possible states of each molecule~\cite[p.~414]{onsager_reciprocal_1931}. Thus, due to the exponentially decreasing probability, all macroscopic processes appear to us irreversible.

A given macroscopic (thermodynamic) state is consistent with  the great variety of
possible microscopic states. 
Strictly speaking, any process may be reversed since it may happen that the macroscopic initial conditions were chosen exactly satisfying the microscopic initial conditions. However, the probability of this event will be incredibly small, exponentially decreasing with the number of particles and their degrees of freedom. 
This reasoning explains why most processes which we observe are irreversible: Mixing substances, cracking  objects, combustion, and even lossy phenomena. Nevertheless, although it sounds somewhat bizarre, all these processes are reversible under time reversal (with correct microscopic initial conditions) since they are governed by time-symmetric physical laws.

Until the middle of the twentieth century, scientists could not find any exception among  physical laws that would be asymmetric with respect to time reversal. Moreover, the discovery of the CPT theorem in quantum field theory~\cite{schwinger_theory_1951}, stating that all physical laws are symmetric under the simultaneous transformations of charge conjugation (C), parity transformation (P), and time reversal (T), became an additional argument for the universality of time reversal symmetry. However,  in
the late 1950s  a violation of parity symmetry by phenomena that involve the weak force was reported. Starting from 1964, a series of experiments on decay of K-meson has demonstrated that even CP~symmetry can be violated. Assuming that the CPT theorem is fundamental, the latter result meant automatically that time symmetry of the weak interactions can be broken. Coming back to the question in the beginning of this section, we now see that  the direction of time in fact can be determined.

\subsection{Time reversal symmetry of Maxwell's equations}

It is reasonable to assume that under time reversal,  microscopic electrodynamic quantities either do not change (time-reversal symmetric or even) or flip sign (time-reversal asymmetric or odd). Moreover, it is reasonable to assume that the electric charge $q$, charge density $\rho$, and coordinate $r$ do not change under time reversal\footnote{Here  we assume  that the electric charge is an even quantity with respect to time reversal, as it is usually done. In fact, an alternative assumption is equally possible~\cite{serdyukov_electromagnetics_2001}}. Under this assumption, speed $\_v$ and electric current $\_j$ are odd with respect to time reversal (see also Fig.~\ref{fig2}). The force is proportional to the acceleration  $\_a$ and, therefore, is symmetric.   Next, it is easy to determine the time-reversal properties of most electrodynamic microscopic quantities. The electric field $\time{\_e}$, being proportional to the force acting on a unit charge, is time-reversal even. From the formula for the Lorentz force $\time{\_F}=q \time{\_e} +q \time{\_v} \times \time{\_b}$, one can deduct that magnetic field $\time{\_b}$ must be time-reversal odd (multiplication of speed and magnetic field must be time-reversal even).
Note that  hereafter we use tilde symbol ``$\sim$''   above physical quantities defined in the time domain. The frequency-domain version of these quantities will be written without tilde.   

We see that under these assumptions the \emph{microscopic} Maxwell's equations 
% \e \begin{array}[c]
% \nabla\times\ve{e}=-\frac{\partial \ve{b}}{\partial t} \\
% \nabla\times\ve{b}=\frac{1}{c^2}\frac{\partial \ve{e}}{\partial t}+\mu_0\ve{j}
% \end{array}
% \f
\e\begin{array}{c}\displaystyle
\nabla\times \time{\ve{e}}=-\frac{\partial \time{\ve{b}}}{\partial t}, \qquad \nabla \cdot \time{\ve{b}} =0  \\\displaystyle
\nabla\times \time{\ve{b}}=\frac{1}{c^2}\frac{\partial \time{\ve{e}}}{\partial t}+\mu_0 \time{\ve{j}}, \qquad \nabla \cdot \time{\ve{e}} =\time{\rho} / \varepsilon_0
\end{array}\l{eq1}
\f
are \emph{invariant} under time reversal\footnote{
As it was mentioned above, $\partial t$ does not flip the sign under time reversal, while  $\partial \time{\ve{e}}$ does. The nabla operator and $\partial \time{\ve{b}}$ are on the contrary time-reversal even. 
}.
This conclusion is in agreement with experimental observations about electromagnetic phenomena, most importantly, with the reciprocity principle, which we will discuss in detail. 
Table~\ref{tab:1} summarizes  the time-reversal properties of microscopic electrodynamic  quantities. 

\begin{table}[tb]
\centering
 \begin{tabularx}{\columnwidth}{p{0.3\columnwidth}p{0.25\columnwidth}p{0.3\columnwidth}}
\toprule
    \centering\textbf{Physical Quantity} 
    &\centering\textbf{Time Reversal} \\ (microscopic quantities) 
    &\centering\textbf{Time Reversal}
    \\ (macroscopic quantities) \cr
\midrule
    Charge density &  \centering ${\time{\rho}}(t) \mapsto +{\time{\rho}}(-t)$ 
    & \centering ${\time{\rho}}(t)\mapsto +{\time{\rho}}(-t)$ \cr
\midrule
   Current density & \centering  $ \time{\ve{j}}(t) \mapsto -\time{\ve{j}}(-t)$  
   & \centering $\time{\ve{J}}(t)\mapsto -\time{\ve{J}}(-t)$  \cr
\midrule
    Displacement & \centering --
   & \centering  $\time{\ve{D}}(t) \mapsto +\time{\ve{D}}(-t)$  \cr
\midrule
   Electric field & \centering  $\time{\ve{e}}(t)\mapsto +\time{\ve{e}}(-t)$ 
   &  \centering  $\time{\ve{E}}(t)\mapsto +\time{\ve{E}}(-t)$ \cr
\midrule
 \addlinespace
    Magnetic field & \centering --
    &  \centering  $\time{\ve{H}}(t) \mapsto -\time{\ve{H}}(-t)$ \cr
\midrule
    Magnetic induction & \centering  $ \time{\ve{b}}(t) \mapsto -\time{\ve{b}}(-t)$
   &  \centering  $ \time{\ve{B}}(t) \mapsto -\time{\ve{B}}(-t)$ \cr
\midrule
   Magnetization & \centering --
   &  \centering   $\time{\ve{M}}(t) \mapsto -\time{\ve{M}}(-t)$ \cr
\midrule
    Polarization density & \centering --
   & \centering $\time{\ve{P}}(t)\mapsto +\time{\ve{P}}(-t)$ \cr
\midrule
   Poynting vector &  \centering  $ \time{\ve{S}}(t) \mapsto -\time{\ve{S}}(-t)$ 
    &  \centering $ \time{\ve{S}}(t) \mapsto -\time{\ve{S}}(-t)$  \cr
\bottomrule
\end{tabularx}
\caption{Transformation properties of electrodynamic quantities under time  reversal}
\label{tab:1}
\end{table}

The macroscopic electromagnetic fields  are obtained through volume averaging of the microscopic ones. In the macroscopic form, Maxwell's equations are usually written as
% \pink{The preceding discussion of the time reversal properties is based on microscopic quantities. Since the macroscopic fields are obtained through averaging of microscopic fields, their properties under time reversion remain the same. 
% % It is easy to demonstrate that the equations of macroscopic electromagnetism are also invariant under time reversal. 
% Macroscopic equations are usually written in the following form:}
\begin{equation}
\begin{array}{c}\displaystyle
\nabla\times\time{\ve{E}}=-\frac{\partial \time{\ve{B}}}{\partial t}, \qquad \nabla \cdot \time{\ve{B}} =0, \\\displaystyle
\nabla\times \time{\ve{H}}=\frac{\partial \time{\ve{D}}}{\partial t}+\time{\ve{J}}_{\rm ext}, \qquad \nabla \cdot \time{\ve{D}} =\time{\rho}_{\rm ext},
\end{array}\l{eq12}
\end{equation}
where $\time{\ve{E}}$ and $\time{\ve{B}}$ are the   electric field  and magnetic flux density averaged over a small macroscopic volume,  $\time{\ve{D}}$ and $\time{\ve{H}}$ are the electric displacement field and magnetic field, respectively, and $\time{\rho}_{\rm ext}$ and $\time{\ve{J}}_{\rm ext}$ are the averaged \emph{external} (free) electric charge and current density. Here, the term ``external'' describes the charges and currents which are not affected by the fields (they are not induced by  the electric and magnetic fields governed by \emph{this} set of Maxwell's equations, being external to this system). The two latter fields  are defined as 
\e \time{\ve{D}}=\varepsilon_0\time{\ve{E}}+\time{\ve{P}},  \qquad \time{\ve{H}}=\frac{1}{\mu_0}\time{\ve{B}}-\time{\ve{M}}, \l{eqm} \f
where $\time{\ve{P}}$ and $\time{\ve{M}}$ are the volume polarization densities of electric and magnetic dipole moments induced in the material. The electric polarization density is defined  as $\nabla \cdot \time{\_P}=-\time{\rho}_{\rm ind}$, where $\time{\rho}_{\rm ind}$ stands for the induced (bound)  electric charge density. The magnetization is defined by $\nabla \times \time{\_M}=\time{\_J}_{\rm ind}-\partial \time{\_P} /\partial t$~\cite[Eq.~(2.45)]{serdyukov_electromagnetics_2001}.

Because the process of volume averaging does not involve the time variable, the same property of time-reversal symmetry is true also for the system of macroscopic Maxwell's equations. This conclusion implies, naturally, the assumption that the time-reversal operation includes inversion of equations which govern also the \emph{external} charges and currents, $\time{\rho}_{\rm ext}$ and $\time{\ve{J}}_{\rm ext}$, i.e. inversion with correct microscopic initial conditions (see discussion in Section~\ref{Loschmidt}).  
% Let us stress again that this time-reversal symmetry of \emph{field equations} does not mean that \emph{electromagnetic processes} are reversible in time: Dissipation (modeled by the constitutive relations of materials where the fields exist) will lead to increasing entropy of all processes, both for the original and time-reversed versions of the field equations. 
%The important question here is whether the external and induced currents are reversible in time or not. If the material where the currents are flowing has dissipation losses, then the currents in the time-reversed material will still obey loss mechanism and, similarly to the considered system with billiard balls, will not be reversible in time (neither even nor odd time-symmetric). In this case, electrodynamic quantities $\_P$, $\_M$, $\_D$, and $\_H$, as well as the \emph{macroscopic} Maxwell equations,  will   not be reversible in time. 
% True time reversal of wave propagation  in a lossy material would require  demonstrated in Fig.~\ref{fig3} (cells~A.1 and A.2 correspond to the direct and reversed propagations).
However, only if the dissipation losses in the system are negligible, then
the time-reversal symmetry of field equations may lead to time-reversibility of electromagnetic processes.  
%macroscopic current densities $\_J_{\rm ind}$ and $\_J_{\rm ext}$ switch signs under time inversion. The same conclusion is applicable to quantities $\_M$ and $\_H$. Thus, 
%equations~\r{eqm} and the Maxwell equations~\r{eq12} become time-symmetric, as can be verified using the transformations in Table~I and Fig.~\ref{fig2}~
%In that case, $\rho$, $\ve{E}$, $\ve{P}$, and $\ve{D}$ are odd with respect to time-reversal, while $\ve{J}$, $\ve{B}$, $\ve{M}$, and $\ve{H}$ are even.}. 
% The former quantity depends on the charges and their relative position, and therefore,  is invariant under time reversal. On contrary, $\ve{M}$ is proportional to the molecular magnetic momentum generated by currents (electron orbiting) and it switches sign under  time reversal.  Finally, it is straightforward to see that the  fields $\ve{D}$ and $\ve{H}$ are, respectively, even and odd quantities
%In order to determine

%\begin{figure}[h!]
%\centering
%\epsfig{file=Figures_conference_1.eps, width=0.35\linewidth}
%\\
% \epsfig{file=Figures_conference_23.eps, width=0.9\linewidth}
%\caption{The time-reversal symmetry of field equations leads to the reciprocity property.}
%\label{rec_concept}
%\end{figure}

% An alternative way to interpret TR is defined as when t goes from –t0 to 0. If we consider the evolution in a system $\Phi(t,r)$.  Explain it with Figures in table 2

\subsection{Time reversal of material relations} \label{trmatrel}

Let us consider time reversal of an arbitrary wave process  in a stationary dielectric material. The dielectric is assumed to be isotropic, possibly nonuniform,  and its magnetization is zero so that $\time{\_P}=\time{\_P}(\_r,t)$ and $\time{\_M}=0$. Assuming spatially local response, the volume electric polarization density $\time{\_P}$    in  the time domain  is related to  the electric field acting on the material by $ \time{\ve{P}}(t)=\varepsilon_0\int_{-\infty}^t \time{\chi}(t-t')\time{\ve{E}}(t') {\rm d}t'$, where $\time{\chi}$ is the electric susceptibility. Note that the upper integration limit is $t$, rather than $+\infty$ to account for  causality of the process.  After the standard replacement, the displacement vector can be written as the convolution integral 
\e \time{\ve{D}}(t)=\int_{0}^\infty \varepsilon_0 \time{\varepsilon}(\tau) \time{\ve{E}}(t-\tau){\rm d}\tau, 
\l{causal}\f
where $\time{\varepsilon}(\tau)=\delta(\tau)+\time{\chi}(\tau)$. 
It is   convenient to simplify the integral expression for the displacement vector using the Fourier transform, which yields the well-known material relations for an isotropic dielectric without spatial dispersion:   
% As we mentioned, the polarization vector represents the response of the material to the macroscopic electric field. If the properties of the material do not
% change with time, 
% the relation between the polarization vector and the electric field can be expressed as
% $ \ve{P}(t)=\varepsilon_0\int_{-\infty}^t \chi(t-t')\ve{E(t')}dt'$. It is important to notice that the polarization at moment $t$   depends not only on the electric field at this particular moment of time but on all previous moments, showing the inertial response of the materials. 
%  Similar approach can be use for obtaining the relation between $\ve{B}$ and $\ve{H}$. The Fourier transform of  these expressions will give us the relations between the fields in frequency domain:
\e \freq{\ve{D}}(\omega)=\varepsilon_0 \freq{\varepsilon}(\omega) \freq{\ve{E}}(\omega),\qquad \freq{\ve{B}}(\omega)=\mu_0 \freq{\ve{H}}(\omega), \l{material}\f 
where 
\e \freq{\varepsilon}(\omega)=\int_0^\infty \time{\varepsilon}(\tau) {\rm e}^{-j\omega \tau} {\rm d}\tau \label{eps_def},\f 
and $\varepsilon_0$ and $\mu_0$ are the permittivity and permeability of vacuum.

By applying the Fourier transform to  both sides of the Maxwell equations~\r{eq12} and substituting~\r{material}, we obtain their frequency-domain version:
\e\begin{array}{c}\displaystyle
\nabla\times \freq{\ve{E}} (\omega, \_r)=- j\omega  \mu_0 \freq{\ve{H}}(\omega, \_r), \quad 
\nabla \cdot \left[ \freq{\varepsilon}(\omega, \_r) \freq{\ve{E}}(\omega, \_r) \right] =0, \\\displaystyle
\nabla\times \freq{\ve{H}}(\omega, \_r)= j\omega \varepsilon_0 \freq{\varepsilon}(\omega, \_r) \freq{\ve{E}}(\omega, \_r), \quad 
\nabla \cdot \freq{\ve{H}}(\omega, \_r) =0,
\end{array}\l{max1}
\f
where  the free currents and charge densities are assumed to be zero. Using conventional algebraic manipulations, we obtain the following wave equation in terms of the magnetic field:
\e \nabla\times \left[\frac{1}{\freq{\varepsilon}(\omega, \_r)} \nabla \times \freq{\_H}(\omega, \_r) \right]= \frac{\omega^2}{c^2} \freq{\_H}(\omega, \_r).  \l{waveeq}\f 
An additional requirement for the magnetic field is dictated by condition $\nabla \cdot \freq{\ve{H}}(\omega, \_r) =0$. Note that condition $\nabla \cdot \left[ \freq{\varepsilon}(\omega, \_r) \freq{\ve{E}}(\omega, \_r) \right] =0$ is satisfied automatically due to Faraday's law in~\r{max1} and the fact that the divergence of a curl is always zero.

Next, let us investigate how   time reversal applies to the  magnetic field $\time{\_H}$ in the frequency domain. The frequency spectrum of the field in the original process is given by  the Fourier transform
\e \freq{\ve{H}}(\omega, \_r)=1/(2\pi)\int_{-\infty}^{+\infty} \time{\_H}(t,\_r) {\rm e}^{-j \omega t} {\rm d}t.
\l{thor}
\f 
Under time reversal, the field in the time domain transforms as $T \{ \time{\_H}(t,\_r) \}=-\time{\_H}(-t,\_r)$ (see Table~I). The Fourier transform of the field in the reversed process is 
\e \begin{array}{c} \displaystyle
T \{ \freq{\ve{H}}(\omega, \_r)\}=1/(2\pi)\int_{-\infty}^{+\infty} -\time{\_H}(-t,\_r) {\rm e}^{-j \omega t} {\rm d}t 
 \vspace{1mm} \\ \displaystyle
= - 1/(2\pi)\int_{-\infty}^{+\infty} \time{\_H}(t,\_r) {\rm e}^{j \omega t} {\rm d}t.
\end{array}
\l{thor2} \f
By comparing \r{thor} and \r{thor2}, we obtain
% we see  that time reversal implies $T \{ \freq{\ve{H}}(\omega, \_r) \}=1/(2\pi)\int_{-\infty}^{+\infty} \time{\_H}(-t,\_r) {\rm e}^{j \omega t} dt$ (here $T$ is the operator of time reversal). Taking into account that $\time{\_H}(-t,\_r)=-\time{\_H}(t,\_r)$ (see Table~I), we observe that   time reversal of the magnetic field  results in the frequency domain  in  complex conjugation of the field amplitude with additional sign flip, that is 
\e T \{ \freq{\ve{H}}(\omega, \_r) \}=-\freq{\ve{H}}(-\omega, \_r)= -\freq{\ve{H}}^\ast (\omega, \_r).
\l{tr}\f 
For time-even fields,  time reversal results in complex conjugate without the sign flip, i.e. for electric field  
\e T \{ \freq{\ve{E}}(\omega, \_r) \}=\freq{\ve{E}}(-\omega, \_r)= \freq{\ve{E}}^\ast (\omega, \_r).
\l{tr101}\f 
Next, applying time reversal to both sides of~\r{material}, we conclude that the  time-reversal symmetry of field equations (here, including material relations)  dictates the following rule for time-reversal of the complex permittivity:
% Likewise, from the definition \eqref{eps_def} we conclude that time reversal \emph{mathematically} corresponds to complex conjugation of the complex permittivity:
\e T \left\{\freq{\varepsilon}(\omega)\right\} =\freq{\varepsilon}^*(\omega). \l{tr11} \f 
It should be noted that the same results were reported in~\cite[\textsection~VIII]{caloz_electromagnetic_2018}.
The expression in \r{tr11} implies that under time reversal lossy media become active and vice versa. This should not be surprising since time reversal involves the global reversal of the process with correct {\it microscopic} initial conditions. If the direct process was lossy, in the reversed process, the phonons of the dielectric lattice will oscillate in such a way that their energy will be transformed back into the energy of electromagnetic waves (similarly to the process in Fig.~\ref{molb}). Naturally, this exact process reversal is impossible in practice, due to the vast number of microscopic conditions to be satisfied. We are able to reverse only the macroscopic conditions,  sending the wave in the opposite direction without reversing the lattice vibrations. Then the dielectric permittivity  will remain lossy and the reversed process will be different from the original one. 
% Here it should be noted that relation~\r{tr11} physically does not hold for lossy materials due to the growing entropy in both original and reversed systems.   
% [THIS PART IS A BIT DISCONNECTED AND EXPLANATIONS ARE TOO FAST. IT CONCERNS $ j\omega$ and ${\rm e}^{-j \_k\cdot  \_r}$ and ${\rm e}^{+j \_k^\ast\cdot  \_r}$.]
%%
% On the other hand, we note that derivative with respect to time corresponds, in the frequency domain, to multiplication by $j\omega$. Thus, in the frequency-domain equations, time reversal corresponds\footnote{Note that for all real (in time domain) physical quantities we have $\tilde{\ve{H}}^\ast(\omega, \_r)= \tilde{\ve{H}} (-\omega, \_r).$} to changing sign of $\omega$.

The original and time-reversed waves described in the frequency domain by magnetic   fields $ \freq{\ve{H}}(\omega, \_r)$ and $T \{ \freq{\ve{H}}(\omega, \_r) \}=-\freq{\ve{H}}^\ast(\omega, \_r) $, respectively,   propagate in the opposite directions. This can be  shown on the example of plane wave propagation so that $ \freq{\ve{H}}(\omega, \_r)= \_H_0 {\rm e}^{-j \_k(\omega) \_r}$, where $\_k(\omega)$ is the wavevector and $\_H_0$ denotes  real vector. The time-harmonic magnetic field for these two waves would be $\time{\_H}(t,\_r)=\_H_0 \Re ({\rm e}^{-j \_k \_r} {\rm e}^{ j \omega t})$ and $T \{\time{\_H}(t,\_r) \}= -\_H_0 \Re ({\rm e}^{j \_k^\ast \_r} {\rm e}^{ j \omega t})$, respectively. Due to the different signs in front of the wavevectors, the propagation directions of these two waves are opposite. 
% since they have different dependencies ${\rm e}^{-j \_k\cdot  \_r}$ and ${\rm e}^{+j \_k^\ast\cdot  \_r}$ (the time dependence is ${\rm e}^{j \omega t}$ for both cases, because both $t$ and $\omega$ flip sign).
Next, we will discuss in detail time reversal in two different characteristic groups of materials:  Dielectric and magneto-optical materials.

\subsection{Time reversal of wave propagation in a dielectric material}\label{diel} 

%\red{Let us consider   wave propagation in the time-reversed version of  a dielectric material. Since time reversal implies that all microscopic time-odd quantities flip sign and time-even quantities remain the same, effectively a lossy medium would be transformed into an active one, and vice versa. Mathematically, it means that the time-reversed version of the dielectric is described by complex conjugate of the original permittivity $\freq{\varepsilon}_{\rm micr}(\omega, \_r)= T\{ \freq{\varepsilon}_{\rm orig}(\omega, \_r) \}= \freq{\varepsilon}_{\rm orig}^\ast(\omega, \_r)$. It is easy to see that if $\freq{\_H}_{\rm orig}(\omega, \_r)$ is the field solution in the original material
%\e \nabla\times \left[\frac{1}{\freq{\varepsilon}_{\rm orig}(\omega, \_r)} \nabla \times \freq{\_H}_{\rm orig}(\omega, \_r) \right]= \frac{\omega^2}{c^2} \freq{\_H}_{\rm orig}(\omega, \_r),  \l{original}
%\f 
%then $\freq{\ve{H}}_{\rm micr}(\omega, \_r) = T \{ \freq{\ve{H}}_{\rm orig}(\omega, \_r) \}=-\freq{\ve{H}}_{\rm orig}^\ast(\omega, \_r) $ is the field solution in the time-reversed material
%\e \nabla\times \left[\frac{1}{\freq{\varepsilon}_{\rm micr}(\omega, \_r)} \nabla \times  \freq{\_H}_{\rm micr}(\omega, \_r)  \right]= \frac{\omega^2}{c^2} \freq{\_H}_{\rm micr}(\omega, \_r).  \l{micro2}
%\f 
%which implies that magnetic field functions $\freq{\_H}_{\rm orig}$ and $\freq{\_H}_{\rm micr}$ are equal up to a constant multiplier.   
%}

\begin{figure*}[tb]
	\centering
	\includegraphics[width=0.98\linewidth]{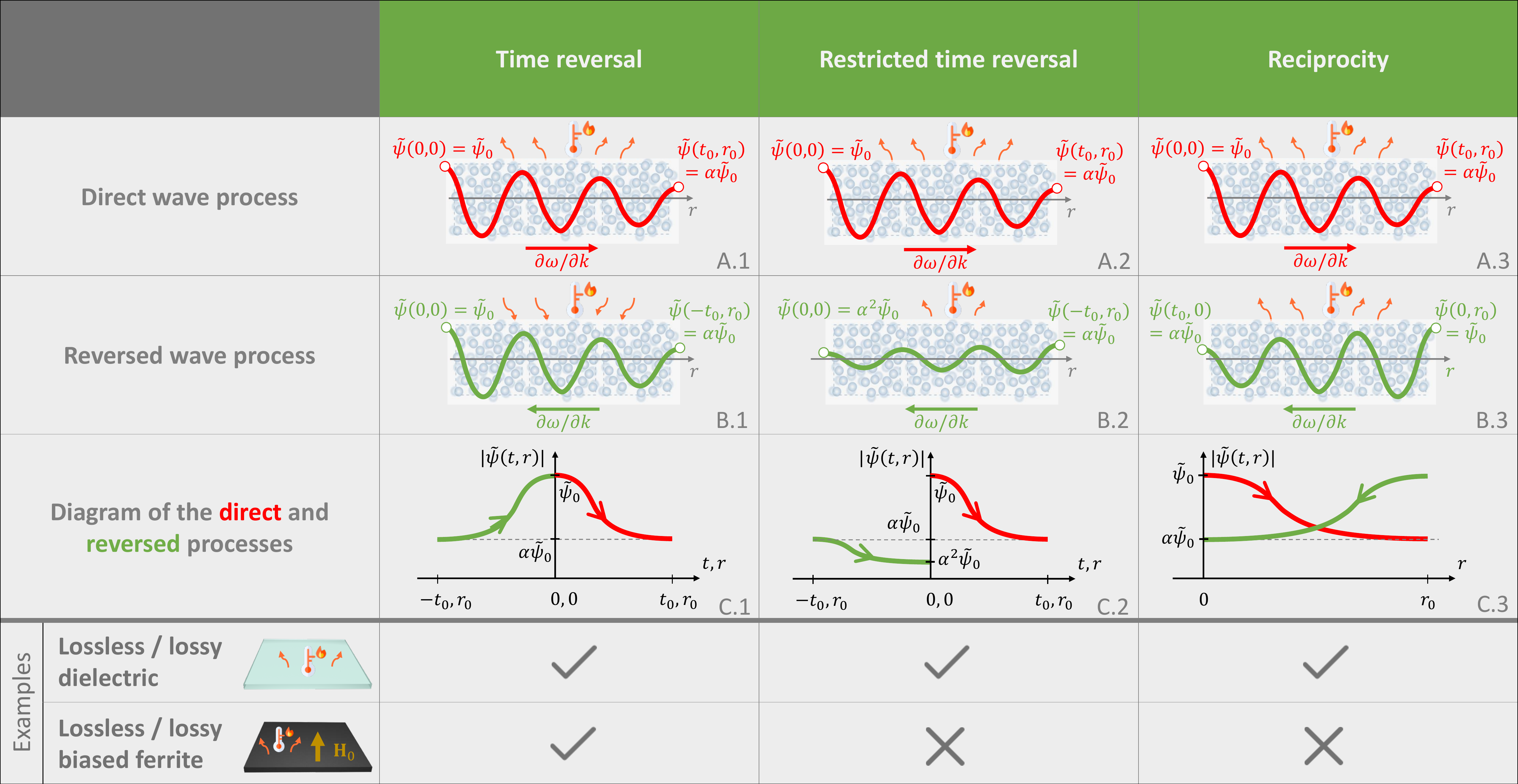}
	\caption{Table comparing the concepts of time reversal, restricted time reversal, and reciprocity for several classes of materials (for simplicity, we assume homogeneous materials). Red and green curves represent the general time-even (e.g., electric field) field function $\psi(t,r)$  of the direct and reversed wave processes, respectively.  For a time-odd field function (e.g., magnetic field), the green curves in cells B.1--B.3 must be additionally flipped along the vertical direction. }
	\label{fig3}
\end{figure*}

Let the original wave propagation process  in a dielectric material with some complex permittivity $\freq{\varepsilon}(\omega, \_r)=\freq{\varepsilon}'-j\freq{\varepsilon}''$ be described  by   wave equation~\r{waveeq} written as  
\e \nabla\times \left[\frac{1}{\freq{\varepsilon}(\omega, \_r)} \nabla \times \freq{\_H}_{\rm orig}(\omega, \_r) \right]= \frac{\omega^2}{c^2} \freq{\_H}_{\rm orig}(\omega, \_r).  \l{original}
\f 
Next, consider   wave propagation in the time-reversed version of the dielectric material. Since time reversal implies that all microscopic time-odd quantities flip sign and time-even quantities remain the same, effectively a lossy medium would be transformed into an active one, and vice versa. Mathematically, it means that the time-reversed version of the dielectric is described by complex conjugate of the original permittivity $\freq{\varepsilon}^\ast(\omega, \_r)$.
 By applying complex conjugate to both sides of~\r{original} and taking into account~\r{tr}, it is  straightforward to see  that the wave solution in the time-reversed medium corresponds to $T\{{\_H}_{\rm orig}\}$. 
% Such wave has the same waveform as the original one but propagates in the opposite direction. 
% Since we are looking for wave propagation in the direction opposite to the original one, we can demand that the obtained  field solution must correspond to some yet unknown time-reversed field  $T \left\{\freq{\_H}_{\rm micr}(\omega, \_r)\right\}$ \red{(next it will be shown that $\freq{\_H}_{\rm micr}$ and $\freq{\_H}_{\rm orig}$ are equivalent):}
% \e \nabla\times \left[\frac{1}{\freq{\varepsilon}^\ast(\omega, \_r)} \nabla \times T \left\{\freq{\_H}_{\rm micr}(\omega, \_r)\right\} \right]= \frac{\omega^2}{c^2} T \left\{\freq{\_H}_{\rm micr}(\omega, \_r)\right\}.  \l{micro}
% \f 
% Here, notice that ${\_H}_{\rm{micr}}$ is  the time reversal of the field $T\{{\_H}_{\rm micr}\}$. The latter one  possesses the two mentioned properties: Firstly, it satisfies the above equation, and secondly, it propagates in the opposite direction (compared to the original field ${\_H}_{\rm{orig}}$).
% By applying complex conjugate to both sides of this equation and using~\r{tr}, we obtain
% \e \nabla\times \left[\frac{1}{\freq{\varepsilon}(\omega, \_r)} \nabla \times  \freq{\_H}_{\rm micr}(\omega, \_r)  \right]= \frac{\omega^2}{c^2} \freq{\_H}_{\rm micr}(\omega, \_r),  \l{micro2}
% \f 
% which implies that magnetic field functions $\freq{\_H}_{\rm orig}$ and $\freq{\_H}_{\rm micr}$ are equal up to a constant multiplier.   
Thus, the original and time-reversed (microscopically) waves propagating in  a dielectric material  have the same waveform  but opposite propagation directions, as illustrated  in cells~A.1 and B.1 of the table in Fig.~\ref{fig3}. Wave function $\time{\psi}(t,r)$ in the table represents a general time-even field quantity (magnetic field is time-odd and has an additional sign flip). Cell~C.1 shows   this function versus time and coordinate. The time-reversed field function is just a mirror copy of the direct field function with respect to the point $t=0$. 

It is important to mention   that  the presented definition of time-reversal symmetry differs from that used in~\cite[\textsection~XII]{caloz_electromagnetic_2018}. In particular, our definition corresponds to microscopic reversal and, therefore, lossy dielectric materials are considered time-reversal symmetric. On the contrary, in~\cite{caloz_electromagnetic_2018} the definition is macroscopic, and lossy dielectric materials break time-reversal symmetry.

\subsection{Time reversal of wave propagation in a magneto-optical material}\label{momat} 

Magneto-optical materials are materials biased by external or internal {\it static} (sometimes, quasi-static) magnetic field, which we denote~$\_H_0$. The bias field can be created by some external magnet or by exchange interactions of the material itself (like in magnetic crystals), which aligns  permanent magnetic moments of atoms. For example, in a magnetized free-electron plasma, owing to electron cyclotron orbiting, the permittivity is described by a second-rank tensor  with non-zero antisymmetric part~\cite[in \S~8.8]{ishimaru_electromagnetic_2017} (see also the phenomenological derivation in Section~\ref{noneffects}):
\e \overline{\overline{\freq{\varepsilon}}}(\omega, \_H_0)=
\begin{pmatrix}
\freq{\varepsilon}_{\rm s}(\omega) & j  \freq{\varepsilon}_{\rm a}(\omega, \_H_0) & 0\\ 
-j \freq{\varepsilon}_{\rm a}(\omega, \_H_0) &  \freq{\varepsilon}_{\rm s}(\omega)& 0\\
0 &  0 & \freq{\varepsilon}_z(\omega)\\
\end{pmatrix}, \l{tensor}\f 
where $\freq{\varepsilon}_{\rm s}$, $\freq{\varepsilon}_{\rm a}$, and $\freq{\varepsilon}_z$ are (real-valued in the lossless case) functions and the external magnetic field is applied along the $z$-axis. Permittivity component $\freq{\varepsilon}_{\rm a}$ is a linear function of $\_H_0$.

The time-reversed version of the magneto-optical material  is described by the complex conjugate of its permittivity as in~\r{tr11} with an additional flip of the magnetic bias field~$\_H_0$, i.e. $T \left\{ \overline{\overline{\freq{\varepsilon}}} (\omega, \_H_0) \right\}=
\overline{\overline{\freq{\varepsilon}}}^\ast (\omega, -\_H_0)$. This practically means changing loss to gain and reversing the direction of the bias field.
The wave equation~\r{waveeq} for magneto-optical media has the  form:
\e \nabla\times \left[\overline{\overline{\freq{\varepsilon}}}^{-1} (\omega,\_H_0) \cdot [\nabla \times \freq{\_H}_{\rm orig}(\omega,  \_H_0)] \right]= \frac{\omega^2}{c^2} \freq{\_H}_{\rm orig}(\omega,  \_H_0).  \l{waveeqmo}\f 
%Note the additional minus sign appearing in the latter equality. This minus sign is due to the fact that time reversal   flips the sign of the static magnetic field bias    $\_H_0 \rightarrow -\_H_0$ and $\freq{\varepsilon}_{\rm a}$ is a linear function of $\_H_0$.  
By applying complex conjugate to both sides of~\r{waveeqmo} with additional flipping the sign of $\_H_0$ and taking into account $ T \{ \freq{\ve{H}}(\omega,   \_H_0) \} = -\freq{\ve{H}}^\ast (\omega, -\_H_0)$ (similar to~\r{tr}), we see  that the wave solution in the time-reversed magneto-optical material corresponds to $T\{{\_H}_{\rm orig}\}$ (same waveform but opposite direction).
% Next, if we consider wave propagation in the time reversed magneto-optical material in the  direction opposite to the original one, similarly to \r{micro} and \r{micro2}, we would obtain that it has the same waveform.
We should stress that the perfect inversion of wave propagation process in a magneto-optical material is the consequence of the definition of time reversal used in this tutorial. According to this definition, time reversal acts ``globally'' on the system and all the sources  external to it.  Nevertheless, in the literature one can find an alternative definition which implies time reversal of only the system itself.
% Indeed, any source of static
% magnetic field must include some circulating direct currents which maintain this field: It can be an electromagnet with a solenoid coil or natural
% magnets consisting of ordered microscopic current loops (magnetic dipole
% moments) formed by rotating electrons. Thus, under time reversal, the currents which form the static magnetic field
% will change their directions and the magnetization field~$\_H_0$ will be reversed.

\section{Restricted time reversal  }\label{restricted}

% As we saw above, the microscopic field equations as well as the macroscopic equations, which connect the four vectors $\_E$, $\_D$, $\_H$, and $\_B$ are symmetric with respect to time reversal. However, electromagnetic fields are coupled to other, external with respect to the considered system of charges, currents, and fields, objects. Energy dissipation in lossy systems means that the electromagnetic fields (which obey our time-even system of equations) excite mechanical oscillations (phonons) of the medium. Mechanical movements of atoms are governed by their own equations, and the complete description of the system should include also equations for all other processes which can couple to our electromagnetic fields. %
% Full time reversal of all equations, both in the Maxwell equations and in the equations for all other coupled  systems will apparently lead to time reversal of all electromagnetic properties (provided that all the other equations are also time-symmetric).

As it was discussed in the previous sections, most physical laws are time-reversal symmetric. Although the traditional definition of time reversal (satisfying the microscopic initial conditions) is crucially important for many branches of physics, especially quantum field theory, it is in practice not easily applicable for classical electrodynamics.  Indeed, the main subject of study in classical electrodynamics are macroscopic systems and processes. The reversal of such processes is usually understood in the macroscopic sense (satisfying only the macroscopic initial conditions). In this framework, lossy materials remain lossy for the reversed process, and  wave processes in lossy materials appear irreversible. 
Therefore, it is useful to consider an alternative  notion of \emph{restricted time reversal}~\cite{altman_generalization_1991,dmitriev_space-time_2004},\cite[in \S~7]{altman_reciprocity_2011}. Under this transformation, it is assumed that  the time in Maxwell's equations of the considered system is reversed, but the time in equations governing all other processes which are coupled to the electromagnetic system under study (such as equations of motion of atoms in materials) is not reversed. Moreover,  the external bias fields are not reversed. In this scenario, considering time-reversed processes, only macroscopic initial conditions are considered and properly reversed. The electromagnetic processes remain dissipative under the restricted time reversal (loss is not transformed into equivalent gain). Most importantly, since all the laws of classical physics are time-symmetric, all the dissipation processes will be governed by exactly the same laws after restricted time reversal, including the formulas for calculating  dissipated power. 
%\blue{Thus, although the restricted time reversal   does not guarantee  reversibility of wave propagation, it implies that the wave returns back with the same waveform and with the same phase, only the field amplitude might differ from the original value.}

As will be mentioned in Section~\ref{onscas}, media which are symmetric (do not change) under restricted time reversal satisfy the same conditions for material parameters as those dictated by the Lorentz reciprocity theorem. Thus, restricted time reversal is strongly connected to the  notion of electromagnetic reciprocity.

\subsection{Restricted time reversal of wave propagation in a dielectric material}\label{dielrestricted} 
Let us consider wave propagation in a dielectric material and show that it is symmetric with respect to the restricted time reversal.  
Under such reversal, the dielectric material remains unchanged with the same dielectric function $\freq{\varepsilon}(\omega, \_r)$. 
% \red{
% Since we are looking for wave propagation in the direction opposite to the original one, we can demand that the obtained  field solution  (in this  opposite direction) must correspond to some yet unknown time-reversed field  $T \left\{\freq{\_H}_{\rm micr}(\omega, \_r)\right\}$ \red{(next it will be shown that $\freq{\_H}_{\rm micr}$ and $\freq{\_H}_{\rm orig}$ are equivalent):}
% \e \nabla\times \left[\frac{1}{\freq{\varepsilon}^\ast(\omega, \_r)} \nabla \times T \left\{\freq{\_H}_{\rm micr}(\omega, \_r)\right\} \right]= \frac{\omega^2}{c^2} T \left\{\freq{\_H}_{\rm micr}(\omega, \_r)\right\}.  \l{micro}
% \f 
% Here, notice that ${\_H}_{\rm{micr}}$ is  the time reversal of the field $T\{{\_H}_{\rm micr}\}$. The latter one  possesses the two mentioned properties: Firstly, it satisfies the above equation, and secondly, it propagates in the opposite direction (compared to the original field ${\_H}_{\rm{orig}}$).
% By applying complex conjugate to both sides of this equation and using~\r{tr}, we obtain
% \e \nabla\times \left[\frac{1}{\freq{\varepsilon}(\omega, \_r)} \nabla \times  \freq{\_H}_{\rm micr}(\omega, \_r)  \right]= \frac{\omega^2}{c^2} \freq{\_H}_{\rm micr}(\omega, \_r),  \l{micro2}
% \f 
% which implies that magnetic field functions $\freq{\_H}_{\rm orig}$ and $\freq{\_H}_{\rm micr}$ are equal up to a constant multiplier.   
% }
Since we are looking for wave propagation in the direction opposite to the original one, we can demand that the obtained  field solution  (in this  opposite direction) must correspond to some yet unknown time-reversed field  $T \left\{\freq{\_H}_{\rm macr}(\omega, \_r)\right\}$ (here the subscript denotes the time reversal operation in which  the macroscopic initial conditions are reversed):
\e \nabla\times \left[\frac{1}{\freq{\varepsilon}(\omega, \_r)} \nabla \times T \left\{\freq{\_H}_{\rm macr}(\omega, \_r)\right\} \right]= \frac{\omega^2}{c^2} T \left\{\freq{\_H}_{\rm macr}(\omega, \_r)\right\}.  \l{macro}
\f 
Here, notice that ${\_H}_{\rm{macr}}$ is  the time reversal of the field $T\{{\_H}_{\rm macr}\}$. The latter one  possesses the two mentioned properties: Firstly, it satisfies the above equation, and secondly, it propagates in the opposite direction (compared to the original field ${\_H}_{\rm{orig}}$).
By applying complex conjugate to both sides of this equation and using~\r{tr}, we obtain
\e \nabla\times \left[\frac{1}{\freq{\varepsilon}^\ast(\omega, \_r)} \nabla \times  \freq{\_H}_{\rm macr}(\omega, \_r)  \right]= \frac{\omega^2}{c^2} \freq{\_H}_{\rm macr}(\omega, \_r).  \l{macro2}
\f 
It is seen that wave equation \r{macro2} differs from the original \r{original}. Let us for simplicity assume that the considered material is homogenious, i.e. the permittivity does not depend on the coordinate~$\_r$.  Then from \r{original} and \r{macro2}, we can readily deduce the wave equations in the traditional form
\e \begin{array}{c}\displaystyle
\left[\nabla^2  + \frac{\omega^2}{c^2} \freq{n}^2(\omega) \right] \freq{\_H}_{\rm orig}(\omega, \_r)=0,   \vspace{1mm} \\ \displaystyle
\left[\nabla^2  + \frac{\omega^2}{c^2} \freq{n}^{\ast 2}(\omega) \right] \freq{\_H}_{\rm macr}(\omega, \_r)=0,
\end{array}\l{macro3}
\f 
where $\freq{n}(\omega)=\freq{n}'-j \freq{n}''$ denotes the complex refractive index for which $\freq{n}^2(\omega)=\freq{\varepsilon}(\omega)$ and $\freq{n}^{\ast 2}(\omega)=\freq{\varepsilon}^\ast(\omega)$. The field solutions of \r{macro3} are given by
\e 
\freq{\_H}_{\rm orig}(\omega, \_r)= {\rm e}^{-\freq{n}'' r \omega/c } {\rm e}^{-j \freq{n}' r \omega/c }, \l{macro4} \f
\e 
\freq{\_H}_{\rm macr}(\omega, \_r)= {\rm e}^{\freq{n}'' r \omega/c } {\rm e}^{-j \freq{n}' r \omega/c },
\l{macro5} \f
where we denoted $r=|\_r|$. 
Recalling that the reversed field was defined as $T \left\{\freq{\_H}_{\rm macr}(\omega, \_r)\right\}$ and using \r{tr}, we obtain
\e 
T \left\{\freq{\_H}_{\rm macr}(\omega, \_r)\right\}= - {\rm e}^{\freq{n}'' r \omega/c } {\rm e}^{j \freq{n}' r \omega/c }.
\l{macro6} \f 
By comparing \r{macro4} and \r{macro6}, we see that the original and reversed waves propagate in the opposite directions with the same phase $\freq{n}' r \omega/c$ and attenuation constant  $\freq{n}''\omega/c$. 
The illustration of this wave propagation is shown in Fig.~\ref{fig3} in cells A.2, B.2, and C.2. It is seen that wave attenuates during propagation from $t=0$ to $t=t_0$ by the same ratio $\alpha$ as during propagation from $t=-t_0$ to $t=0$. The phase and polarization of the reversed and original waves are equal at $t=0$. Thus, lossy dielectric materials are symmetric under restricted time reversal.

\subsection{Restricted time reversal of wave propagation in a magneto-optical material}\label{morestricted} 
Let us consider wave propagation in a magneto-optical material and show that it is asymmetric with respect to the restricted time reversal.  
Under such reversal (only the macroscopic initial conditions are satisfied), the dielectric material remains unchanged with the same dielectric function $ \overline{\overline{\freq{\varepsilon}}}(\omega, \_H_0)$. Note that the direction of $\_H_0$ is not reversed. Using the same procedure as in \r{macro} and \r{macro2}, we obtain the following wave equation for the reversed propagation: 
\e \nabla\times \left[(\overline{\overline{\freq{\varepsilon}}}^\ast)^{-1} (\omega,\_H_0) \cdot [\nabla \times \freq{\_H}_{\rm macr}(\omega, \_r)] \right]= \frac{\omega^2}{c^2} \freq{\_H}_{\rm macr}(\omega, \_r).  \l{waveeqmo5}\f 
Comparing \r{waveeqmo} and \r{waveeqmo5}, one can see that field functions $\freq{\_H}_{\rm orig}$ and $\freq{\_H}_{\rm macr}$ are eigenfunction of different equations and, therefore, have different waveform. Importantly, even assuming the lossless magneto-optical material ($\freq{\varepsilon}_{\rm s}$ and $\freq{\varepsilon}_{\rm a}$ are purely real), the dielectric function are not equal $\overline{\overline{\freq{\varepsilon}}} \neq \overline{\overline{\freq{\varepsilon}}}^\ast$, resulting in  $\freq{\_H}_{\rm orig}$ and $\freq{\_H}_{\rm macr}$ having different waveform.  Thus, magneto-optical materials are  asymmetric under restricted time reversal.

\section{Reciprocity and nonreciprocity}
In the two previous sections, we have described the concepts of time reversal and restricted time reversal and demonstrated their applicability on several example materials. As it will be shown below, the concept of reciprocity  is closely related  with that of restricted time reversal. For time-invariant systems (whose properties do not change with time), pointwise (i.e., at each point) reciprocity holds if   the restricted time reversal does not change the system, and vice versa.
%\red{On the other hand, the reverse statement is not always true, as will be demonstrated below for systems which are overal reciprocal but pointwise nonreciprocal.  }
However, while the time-inversion concept is intrinsically theoretical and implies process inversion with correct macroscopic initial conditions, the reciprocity principle  can be easily  applied to real systems and requires only  interchanging of the source and detector locations.

\subsection{The Onsager reciprocal relations } \label{onsrec}

As it was discussed in Section~\ref{Loschmidt}, due to the time symmetry of most physical laws, all   processes governed by these laws are time-reversal symmetric on the microscopic level. In 1931, L.~Onsager, using this microscopic reversibility, derived his famous reciprocal relations for lossy linear structures (where the processes are irreversible)~\cite{onsager_reciprocal_1931, onsager_reciprocal_1931-1}. These relations, referred  sometimes as ``the fourth law of thermodynamics'' due to their universality, can be applied to the enormous variety of physical phenomena since they were derived using  only four basic assumptions: \textit{Microscopic reversibility} (holds even in lossy systems; equivalent to the definition of time reversal given above), \textit{linearity}, \textit{causality}, and \textit{thermodynamic  quasi-equilibrium}. Below we shall outline the derivation of the Onsager reciprocal relations, their generalization by other authors, and applications to several phenomena. 

\begin{figure}[tb]
\centering
   \includegraphics[width=0.62\columnwidth]{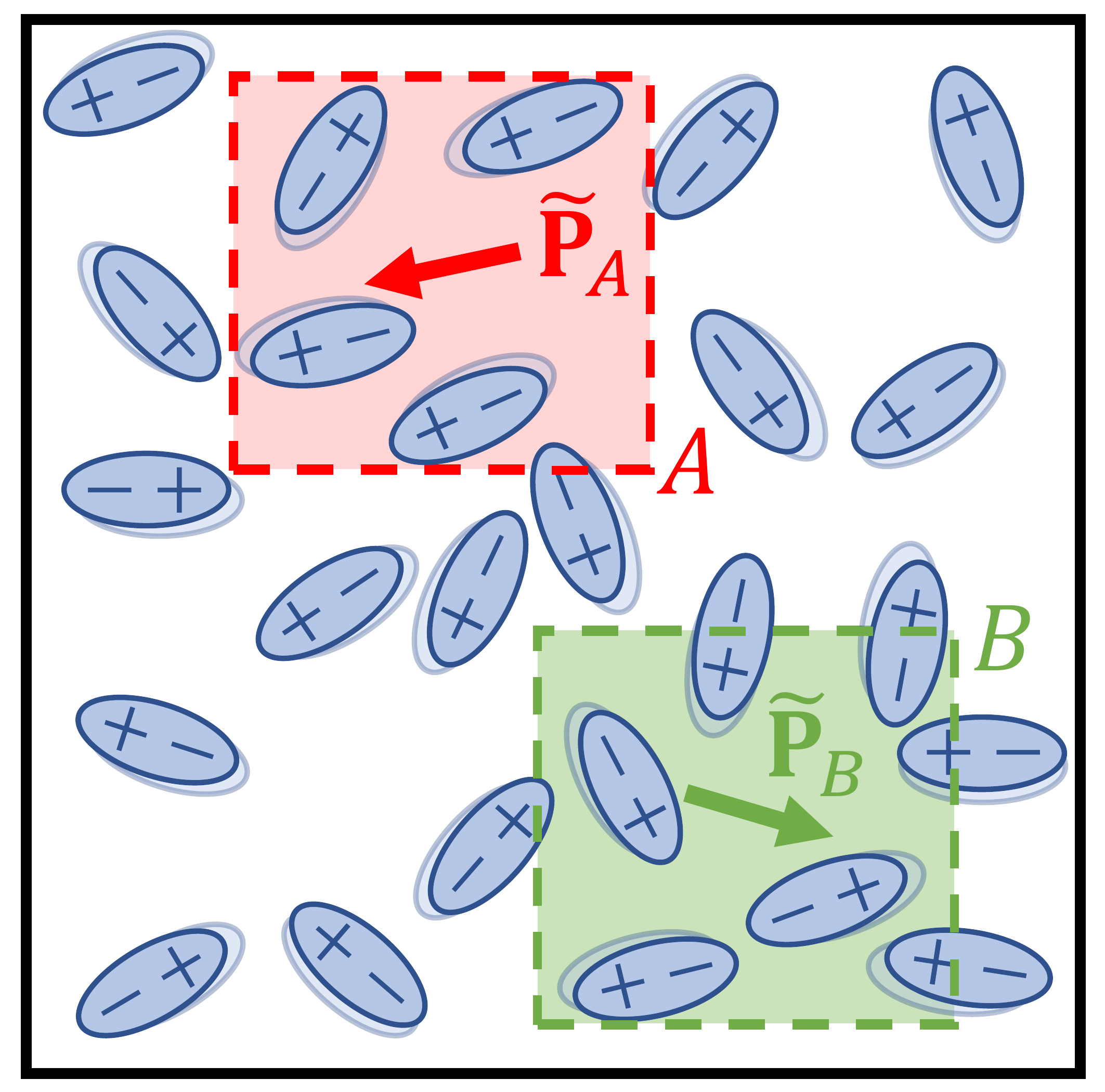}
\caption{ Fluctuations of electric polarization $\time{\_P}$ in   bulk dielectric. Although there is no external electric field, the polarization continuously and randomly changes due to the thermal jiggling of polar molecules. Two arbitrary regions $A$ and $B$ are shown.   }
\label{dielectric}
\end{figure}

According to quantum statistical mechanics, any system   in the equilibrium state undergoes fluctuations (small deviations from the mean values) of its macroscopic parameters. As an example of such macroscopic system, let us consider a polar dielectric without   external applied fields  at equilibrium, illustrated in Fig.~\ref{dielectric}. Due to the continuous  jiggling motion of  molecules, electric polarization defined for an arbitrary macroscopic region~$A$   fluctuates over time around zero value (dielectric is neutral and no electric field is applied). The polarization fluctuations  $\time{P}_A$ at region~$A$ are different at each moment from the polarization fluctuations $\time{P}_B$ at region~$B$ (here we consider  the polarization along some arbitrary direction).    
Importantly, the  fluctuations in regions~$A$ and $B$ are not independent, due to  electrostatic interactions of polar molecules. Indeed, if we consider a single molecule, it can have  any orientation with equal  probability. However, when we consider two molecules, then for a given orientation of the first, the various orientations of the second will be not equally probable (with a higher probability it will orient so that the potential  energy of interaction is minimized). This correlation of polarization fluctuations at different locations is conventionally characterized by the correlation function $\langle \time{P}_A \time{P}_B \rangle$, which implies, basically,  averaging with respect to probabilities of various values of $ \time{P}_A$ and $\time{P}_B$~\cite[\textsection~116]{landau_course_1980}. To verify that this correlation function makes sense, one can consider the case when  $ \time{P}_A$ and $\time{P}_B$ can have arbitrary values independently. Then for any given $\time{P}_{A,i}$,  $\time{P}_B$ can be positive and negative with the same probability and summation  $ \sum_{j}  (\time{P}_{A,i} \time{P}_{B,j})=\time{\Xi}_i$ will be zero (here and below, repeating indices imply summation according to the Einstein notation). The correlation function in this case $\langle \time{P}_A \time{P}_B \rangle = \sum_{i}  \time{\Xi}_i = 0$.

In addition to the spatial  correlation of fluctuations, one can analogously define temporal correlations. Moreover, correlation can be between fluctuations of different macroscopic quantities, e.g. electric polarization and displacement of heat:  $\langle \time{P}_A(t) \time{\gamma}_B(t+ \tau) \rangle \neq 0$ (here   $\time{\gamma}_B$ defines the   deviation from equilibrium along a given direction in region~$B$, such as fluctuation of temperature in space~\cite[Eq.~(4.3)]{onsager_reciprocal_1931} and $\tau$ is the time delay between the two fluctuations).   Lars Onsager recognized  the fact that due to microscopic reversibility, some specific  polarization $\time{P}=\time{P}_0$, followed  $\tau$ later by some specific heat displacement $\time{\gamma}=\time{\gamma}_0$, must occur just as often as the displacement $\time{\gamma}=\time{\gamma}_0$, followed $\tau$ later by the  polarization $\time{P}=\time{P}_0$~\cite[Eq.~(4.10)]{onsager_reciprocal_1931}:
\e 
\langle \time{P}_A(t) \time{\gamma}_B(t+\tau) \rangle = \langle \time{P}_A(t+\tau) \time{\gamma}_B(t) \rangle.
\nonumber 
%\l{ons0} 
\f
The same equation written for  fluctuations of   general  macroscopic quantities $\time{x}_i$ and $\time{x}_k$ read
\e 
\langle \time{x}_i(t) \time{x}_k(t+\tau) \rangle = \sigma \langle \time{x}_i(t+\tau) \time{x}_k(t) \rangle,
\l{ons1} 
\f
where $\sigma=1$ if   quantities $\time{x}_i$ and $\time{x}_k$ have the same time-reversal  symmetry (see Table~\ref{tab:1}) and $\sigma=-1$ if they have the opposite symmetry~\cite[\textsection~119]{landau_course_1980}. In the frequency domain, relation~\r{ons1}  can be written as~\cite[see Eq.~(122.11)]{landau_course_1980}
\e 
\freq{x}_i  \freq{x}_k = \sigma \freq{x}_k  \freq{x}_i,
\l{ons2} 
\f
where definition $\freq{x}_i  \freq{x}_k =  \int_{-\infty}^{\infty}  \langle \time{x}_i(t+\tau) \time{x}_k(t) \rangle  {\rm e}^{-j\omega \tau} {\rm d}\tau$ was used.

Relations \r{ons1}  and \r{ons2}  indicate constraints on {\it fluctuations of physical quantities in the equilibrium}  imposed by the microscopic time reversibility. Next, we need to  determine what constraints are imposed by microscopic reversibility on {\it stationary  processes   under small external perturbations}. Stationary processes are processes during which the system can be considered  near thermodynamic equilibrium, i.e. in quasi-equilibrium; there are no net macroscopic flows of energy) at each moment of time. In the presence of an external perturbation, a physical quantity  $\time{x}_i(t)$ in addition to the fluctuations  acquires some nonzero mean value $\overline{\time{x}}_i(t)$:
\e 
\overline{\time{x}}_i(t)  =\int_{0}^{\infty}  \time{\alpha}_{i k} (\tau) \time{f}_k (t-\tau) {\rm d}\tau.
\l{ons3} 
\f
In this relation  $\time{\alpha}_{i k}$ is the so-called generalized susceptibility tensor which relates the response of the system $\overline{\time{x}}_i(t)$ to the generalized forces $\time{f}_{k}(t)$~\cite[Eq.~(125.2)]{landau_course_1980}. Note that integration in \r{ons3} extends from 0 to $+\infty$, rather than from $-\infty$ to $+\infty$, due to the causality principle applicable to all physical processes (see the beginning of Section~\ref{trmatrel}). One can see that one special case described by  \r{ons3}  is material relation~\r{causal}, where the role of the generalized forces is played by the three vectorial components of the  electric field and the electric displacement vector is  the response function. Relation~\r{ons3} is applicable to all linear causal perturbation processes.  
%in the equilibrium state in the absence of perturbations, $\overline{x}_i(t)=0$ 

The relation between fluctuations and perturbation processes is given by the
fluctuation-dissipation theorem~\cite{callen_irreversibility_1951}, \cite[\textsection~125]{landau_course_1980}:
\e 
\freq{x}_i  \freq{x}_k  = \frac{-j \hbar}{2} (\freq{\alpha}_{k i}^\ast-\freq{\alpha}_{i k}) \coth{\frac{\hbar \omega}{2 k_{\rm B} T}},
\l{ons4} 
\f
where $k_{\rm B}$ is the Boltzmann constant, $\hbar$ is the reduced Planck constant, and $T$ is the temperature.  
The theorem  states that thermal fluctuations of some macroscopic quantity in a system \textit{in thermal equilibrium} (the left-hand side) have the same nature as the dissipation processes  related to this quantity in the system \textit{in thermal quasi-equilibrium} (the brackets on the right-hand side). Applied to the electric polarization, the theorem implies that the intensity of the polarization fluctuations in the material   is proportional to the imaginary part of its permittivity which is responsible for dissipation of energy in the material. Thus, if there is a process accompanied by energy dissipation into heat, there should exist  a reversed process  which converts    heat into thermal fluctuations.
Other examples of such dual processes include loss in electrical resistance and Johnson noise, air resistance and Brownian motion, etc.

Substituting \r{ons4} into   both sides of Eq.~\r{ons2}, one obtains relation
\e 
 \freq{\alpha}_{k i}^\ast-\freq{\alpha}_{i k}= \sigma (\freq{\alpha}_{i k }^\ast-\freq{\alpha}_{k i }),
\l{ons5} 
\f
which   together with the Kramers-Kronig formulae results in~\cite[Eq.~(125.13)]{landau_course_1980}
\e 
 \freq{\alpha}_{k i}(\omega)= \sigma \freq{\alpha}_{i k}(\omega).
\l{ons6} 
\f
Note that in Eqs.~\r{ons5}--\r{ons6},   for every combination of $i$ and $k$ indices, parameter $\sigma$ should be chosen either $+1$  when the response quantities $\time{x}_i$ and $\time{x}_k$ have the same symmetry   under time-reversal or $-1$ when they  have the opposite symmetry. 
Relations~\r{ons6}, stemming from microscopic reversibility conditions~\r{ons1}, are referred to as the Onsager reciprocal relations. They impose a fundamental restriction on the generalized susceptibility tensors of {\it arbitrary nature}. If the relations are satisfied, the system is called {\it reciprocal}. When they do not hold, it is said that the system is {\it nonreciprocal}.

Subsequently, H.~Casimir pointed out that although  nonreciprocal systems, i.e. systems with external time-odd bias, such as the magnetic field, are not constrained by relations~\r{ons6}, there is another relation which they must obey. This relation reads~\cite{casimir_onsagers_1945}
% the applicability of relations~\r{ons6} can be extended also to non-reciprocal systems, i.e. systems with external time-odd bias, such as the magnetic field  (for the sake of compactness, we denote all the bias parameters as a single time-odd parameter, the magnetic field vector~$\_H_0$)~\cite{casimir_onsagers_1945}:
\e 
 \freq{\alpha}_{k i}(\omega,\_H_0)= \sigma \freq{\alpha}_{i k}(\omega,-\_H_0).
\l{ons7} 
\f
Here for the sake of compactness, we denote all the bias parameters as a single time-odd parameter, the magnetic field vector~$\_H_0$. 
Relations~\r{ons7}  are referred to as the Onsager-Casimir  relations. They cannot be used to determine whether a system is reciprocal or nonreciprocal since they   hold for either of these  cases. These relations can be applied to a variety of irreversible physical processes of different nature~\cite{miller_thermodynamics_1960}: Acoustic, electromagnetic, mechanical, thermoelectric, diffusion, etc.
In what follows,  we consider two examples of application of  the Onsager reciprocal  relations~\r{ons6}  to electromagnetic processes.

% \pink{
% He established that, in linear systems and assuming the microscopic reversibility, the coefficients relating the thermodynamic forces and  thermodynamic fluxes are symmetric. 
% In the most general terms, this condition can be expressed in term of the so-called cross kinetic coefficients; defined by the following linear system the relation
% \e \mbox{Response}_i=\alpha_{ij}\mbox{Force}_j\f 
% The symmetry of kinetic coefficients is expressed by the relation $  \alpha_{ij}=\alpha_{ji}$. This condition assumes that the parameters which characterizes the system are invariant under time inversion (see Table \ref{tab:1}).
% This principle was generalized by Casimir who extended Onsager’s considerations to parameters that change their sign under time inversion (for example in the presence of external  magnetic field or if the system rotates with an angular velocity) \cite{casimir_onsagers_1945}. With this consideration the symmetry of kinetic coefficients can be expressed in a more general way as $ \alpha_{ij}=\pm \alpha_{ji}$. The sign in this relation is defined by the direction of the external magnetic vector or the  angular velocity. The symmetry of kinetic coefficients was initially formulated for instantaneous phenomena and latter, due to the development of the fluctuation-dissipation theorem, extended more general cases. In the electromagnetic theory, this principle has important consequences in the for microscopic probabilities and macroscopic constitutive relations. }

As the first example of a physical process subject to the Onsager reciprocal relations, we examine  radiation from electromagnetic  sources. Here we assume that the sources are represented by some electric current density distribution  $\freq{\_J}(\_r)$  with the dimensions of A/${\rm m}^2$ in a general non-homogeneous and anisotropic medium. The electric field  radiated by the sources (in the frequency domain) is given by the volume integral equation
\e 
 \freq{\_E}(\_r) = \int_V \dya{\freq{G}}(\_r,\_r')\cdot \freq{\_J}(\_r') {\rm d}V',
\l{green1} 
\f
where $\dya{\freq{G}}(\_r,\_r')$ is   dyadic\footnote{A dyadic   is a second order tensor  written in a notation that fits in with vector algebra.} Green's function. For example, for an isotropic homogeneous medium with relative permeability $\freq{\mu}$ it has simple form~\cite[p.~30]{novotny_principles_2006}:
\e 
\dya{\freq{G}}(\_r,\_r') = \dya{\freq{G}}(\_r',\_r) =  -j \omega \freq{\mu} \mu_0 \left[\dya{I}+ \frac{1}{k^2} \nabla \nabla \right] \frac{{\rm e}^{-j k |\_r-\_r'|}}{4\pi |\_r-\_r'|}.
\l{green2} 
\f
% Note that previously we used tilde symbol ``$\sim$'' like in~(\ref{eps_def})  above physical quantities to distinguish the frequency-domain functions from time-domain ones. From now on, for the sake of compactness,  we omit the use of the tilde symbol  and will specify separately where the corresponding time-domain quantities are used. 
Green's function  describes how strong is an elementary electric field ${\rm d} \freq{\_E}(\_r)$ at   point $\_r$ created by an elementary  single point source $\freq{\_J}(\_r') {\rm d}V'$   at  point $\_r'$:
\e 
{\rm d} \freq{\_E}(\_r) =  \dya{\freq{G}}(\_r,\_r')\cdot \freq{\_J}(\_r') {\rm d}V'.
\l{green3} 
\f
By integrating \r{green3} over the overall volume of the source currents~$V$, one obtains~\r{green1}.
Relation~\r{green3} implies the linear (the electric field is a linear function of the current density) and  causal (electric field is the response function of the system depending on the current radiation in the past only)   process. Taking into account macroscopic reversibility of the process (no weak interactions occur in the process) and assuming that  it evolves in thermodynamic quasi-equilibrium, one can see that   Green's function satisfies all the conditions of a generalized susceptibility in \r{ons3}. Let us assume now that there are no external bias fields in the system, i.e. $\_H_0=0$ (the opposite case will be considered below; see relation~\r{green100}). 
Applying the Onsager reciprocal relations~\r{ons6} for the two infinitesimal  current sources  positioned at $\_r$ and $\_r'$, one can obtain\footnote{
In this system, the response function is the electric field~$\time{\_E}$. Interestingly, similarly to quantity $\time{x}$ in~\r{ons1}, the electric field fluctuates around the zero value  in the {\it absence} of perturbation currents~$\time{\_J}$. In this case, if we probed  the electric field with a lossless (to avoid thermal noise of the antenna itself) receiving antenna, we would measure a nonzero  fluctuating   voltage at the terminals. The source of these fluctuations   is the thermal noise of the radiation resistance on which the antenna is loaded, i.e. the ``temperature'' of infinite surrounding space. These   fluctuations include also  the quantum fluctuations (present even at zero temperature).
}
\e 
  \dya{\freq{G}}(\_r,\_r')=  \dya{\freq{G}}^T(\_r',\_r),
\l{green4} 
\f
where   $T$ denotes the transpose operator. In the derivations, the response functions $\overline{\freq{x}}_i$ and $\overline{\freq{x}}_k$ were replaced by  ${\rm d} \freq{\_E}(\_r) =  \dya{\freq{G}}(\_r,\_r')\cdot \freq{\_J}(\_r') {\rm d}V'$ and ${\rm d} \freq{\_E}(\_r') =  \dya{\freq{G}}(\_r',\_r)\cdot \freq{\_J}(\_r) {\rm d}V$, respectively.  
Note that we used the fact that parameter $\sigma$ in~\r{ons7} is equal to $+1$ since all the response quantities (the components of the electric field $\time{E}_x$, $\time{E}_y$, and  $\time{E}_z$) are time-even under time reversal. 
It can be checked that    dyadic Green's function in the   form~\r{green2}    satisfies the reciprocity relation~\r{green4}. This fact implies that the process of radiation from electromagnetic sources in a homogeneous medium   described by scalar permittivity~$\freq{\varepsilon}$ and permeability~$\freq{\mu}$ 
%(such a medium was assumed in the derivations of \r{green1}--\r{green4})
is reciprocal. 
%In the general case when the permittivity or permeability of the medium are anisotropic, relation \r{green4} must be modified. 

\begin{figure}[tb]
\centering
\subfigure[]{\includegraphics[height=0.43\columnwidth]{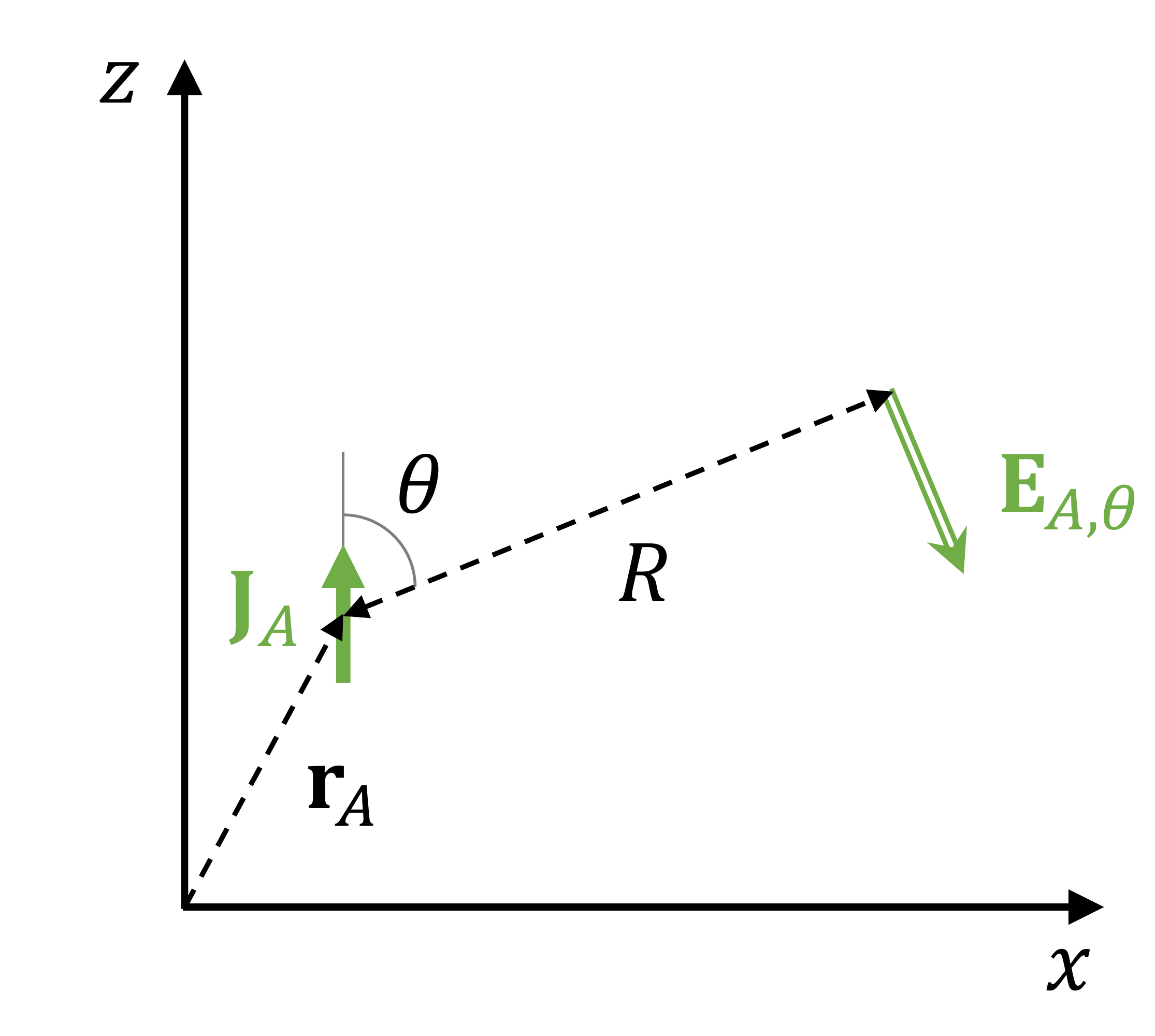} \label{fig4a}} 
\subfigure[]{\includegraphics[height=0.43\columnwidth]{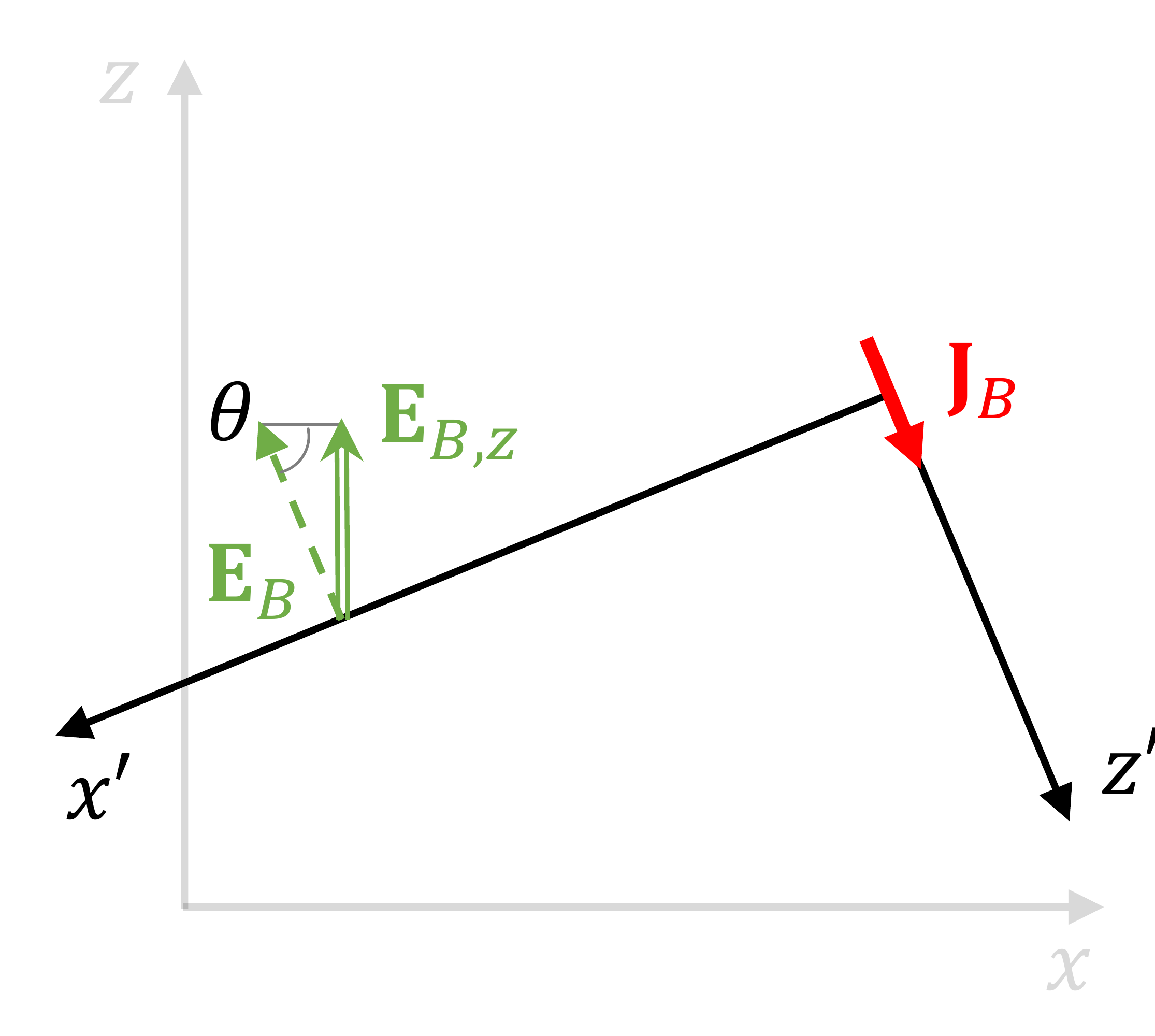} \label{fig4b}} 
\caption{A system of two sources with current densities~$\freq{\_J}_A$ and $\freq{\_J}_B$ separated by a distance $R$. (a) Radiation by source $\freq{\_J}_A$.  (b) Radiation by source $\freq{\_J}_B$.   }
\label{dipoles}
\end{figure}

It is interesting to see what kind of symmetry on the sources and their   fields is imposed by the relation for dyadic Green's function~\r{green4}. In order to do that, we consider the simplest electromagnetic system consisting of two sources $\freq{\_J}_A$ and $\freq{\_J}_B$ whose locations are described by vectors $\_r_A$ and $\_r_B$ (in fact $\_r_A$ and $\_r_B$ define a manifold of vectors which indicate directions to all possible point sources in~$A$ and $B$), respectively. Similar considerations can be made for a system of three and more sources. Let us assume that the sources are located at different positions and have arbitrary orientations in the $xz$-plane (the current densities have only the $x$ and $z$ components), as shown in Fig.~\ref{dipoles}. Using \r{green3}, one can find the   electric fields created by  elementary single points belonging to $A$ and $B$ current sources:
\e \begin{array}{cc} \displaystyle
\begin{pmatrix} 
{\rm d}\freq{E}_{A,x} \\
{\rm d}\freq{E}_{A,z}
\end{pmatrix} 
=  
\begin{pmatrix} 
\freq{G}_{xx}  & \freq{G}_{xz}  \\
\freq{G}_{zx}  & \freq{G}_{zz} 
\end{pmatrix}
\cdot \begin{pmatrix} 
\freq{J}_{A,x} {\rm d}V_A \\
\freq{J}_{A,z} {\rm d}V_A
\end{pmatrix}, \vspace{5pt} \\ \displaystyle 
\begin{pmatrix} 
{\rm d}\freq{E}_{B,x} \\
{\rm d}\freq{E}_{B,z}
\end{pmatrix} 
=  
\begin{pmatrix} 
\freq{G}_{xx}  & \freq{G}_{xz}  \\
\freq{G}_{zx}  & \freq{G}_{zz} 
\end{pmatrix}
\cdot \begin{pmatrix} 
\freq{J}_{B,x} {\rm d}V_B \\
\freq{J}_{B,z} {\rm d}V_B
\end{pmatrix},
\end{array}
\l{green5} 
\f
where ${\rm d}\freq{\_E}_{A}$ and ${\rm d}\freq{\_E}_{B}$ correspond to the electric fields at positions $\_r_B$ and $\_r_A$ created by sources $\freq{\_J}_{A} {\rm d}V_A$ and $\freq{\_J}_{B} {\rm d}V_B$, respectively. System~\r{green5}  comprises four equations with respect to four components of dyadic Green's function. One can rewrite it as
\e \begin{array}{l} \displaystyle
\begin{pmatrix} 
\freq{G}_{xx}  & \freq{G}_{xz}  \\
\freq{G}_{zx}  & \freq{G}_{zz} 
\end{pmatrix} 
\vspace{5pt} \\ \displaystyle 
\hspace{20pt} = \begin{pmatrix} 
{\rm d}\freq{E}_{A,x} & {\rm d}\freq{E}_{B,x}\\
{\rm d}\freq{E}_{A,z} & {\rm d}\freq{E}_{B,z}
\end{pmatrix} 
\cdot 
\begin{pmatrix} 
\freq{J}_{A,x} {\rm d}V_A & \freq{J}_{B,x} {\rm d}V_B \\
\freq{J}_{A,z} {\rm d}V_A & \freq{J}_{B,z} {\rm d}V_B
\end{pmatrix}^{-1}.
\end{array}
\l{green6} 
\f
For reciprocal systems, relation~\r{green4} requires that $\freq{G}_{xz}=\freq{G}_{zx}$, and, therefore, from \r{green6}  it follows that 
\e  \begin{array}{ll} \displaystyle
 {\rm d}\freq{E}_{B,x} \freq{J}_{A,x} {\rm d}V_A +
 {\rm d}\freq{E}_{B,z} \freq{J}_{A,z} {\rm d}V_A 
 \vspace{5pt} \\ \displaystyle 
\hspace{2cm} =
 {\rm d}\freq{E}_{A,x} \freq{J}_{B,x} {\rm d}V_B +
 {\rm d}\freq{E}_{A,z} \freq{J}_{B,z} {\rm d}V_B.
 \end{array}
\l{green7} 
\f
Next, one can likewise repeat the derivations~\r{green5}--\r{green7} for two other scenarios, when the current densities have only the $y$, $z$ and only the $x$, $y$ components. Then one obtains two other equations similar to~\r{green7} but with the interchanged component indices. Summing up these two equations together with~\r{green7}, we derive
\e  
 {\rm d}\freq{\_E}_{B} \cdot \freq{\_J}_{A} {\rm d}V_A  = {\rm d}\freq{\_E}_{A} \cdot \freq{\_J}_{B} {\rm d}V_B.
\l{green8} 
\f
The integration of equation~\r{green8} over the volume occupied by the current source~$A$ (over all possible $\_r_A$) for a fixed point $\_r_B$ yields
\e  
\int\limits_{V_A}   {\rm d}\freq{\_E}_{B} \cdot \freq{\_J}_{A} {\rm d}V_A  =  \freq{\_E}_{A} \cdot \freq{\_J}_{B} {\rm d}V_B.
\l{green9} 
\f
Next, we  integrate  the last equation  over the volume occupied by the current source~$B$ (over all possible $\_r_B$) for a fixed point $\_r_A$:
\e  
\int\limits_{V_A}   \freq{\_E}_{B} \cdot \freq{\_J}_{A} {\rm d}V_A  =  \int\limits_{V_B} \freq{\_E}_{A} \cdot \freq{\_J}_{B} {\rm d}V_B.
\l{green10} 
\f
The obtained relation for two electromagnetic sources and their fields in an anisotropic non-homogeneous medium is a reciprocity condition which is called the Lorentz reciprocity relation~\cite{lorentz_theorem_1896}. Notably, it was discovered by H.~Lorentz in 1896, long  before the Onsager reciprocal relations, which we used for our derivation, were known. 
The relation~\r{green10}     can be   extended to the case  of three and more sources. 
It is worthwhile to note that the above mentioned derivation  of the Lorentz reciprocity relation does not require any prior knowledge. On contrary, the conventional derivation (shown shortly below) is based on the    preceding knowledge that the Lorentz reciprocity relation   relates scalar products of the corresponding electric fields and currents.

Likewise, one can derive similar relation which will hold for radiation in both reciprocal and nonreciprocal media. Indeed, applying the Onsager-Casimir relation~\r{ons7} in our geometry with two sources at $\_r_A$ and $\_r_B$, one obtains that 
\e
\dya{\freq{G}}(\_r_A,\_r_B, \_H_0)=  \dya{\freq{G}}^T(\_r_B,\_r_A, -\_H_0).
\l{eq104}\f
Here argument $-\_H_0$ implies that the corresponding quantity should be considered in the medium with all the  external bias fields reversed. 

Next, let us find the final relation connecting the currents with the electric fields  in a different and more general way than was given by derivations~\r{green5}--\r{green10}. Rewriting~\r{green3} for sources at $\_r_A$ and $\_r_B$, we obtain:
\e \begin{array}{c} \displaystyle
{\rm d} \freq{\_E}_B(\_r_A, \_H_0) =  \dya{\freq{G}}(\_r_A,\_r_B, \_H_0)\cdot \freq{\_J}_B(\_r_B) {\rm d}V_B,
 \vspace{5pt} \\ \displaystyle 
{\rm d} \freq{\_E}_A(\_r_B, -\_H_0) =  \dya{\freq{G}}(\_r_B,\_r_A, -\_H_0)\cdot \freq{\_J}_A(\_r_A) {\rm d}V_A.
 \end{array}
\l{green103} 
\f
Using~\r{eq104} and \r{green103}, it is easy to show that ${\rm d} \freq{\_E}_B(\_r_A, \_H_0) \cdot  \freq{\_J}_A(\_r_A) {\rm d}V_A$ equals
\e   \begin{array}{ll} \displaystyle
\left[ \dya{\freq{G}}(\_r_A,\_r_B, \_H_0) \cdot \freq{\_J}_{B}(\_r_B) \right]^T \freq{\_J}_{A}(\_r_A) {\rm d}V_A   {\rm d}V_B   
 \vspace{5pt} \\ \displaystyle 
 =  \freq{\_J}_{B}^T(\_r_B)    \dya{\freq{G}}^T(\_r_A,\_r_B, \_H_0)   \freq{\_J}_{A}(\_r_A) {\rm d}V_A   {\rm d}V_B 
  \vspace{5pt} \\ \displaystyle 
  =  \freq{\_J}_{B}(\_r_B)  \cdot  \left[ \dya{\freq{G}}(\_r_B,\_r_A, -\_H_0)   \freq{\_J}_{A}(\_r_A) {\rm d}V_A  \right] {\rm d}V_B 
    \vspace{5pt} \\ \displaystyle 
    = {\rm d} \freq{\_E}_A(\_r_B, -\_H_0) \cdot \freq{\_J}_{B}(\_r_B)  {\rm d}V_B.
 \end{array}
\l{green110} 
\f
Integrating this equality like it was done in~\r{green9}--\r{green10}, we obtain
\e  
\int\limits_{V_A}   \freq{\_E}_{B} (\_H_0) \cdot \freq{\_J}_{A} {\rm d}V_A  =  \int\limits_{V_B} \freq{\_E}_{A}(-\_H_0)  \cdot \freq{\_J}_{B} {\rm d}V_B.
\l{green100} 
\f
This relation can be referred to as Onsager-Casimir theorem which is applied to both reciprocal and nonreciprocal linear time-invariant (LTI) systems. 
Naturally, for reciprocal media (with $\_H_0=0$) relation~\r{green100} simplifies to the Lorentz reciprocity relation~\r{green10}.

% \e   \begin{array}{ll} \displaystyle
% \int\limits_{V_A} \left[  \freq{\_E}_{B}(\_H_0) \cdot \freq{\_J}_{A}(\_H_0) +\freq{\_E}_{B}(-\_H_0) \cdot \freq{\_J}_{A}(-\_H_0) \right] {\rm d}V_A  
%  \vspace{5pt} \\ \displaystyle 
%  =
% \int\limits_{V_B} \left[ \freq{\_E}_{A}(\_H_0) \cdot \freq{\_J}_{B}(\_H_0) + 
% \freq{\_E}_{A}(-\_H_0) \cdot \freq{\_J}_{B}(-\_H_0) \right] {\rm d}V_B.
%  \end{array}
% \l{green100} 
% \f

Next, let us verify that relation~\r{green10} is in fact valid for two dipole sources  in a homogeneous isotropic medium illustrated in Fig.~\ref{dipoles}. The sources have infinitesimal thickness and equal lengths~$l$. Without loss of generality, we assume that the sources are located far from one another at a distance $R \gg \lambda$.
%The electric field $\_E_A$ generated by source $\_J_A$  
First, we measure the radiation from source~$A$, located at $\ve{r}_{\rm A}$ and oriented along $z$. The $\theta$-component (in the spherical coordinate system with the center at $\_r_A$) of the electric field $\freq{\_E}_A$ generated at point $\ve{r}_{\rm B}=\ve{r}_{\rm A}+\ve{R}$   can be written as~\cite[Eq.~(1.72a)]{stutzman_antenna_1998}: 
\e 
\freq{E}_{A, \theta}=j \omega \freq{\mu} \mu_0  \frac{\freq{I}_A l}{4\pi} \frac{{\rm e}^{-jkR}}{R}  \sin\theta,
\l{green11}
\f 
where $\freq{I}_A$ is the electric current flowing through dipole~$A$. In the second  scenario, we measure the radiation from source~$B$ which is oriented along the $\theta$-direction with respect to the initial $xyz$ basis for simplifying the calculations. The electric field in the position of   dipole $A$ can be written as:
\e 
\freq{E}_{B}=j \omega \freq{\mu} \mu_0  \frac{\freq{I}_B l}{4\pi} \frac{{\rm e}^{-jkR}}{R}.
\l{green12}
\f 
The projection of this field to the   $z$-axis (the direction in which dipole~$A$ is oriented) is  $ \freq{E}_{B,z}= \freq{E}_B\sin\theta$. Finally, substituting $\freq{E}_{A, \theta}$ and $ \freq{E}_{B,z}$ in \r{green1} and applying scalar product, we get:
\e  
    \freq{E}_{B,z}  \freq{I}_{A}  l  =  \freq{E}_{A,\theta}  \freq{I}_{B}  l,
\l{green13} 
\f
which is obviously an equality.

As the second example of a physical process subject to the Onsager reciprocal relations, we consider  polarization of  a general bianisotropic dipolar scatterer. Incident electric (or magnetic) field induces electric {\it and} magnetic dipole moments~\cite{serdyukov_electromagnetics_2001,Asadchy2018}.  
The derivation below are based on~\cite[\textsection~3.3.1]{serdyukov_electromagnetics_2001}. 
The electric $\freq{\_p}$ and magnetic $\freq{\_m}$ dipoles induced in a bianisotropic scatterer are related to the incident fields through electric $\dya{\freq{\alpha}}_{\rm ee}$, magnetic $\dya{\freq{\alpha}}_{\rm mm}$,   magnetoelectric $\dya{\freq{\alpha}}_{\rm me}$, and electromagnetic $\dya{\freq{\alpha}}_{\rm em}$ polarizability tensors:
\e
\begin{pmatrix} 
\freq{\_p} \\
\freq{\_m}
\end{pmatrix} 
=  
\begin{pmatrix} 
\dya{\freq{\alpha}}_{\rm ee}  & \dya{\freq{\alpha}}_{\rm em}  \\
\dya{\freq{\alpha}}_{\rm me}  & \dya{\freq{\alpha}}_{\rm mm} 
\end{pmatrix}
\cdot \begin{pmatrix} 
\freq{\_E} \\
\freq{\_H}
\end{pmatrix}.
\l{pm1} 
\f
This equations can be written using the six-vector notations as
\e
\overline{\freq{p}}_i (\omega)= \freq{A}_{i k }(\omega) \,\freq{e}_k (\omega),
\l{pm2} 
\f
where $\overline{\freq{p}}_i$ is a vector including six components of the electric and magnetic dipole moments, $\freq{e}_k$ includes, likewise, components of the electric and magnetic fields, and  $\freq{A}_{i k }$ is a $6\times 6$ tensor consisting of all the polarizability components. Equation~\r{pm2}  is analogous to the frequency-domain form of equation~\r{ons3}. Polarizability tensor $\freq{A}_{i k }$ satisfies all the conditions of a generalized susceptibility, and, therefore, one can apply the Onsager reciprocal relations~\r{ons6} and obtain $\freq{A}_{k i  } = \sigma \freq{A}_{i k   }$ for reciprocal bianisotropic scatterers. As it was mentioned earlier, for every combination of $i$ and $k$, parameter  $\sigma$ should be chosen equal either $+1$ if $\time{p}_i$ and $\time{p}_k$ have the same symmetry under time reversal (when both of them are components of $\time{\_p}$ or $\time{\_m}$)  or $-1$ in the opposite case  (when one of them is component of   $\time{\_p}$ and another is component of $\time{\_m}$).  Thus, the Onsager reciprocal relations for polarization of a bianisotropic scatterer read
\e 
\dya{\freq{\alpha}}_{\rm ee} = \dya{\freq{\alpha}}_{\rm ee}^T, \quad
\dya{\freq{\alpha}}_{\rm mm} = \dya{\freq{\alpha}}_{\rm mm}^T, \quad 
\dya{\freq{\alpha}}_{\rm me} = - \dya{\freq{\alpha}}_{\rm em}^T.
\l{pm3} 
\f
These relations can be used in order to determine whether a scatterer with given polarizability tensors is reciprocal or nonreciprocal. The Onsager-Casimir constraints~\r{ons7}, applicable for both reciprocal and nonreciprocal scatterers, result in
\e \begin{array}{ll} \displaystyle
\dya{\freq{\alpha}}_{\rm ee}(\_H_0) = \dya{\freq{\alpha}}_{\rm ee}^T(-\_H_0), \quad
\dya{\freq{\alpha}}_{\rm mm}(\_H_0) = \dya{\freq{\alpha}}_{\rm mm}^T(-\_H_0),
 \vspace{5pt} \\ \displaystyle 
\hspace{2cm} 
\dya{\freq{\alpha}}_{\rm me}(\_H_0) = - \dya{\freq{\alpha}}_{\rm em}^T(-\_H_0).
 \end{array}
\l{pm4} 
\f

Similarly to derivations~\r{pm1}--\r{pm4}, one can apply the Onsager reciprocal relations to bulk polarization of  bianisotropic  materials. The reader is referred to works~\cite{rado_reciprocity_1973},\cite[Eq.~(4)]{bokut_phenomenological_1974},\cite[\textsection~3.3.1]{serdyukov_electromagnetics_2001} for detailed derivations.

% \pink{In a more general way, one can consider electric fields at spatial point $r_A$ induced by source currents at  point $r_B$ which are related by $\_E(\_r_B)=\=G(\_r_B,\_r_A)\cdot \_J(\_r_A)$, where $\=G(\_r_B,\_r_A)$ represents the dyadic Green function. If the field equations are time-symmetric,  the symmetry relation can be written as  
%  \e \=G(\_r_B,\_r_A)=\=G^T(\_r_A,\_r_B).\f
% The symmetry property is the mathematical expression of the reciprocity theorem. }

\subsection{The Lorentz lemma and reciprocity theorem   } \label{onscas}

\begin{figure}[tb]
\centering
   \includegraphics[width=0.52\columnwidth]{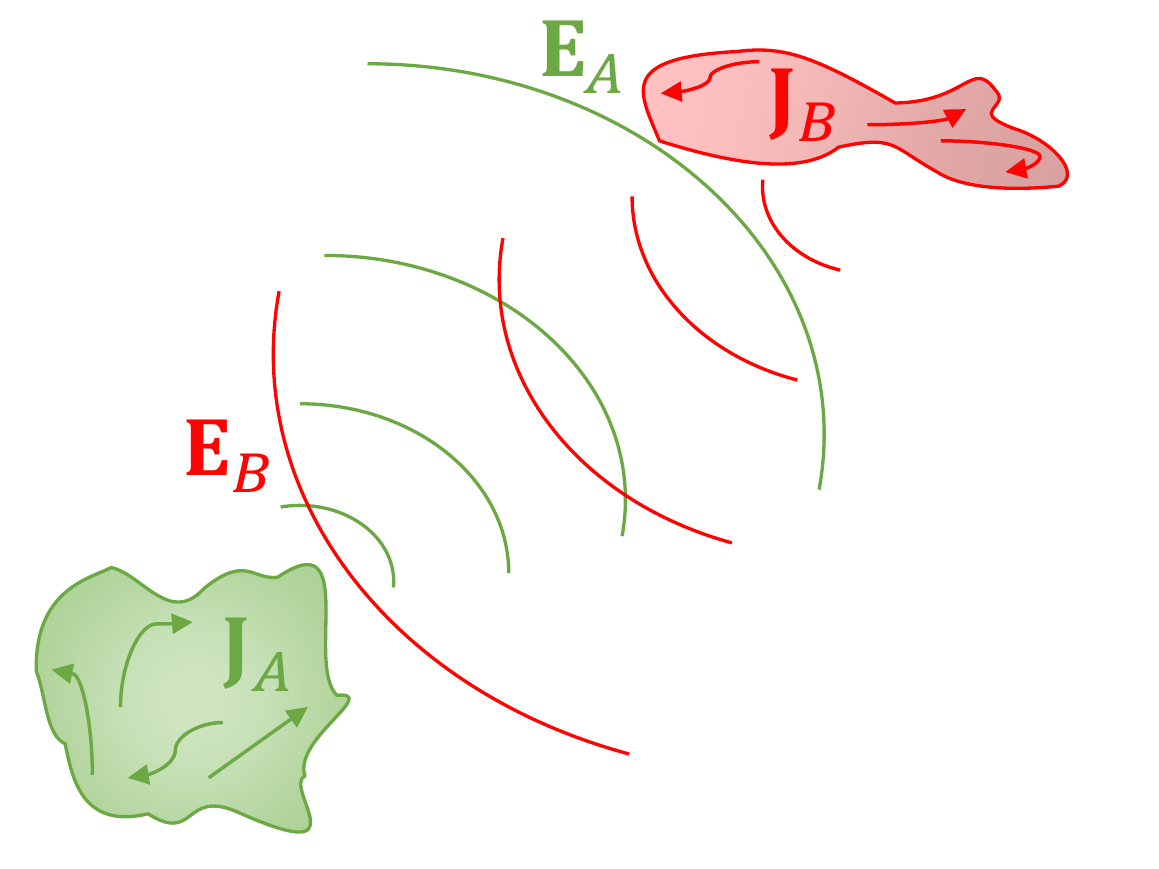}
\caption{Electromagnetic interaction between a pair of sources with the current densities $\freq{\_J}_{A}$ and $\freq{\_J}_{B}$. }
\label{fig3_8}
\end{figure}

The Lorentz reciprocity theorem (or reciprocity relation) is   formulated
for a pair of sources with current densities  $\freq{\_J}_A$ and $\freq{\_J}_B$  which create fields $\freq{\_E}_A$ and $\freq{\_E}_B$~\cite{lorentz_theorem_1896},\cite[\textsection~5.5]{kong_electromagnetic_1986},\cite[\textsection~3.6.2]{lindell_methods_1992} (see illustration in Fig.~\ref{fig3_8}). 
The derivation of this theorem for the general case of a linear homogeneous   medium was described in the previous section.  Here, we formulate this theorem for the special case of  bianisotropic media and discuss its implications on the material tensors.
Before we proceed to the theorem statement, let us declare an auxiliary quantity called \emph{reaction} and introduced in~\cite{rumsey_reaction_1954}. The reaction of field~$\freq{\_E}_B$ on a source with current density $\freq{\_J}_A$ is defined by the following volume integral in the frequency domain:
\e  \langle A,B \rangle=\int\limits_{V_A} \freq{\_J}_A \cdot \freq{\_E}_B {\rm d}V_A,
\l{eq3_38}\f
where the volume $V_A$ contains the source~$A$, and ${\rm d}V_A$ is the volume element. Likewise, the reaction of field~$\freq{\_E}_A$ on the source with current density $\freq{\_J}_B$ is given by
\e  \langle B,A \rangle=\int\limits_{V_B} \freq{\_J}_B \cdot \freq{\_E}_A {\rm d}V_B.
\l{eq3_38plus}\f
Note that we met   this reaction quantity in~\r{green10}. 
The reaction should be distinguished from  the rate of work done on a given charge distribution, despite the fact that these quantities have the same units. The latter one describes the dot product of the electric field and the current $\freq{\_J}_{\rm ind} $ that it induces in the material, i.e. $\int_{V} \freq{\_J}_{\rm ind} \cdot \freq{\_E} {\rm d}V$.

Using the assumed linearity of the system, we can express the fields created by sources by corresponding Green's functions. We stress that these Green's functions are not just simple free-space Green's functions, they take into account possibly very complicated topology of inhomogeneous bianisotropic media.
%We can define such a general Green's function $\dya{G}_{\rm g}$ similarly to~\r{green1}  as
% \e 
%  \freq{\_E}(\_r) =  \int_V \dya{G}_{\rm g}(\_r,\_r')\cdot \_J(\_r') {\rm d}V'.
% \l{green50} \f
Substituting the electric field from~\r{green1}   in~\r{eq3_38} and \r{eq3_38plus}, we obtain
% If the Green function is symmetric, then these two reactions are equal. Indeed,
\e\langle A,B\rangle
= \int_{V_A} \int_{V_B} \ve{\freq{J}}_A\cdot  \dya{\freq{G}}(\_r_A,\_r_B)\cdot  \ve{\freq{J}}_B\, {\rm d}V_B {\rm d}V_A,   \l{ab1}\f
\e \begin{array}{c}\displaystyle
\langle B,A\rangle
= \int_{V_B} \int_{V_A} \ve{\freq{J}}_B\cdot  \dya{\freq{G}}(\_r_B,\_r_A)\cdot  \ve{\freq{J}}_A\, {\rm d}V_A {\rm d}V_B   \vspace{2mm} \\ \displaystyle
= \int_{V_B} \int_{V_A} \ve{\freq{J}}_A\cdot \dya{\freq{G}}^T  (\_r_B,\_r_A)\cdot  \ve{\freq{J}}_B\, {\rm d}V_A {\rm d}V_B, \end{array} \l{ab2}  \f
where we replaced scalar $ \ve{\freq{J}}_B\cdot  \dya{\freq{G}}\cdot  \ve{\freq{J}}_A$ by its transpose $ (\ve{\freq{J}}_B\cdot  \dya{\freq{G}}\cdot  \ve{\freq{J}}_A)^T$ and used tensor identity $ (\ve{\freq{J}}_B\cdot  \dya{\freq{G}}\cdot  \ve{\freq{J}}_A)^T=\ve{\freq{J}}_A\cdot  \dya{\freq{G}}^T \cdot  \ve{\freq{J}}_B$.
If the system satisfies the assumption made in the derivation of the Onsager symmetry relation~\r{ons6},  Green's function is symmetric, i.e.  
% By definition {\bf I think, here we really make use of the Onsager reciprocity relation, that is, I would write  "If the system satisfies the assumption made in the derivation of the Onsager symmetry relation (symmetry of kinetic coefficients),  Green's function is symmetric"} 
$\dya{\freq{G}}(\_r_A,\_r_B)=\dya{\freq{G}}^T(\_r_B,\_r_A)$, which implies that  the two reactions are equal:
\e  \langle A,B \rangle - \langle B,A \rangle=\int\limits_{V_{A }} \freq{\_J}_A \cdot \freq{\_E}_B {\rm d}V_A - \int\limits_{V_{B}} \freq{\_J}_B \cdot \freq{\_E}_A {\rm d}V_B=0.
\l{ab3}\f
Equation~\r{ab3} represents the \emph{Lorentz reciprocity theorem } (or relation) in the frequency domain.  It states that in reciprocal systems, the reaction of field~$\freq{\_E}_A$  on a source  with current density $\freq{\_J}_B$  should be the same as that of field~$\freq{\_E}_B$  on a source  with   $\freq{\_J}_A$.  In other words, interactions between any pair of  electromagnetic sources are reciprocal. Relation \r{ab3} can be considered as the definition of reciprocal electromagnetic systems. This formulation, in fact, does not imply  time  reversibility $t \rightarrow -t$  of electromagnetic processes in the medium.  Instead, it is based on the notion of restricted time reversal and just \textit{emulates} time reversibility by interchanging the locations of the  sources and the field probe.

Let us find the restriction on material properties dictated by the Lorentz reciprocity, i.e. conditions on material parameters which determine whether a given material is reciprocal or not. 
First, we can write the Maxwell equations in the frequency domain applied to each of the two volumetric sources:
\e \begin{array}{c}\displaystyle
\nabla\times \freq{\_E}_A=- j\omega \ve{\freq{\_B}_A}, \quad \nabla\times \_H_A= j\omega \ve{\freq{\_D}_A}  +\freq{\_J}_A, \vspace{1mm} \\ \displaystyle
\nabla\times \freq{\_E}_B=- j\omega \ve{\freq{\_B}_B}, \quad \nabla\times \_H_B=j\omega \ve{\freq{\_D}_B}  +\freq{\_J}_B.
\end{array} \l{max21}\f
Using \r{max21}, we obtain the following relation for the difference of reactions $\langle A,B \rangle - \langle B,A \rangle$: 
\e \begin{array}{l}\displaystyle
\int \limits_{V}\freq{\_J}_A\cdot \freq{\_E}_B\,{\rm d}V-\int \limits_{V} \freq{\_J}_B\cdot \freq{\_E}_A\,{\rm d}V  \vspace{2mm} \\\displaystyle
=\oint \limits_S\left(\freq{\_E}_A\times\freq{\_H}_B-\freq{\_E}_B\times\freq{\_H}_A\right)\cdot d\_S  \vspace{2mm} \\ \displaystyle
-j\omega\int \limits_V\left(\freq{\_E}_B\cdot\freq{\_D}_A-\freq{\_E}_A\cdot\freq{\_D}_B+\freq{\_H}_A\cdot\freq{\_B}_B
-\freq{\_H}_B\cdot\freq{\_B}_A\right){\rm d}V,
%  -\int \limits_V(\_E_B\cdot \frac{\partial }{\partial t} \_D_A-\_E_A\cdot\frac{\partial }{\partial t}\_D_B \vspace{2mm} \\\displaystyle
%  +\_H_A\cdot\frac{\partial }{\partial t}\_B_B
% -\_H_B\cdot\frac{\partial }{\partial t}\_B_A)dV,
\end{array} \l{max22}\f
% \int_{V} \left[\_E_B\cdot ( \nabla\times \_H_A-\frac{\partial \ve{D_A}}{\partial t}) - \_E_A\cdot ( \nabla\times \_H_B-\frac{\partial \ve{D_B}}{\partial t}) \right] \,dV
% \vspace{2mm} \\\displaystyle
which represents the so-called \emph{Lorentz lemma}. Note that lemma~\r{max22} is just a mathematically derived  equation from the Maxwell equations, and it does not imply any reciprocity conditions, being applicable for both reciprocal and nonreciprocal time-invariant systems. Here, we have used the Gauss theorem and an identity from vector calculus $\_F \cdot (\nabla \times \_G)= \nabla \cdot (\_G \times \_F) + \_G \cdot (\nabla \times \_F)$. Volume $V$ and its closed surface area $S$ include both sources $A$ and $B$.

We  stress that the only condition for the validity of the Lorentz lemma (not Lorentz reciprocity theorem) is that both sets of fields satisfy Maxwell's equations and that the involved integrals exist. The two sets of sources can act in two different media, which can have arbitrary electromagnetic properties including  nonlinear. Below we consider the Lorentz lemma for three different scenarios: Reciprocal and nonreciprocal   time-invariant media, as well as   time-varying media.

\subsubsection{Reciprocal time-invariant media}
For monochromatic fields (restricting the generality to sources at the same frequency in linear time-invariant media), the Lorentz lemma~\r{max22} together with the Lorentz reciprocity relation~\r{ab3} result in 
\e \begin{array}{l}\displaystyle
0 =\oint \limits_S\left(\freq{\_E}_A\times\freq{\_H}_B-\freq{\_E}_B\times\freq{\_H}_A\right)\cdot {\rm d}\_S \vspace{2mm}\\\displaystyle
 -j\omega\int \limits_V\left(\freq{\_E}_B\cdot\freq{\_D}_A-\freq{\_E}_A\cdot\freq{\_D}_B+\freq{\_H}_A\cdot\freq{\_B}_B
-\freq{\_H}_B\cdot\freq{\_B}_A\right){\rm d}V.
\end{array}\l{max24}\f
The surface integral in~\r{max24}  vanishes since the surface of integration can be  always extended to infinity from the sources where the electric and magnetic fields are related through $\freq{\_H}_{A,B}=\_n \times \freq{\_E}_{A,B} /\freq{\eta}$ and $\_n \cdot \freq{\_E}_{A,B}=0$, where $\freq{\eta}$ is the impedance of the surrounding medium and $\_n$ is the unit normal vector to the integration surface pointing outwards. Indeed, the expression in the surface integral becomes zero   since $\freq{\_E}_A\times\freq{\_H}_B-\freq{\_E}_B\times\freq{\_H}_A=\_n (\freq{\_E}_A \cdot \freq{\_E}_B)-\_n (\freq{\_E}_B \cdot \freq{\_E}_A)=0$. 
This argument, resulting into vanishing of the surface integral, can be applied only for the case when the medium is homogeneous and isotropic at the considered boundary. Nevertheless, it is possible to proof that the surface integral tends to zero even in the case of a general medium. This proof is conventionally made based on the so-called limiting absorption principle~\cite{schulenberger_limiting_1971,kirsch_limiting_2018}. 
One can assume a tiny absorption everywhere. In this case, the fields will exponentially decay, and hence the surface integral vanishes as the boundary goes to infinity. Next, one can take the limit of the absorption going to zero. Thus, this principle yields vanishing surface integral even in the lossless case.  

Assuming that the integration space is filled with a nonuniform  bianisotropic  medium (general linear medium whose parameters arbitrarily vary in space)~\cite{kong_electromagnetic_1986,serdyukov_electromagnetics_2001} with macroscopic material relations 
\e \begin{array}{l}\displaystyle
\ve{\freq{\_D}}_{A,B}=\dya{\freq{\varepsilon}}(\ve{r},\omega)\cdot\ve{\freq{E}}_{A,B}
+\dya{\freq{\xi}}(\ve{r},\omega)\cdot\ve{\freq{H}}_{A,B}, \vspace{1mm} \\\displaystyle
\ve{\freq{\_B}}_{A,B}=\dya{\freq{\zeta}}(\ve{r},\omega)\cdot\ve{\freq{E}}_{A,B}
+\dya{\freq{\mu}}(\ve{r},\omega)\cdot\ve{\freq{H}}_{A,B},
\end{array}\l{matrel1}\f
equation~\r{max24} yields (here, $\dya{\freq{\xi}}$ and $\dya{\freq{\zeta}}$ are the bianisotropy parameters describing effects of weak spatial dispersion~\cite{asadchy_spatially_2017})
\e \begin{array}{l}\displaystyle
-j\omega\int \limits_V\Big\{\ve{\freq{E}}_B\cdot\left[\dya{\freq{\varepsilon}}-\dya{\freq{\varepsilon}}^T\right]\cdot\ve{\freq{E}}_A 
+\ve{\freq{H}}_A\cdot\left[\dya{\freq{\mu}}-\dya{\freq{\mu}}^T\right]\cdot\ve{\freq{H}}_B \vspace{2mm}\\ \displaystyle
+\ve{\freq{E}}_B\cdot\left[\dya{\freq{\xi}} +\dya{\freq{\zeta}}^T \right]\cdot\ve{\freq{H}}_A 
-\ve{\freq{H}}_B\cdot\left[\dya{\freq{\zeta}} +\dya{\freq{\xi}}^T \right]\cdot \ve{\freq{E}}_A\Big\}{\rm d}V=0.
\end{array}\l{max25}\f
\begin{figure}[tb!]
	\centering
	\subfigure[]{\includegraphics[height=0.45\linewidth]{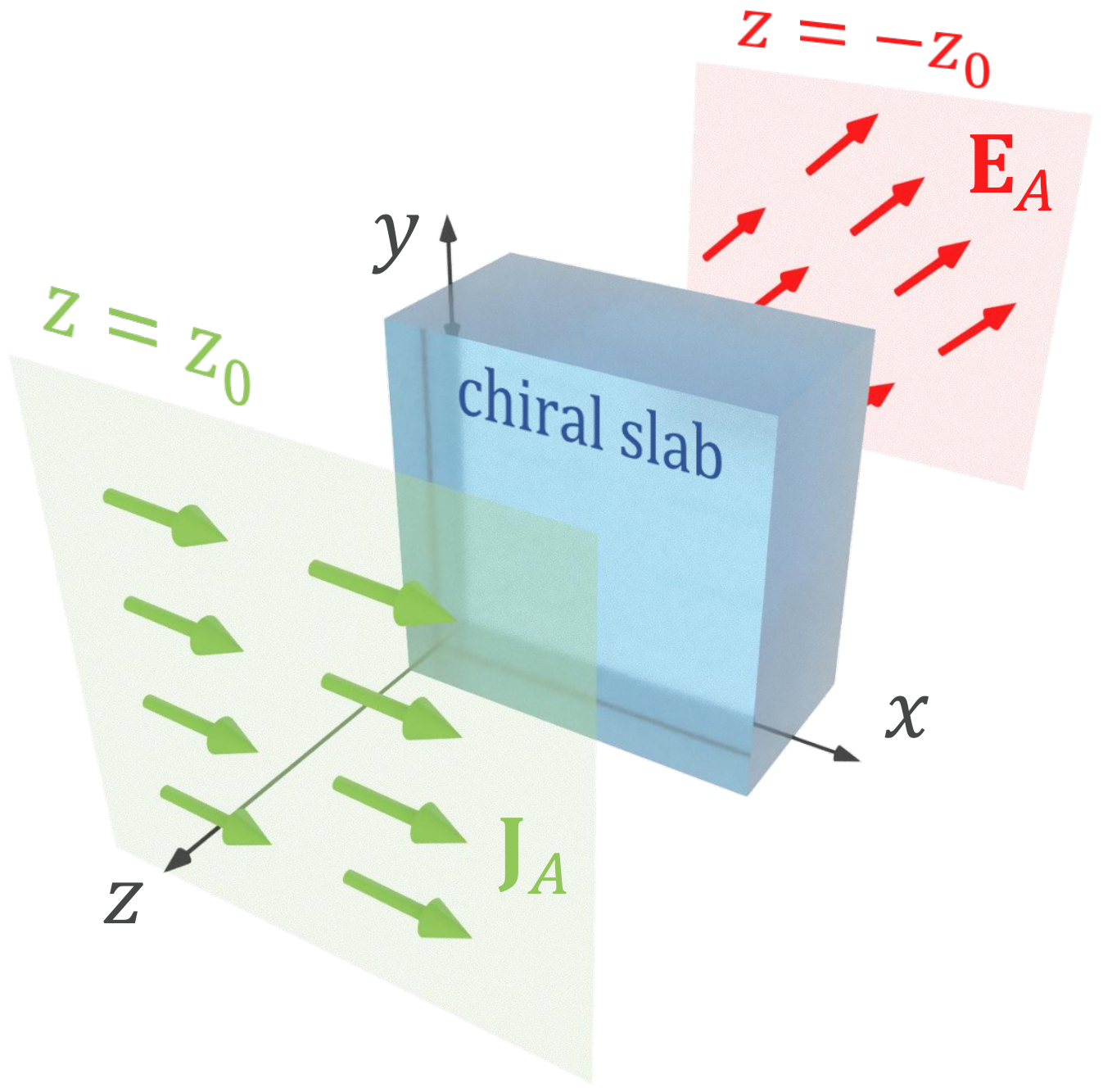} \label{fig6a}} 
	\subfigure[]{\includegraphics[height=0.45\linewidth]{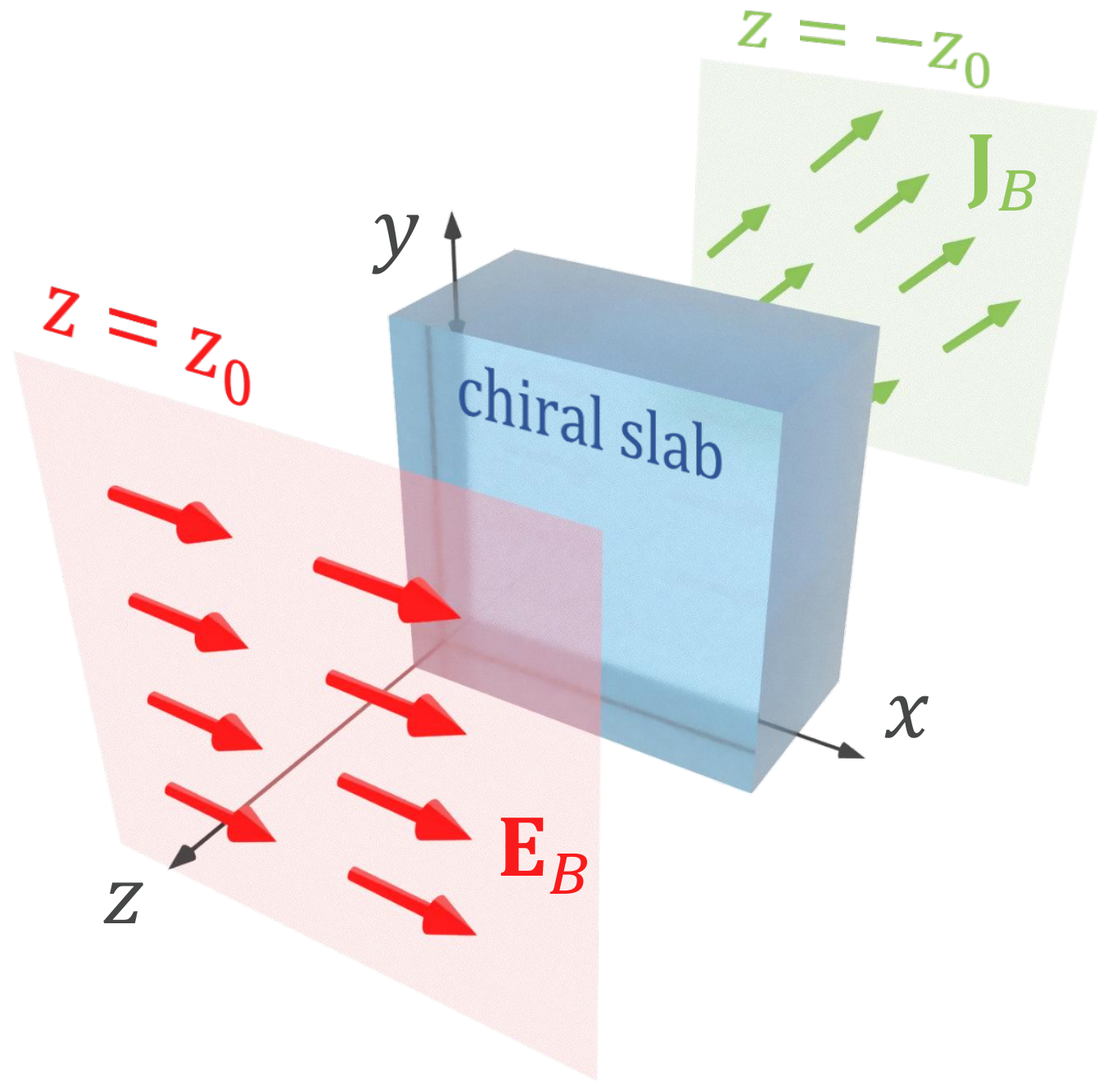}\label{fig6b}}\\
	\subfigure[]{\includegraphics[height=0.45\linewidth]{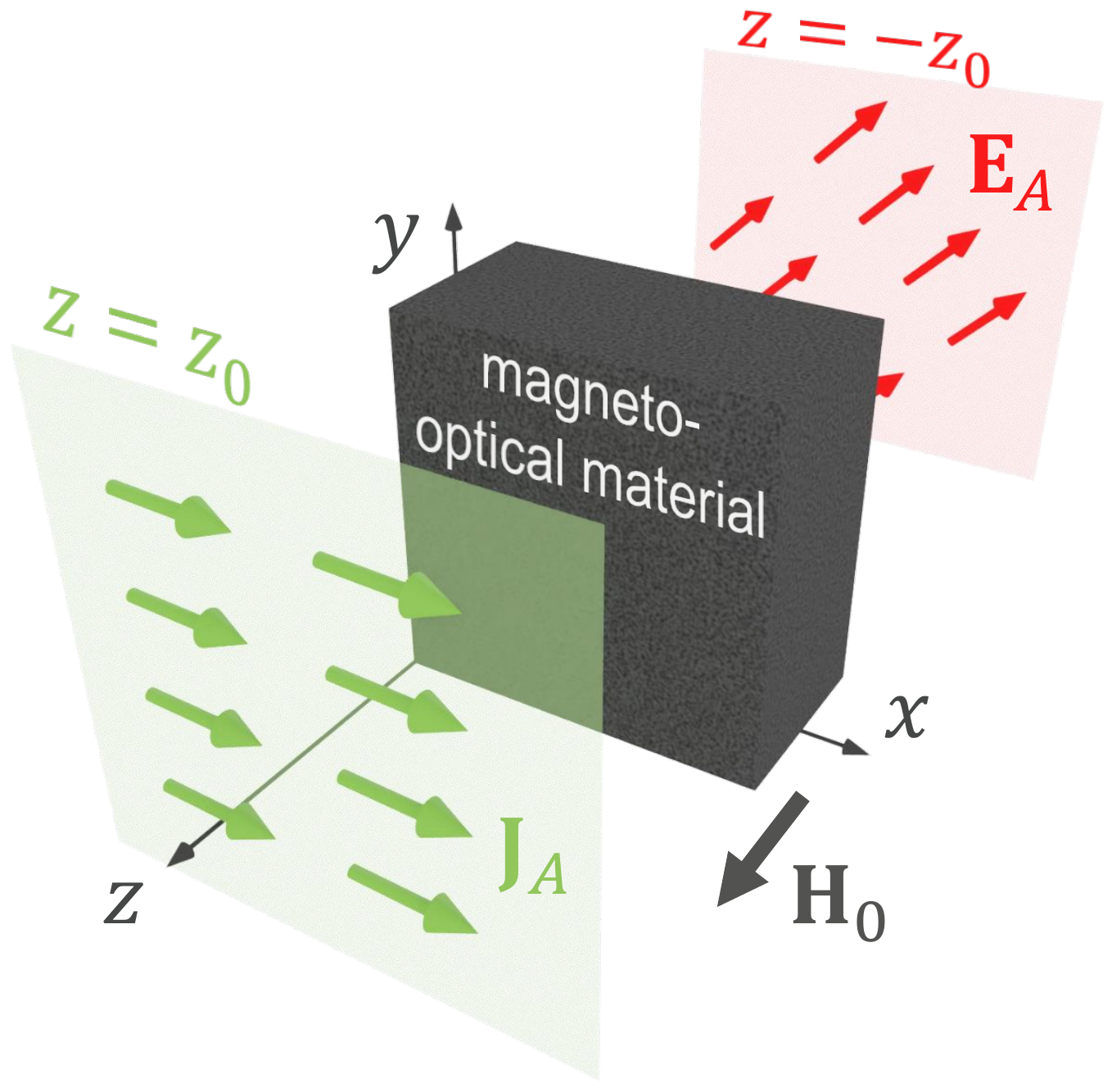} \label{fig6c}}
    \subfigure[]{\includegraphics[height=0.45\linewidth]{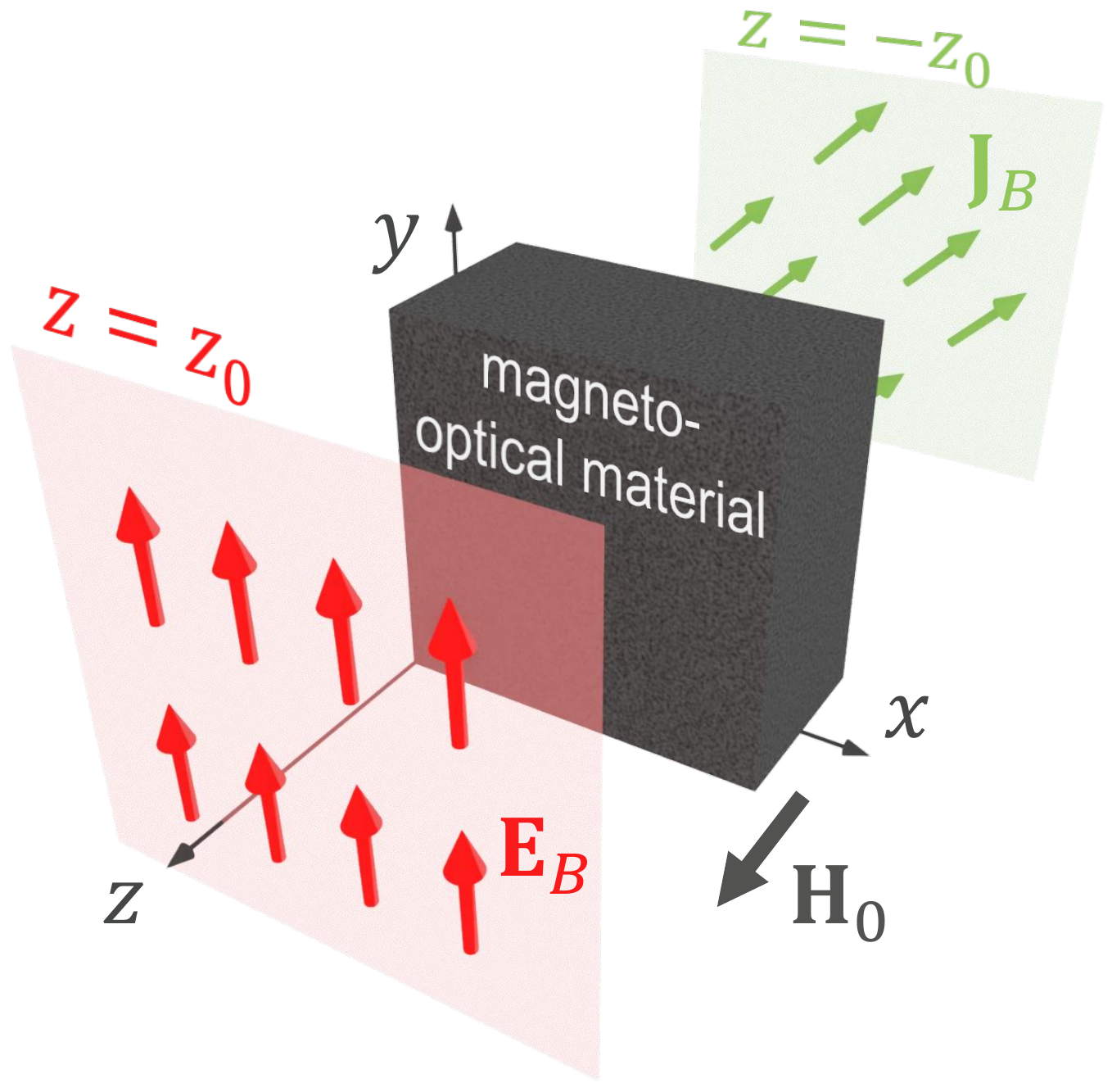} \label{fig6d}}
	\caption{ Application of the Lorentz reciprocity theorem for (a--b) a chiral slab   and (c--d) a slab of magneto-optical material. The theorem holds only for the reciprocal chiral slab.  }
	\label{fig6}
\end{figure}
To obtain this equation, we have used the same tensor identity as in~\r{ab2}. Since this equation is satisfied  for arbitrary fields  $\freq{\_E}_{\rm A,B}$ and $\freq{\_H}_{\rm A,B}$, the expressions in the square brackets in~\r{max25} must be equal to zero, which results in
\e \dya{\freq{\varepsilon}}=\dya{\freq{\varepsilon}}^T, 
\quad
\dya{\freq{\mu}}=\dya{\freq{\mu}}^T,\quad 
\dya{\freq{\xi}}=-\dya{\freq{\zeta}}^T. \l{onscas1}\f
Equations \r{onscas1} are the Onsager reciprocal relations applied on material parameters of   general bianisotropic  reciprocal media. Materials for which these conditions are not satisfied are \emph{nonreciprocal}. Note that these relations are similar to those for polarizabilities of a single bianisotropic scatterer given by~\r{pm3}. In fact, relations \r{onscas1} can be alternatively achieved using derivations similar to \r{pm1}--\r{pm3}.
It should be mentioned that not all time-reversible (in microscopic sense) systems are reciprocal, while all reciprocal systems are time-reversible.
 On the other hand, as was shown in~\cite[p.~697]{altman_generalization_1991},
the  restricted time reversibility of a medium has the same conditions on material tensors as in~\r{onscas1}. Therefore,   reciprocity and restricted time-reversal symmetry  apply in the same  way to different materials (see the bottom rows of the table in Fig.~\ref{fig3}).

Let us consider the applicability of the Lorentz reciprocity theorem for two simple examples of isotropic  materials. In the first example, we consider a bianisotropic chiral  slab whose structural units (molecules or meta-atoms)  have broken mirror symmetry\footnote{The structural unit and its mirror image cannot be superposed onto one another (similarly to a human hand).}. We position an infinite current sheet with $\freq{\_J}_A$ in front of the slab at $z=z_0$ and probe the electric field $\freq{\_E}_A$ which was radiated by the sheet and passed through the slab at the plane  $z=-z_0$, as shown in  Fig.~\ref{fig6a}.  
The wave passed through the chiral slab experienced polarization rotation by an azimuthal angle  $\phi=+45^\circ$. When we interchange the plane of the source with the observation plane (see Fig.~\ref{fig6b}), the wave radiated by the sheet with current density $\freq{\_J}_B$ (tilted at $45^\circ$) is transmitted through the chiral slab   with opposite polarization rotation at an angle $\phi=-45^\circ$. Now, it is clear that the surface integral of the reaction $\int_{S} \freq{\_J}_A \cdot \freq{\_E}_B {\rm d}S$ is equal to $\int_{S} \freq{\_J}_B \cdot \freq{\_E}_A {\rm d}S$, which means that the chiral slab is reciprocal.

In the second example, let us consider a slab of magneto-optical material biased by a static magnetic field $\_H_0$. Such slab rotates polarization of a wave passed through it at the same angle $+45^\circ$ regardless the propagation direction. Therefore, repeating the same thought experiment shown in Figs.~\ref{fig6c} and \ref{fig6d}, one can observe that $\int_{S} \freq{\_J}_A \cdot \freq{\_E}_B {\rm d}S=0$ ($\freq{\_J}_A$ and $\freq{\_E}_B$ are orthogonal), while $\int_{S} \freq{\_J}_B \cdot \freq{\_E}_A {\rm d}S \neq 0$. This result confirms that biased magneto-optical materials are nonreciprocal.

Thus, in the simplified formulation, reciprocity of a system implies that under interchanging the positions of the source and the observation point, the detected field does not change (regardless of losses in the system). This statement   follows from the Lorentz lemma~\r{ab3}, assuming $\freq{\_J}_A=\freq{\_J}_B$. Graphically such principle is depicted in Fig.~\ref{fig3} in cells A.3, B.3, and C.3.
 Observing Fig.~\ref{fig3}, one can conclude that pointwise   reciprocity of a linear time-invariant system  implies that it is locally time-reversible in the \emph{restricted sense}, and vice versa.

\subsubsection{Nonreciprocal time-invariant media}
It should be mentioned that the Onsager relations can be extended to nonreciprocal materials~\cite{rado_reciprocity_1973,caloz_nonreciprocal_2016}. By reversing time of the whole system (globally, including time of the external  sources), we obtain the time-reversed process. 
%Thus, using generalized version of the Lorentz lemma, one can find restrictions on material parameters also for nonreciprocal  media. 
\begin{figure}[tb]
	\centering
\includegraphics[width=.9\linewidth]{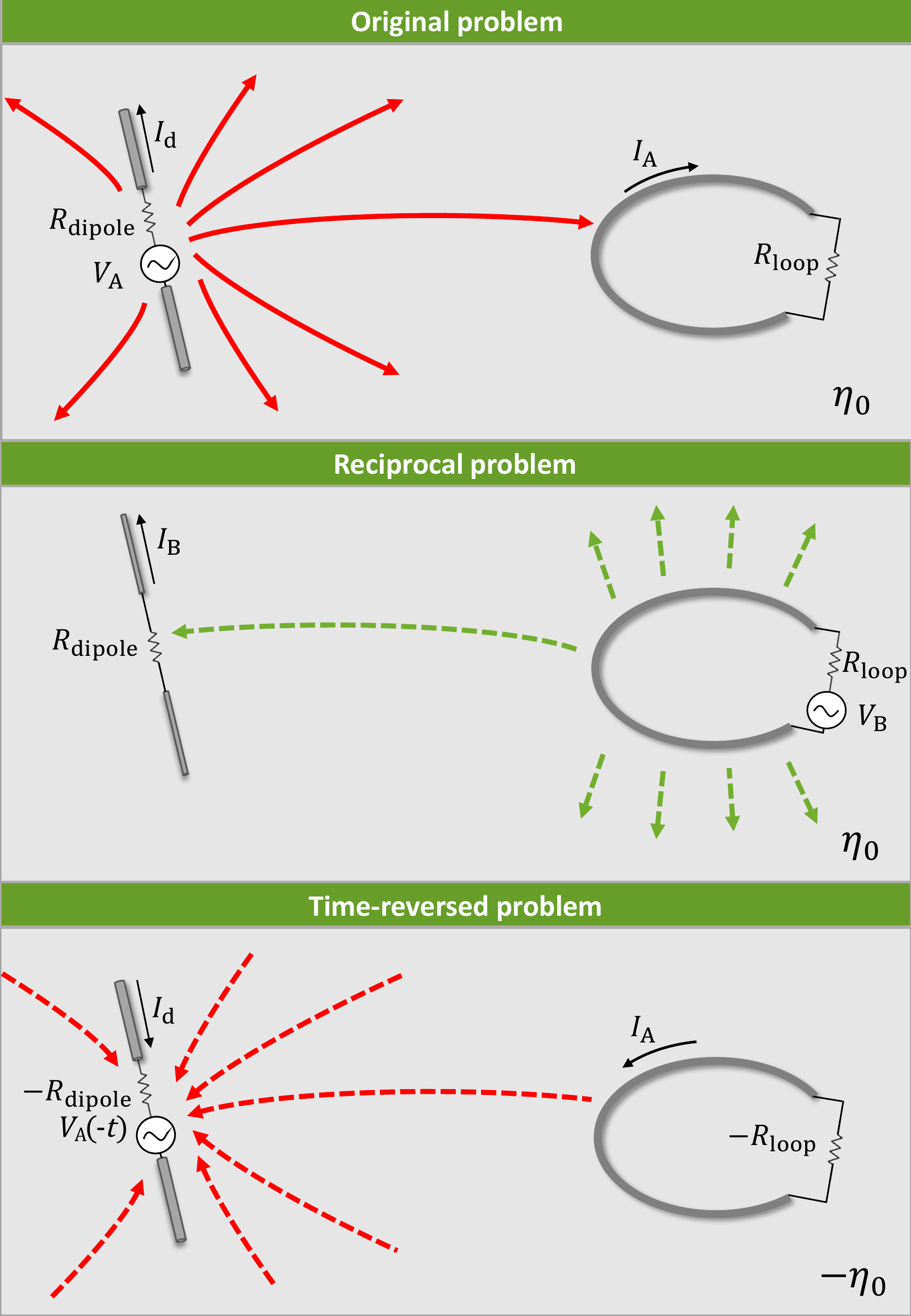} 	\caption{Reciprocity and time-reversibility applied to antenna problems. Original problem is presented in the top figure. }\label{fig_antenna}
\end{figure}
Let us assume that the field $\freq{\_E}_B$ generated by the sour\-ce~$B$ is calculated in the same material but with  \textit{reversed bias fields} $\_H_0$  (to emulate global time reversibility). In this  reversed material,  the material relations~\r{matrel1} for the case of excitation by source with $\freq{\_J}_B$ can be written as
\e \begin{array}{l}\displaystyle
\ve{\freq{\_D}}_B=\dya{\freq{\varepsilon}}(\ve{r},\omega,-\ve{H}_0)\cdot\ve{\freq{E}}_B
+\dya{\freq{\xi}}(\ve{r},\omega,-\ve{H}_0)\cdot\ve{\freq{H}}_B, \vspace{1mm} \\ \displaystyle
\ve{\freq{\_B}}_B=\dya{\freq{\zeta}}(\ve{r},\omega,-\ve{H}_0)\cdot\ve{\freq{E}}_B
+\dya{\freq{\mu}}(\ve{r},\omega,-\ve{H}_0)\cdot\ve{\freq{H}}_B.
\end{array}\l{matrel3}\f
Substituting the material relations to~\r{max24}, one can obtain
\e
\begin{split}
-j\omega&\int_V\Big\{\ve{\freq{E}}_B\cdot\left[\dya{\freq{\varepsilon}}(\ve{H}_0)-\dya{\freq{\varepsilon}}^T(-\ve{H}_0)\right]\cdot\ve{\freq{E}}_A\\
&\qquad
-\ve{\freq{H}}_A\cdot\left[\dya{\freq{\mu}}^T(\ve{H}_0)-\dya{\freq{\mu}}(-\ve{H}_0)\right]\cdot\ve{\freq{H}}_B\\
&\qquad
+\ve{\freq{E}}_B\cdot\left[\dya{\freq{\xi}}(\ve{H}_0)+\dya{\freq{\zeta}}^T(-\ve{H}_0)\right]\cdot\ve{\freq{H}}_A\\
&\qquad
-\ve{\freq{H}}_B\cdot\left[\dya{\freq{\zeta}}(\ve{H}_0)+\dya{\freq{\xi}}^T(-\ve{H}_0)\right]\cdot \ve{\freq{E}}_A\Big\}{\rm d}V=0,
\end{split}
\l{max27}\f
from where the \emph{Onsager-Casimir relations for material parameters} read
\e \begin{array}{c}\displaystyle
\dya{\freq{\varepsilon}}(\ve{H}_0)=\dya{\freq{\varepsilon}}^T(-\ve{H}_0),
\qquad
\dya{\freq{\mu}}(\ve{H}_0)=\dya{\freq{\mu}}^T(-\ve{H}_0), \vspace{1mm}\\\displaystyle
\dya{\freq{\xi}}(\ve{H}_0)=-\dya{\freq{\zeta}}^T(-\ve{H}_0).
\end{array}\l{got}\f
These conditions on material parameters can be applied to both reciprocal and nonreciprocal media. For the former case, the conditions simplify to~\r{onscas1}. Note that these relations are similar to those for polarizabilities of a single bianisotropic scatterer given by~\r{pm4} and can be alternatively derived likewise.

A classical application of reciprocity in time-invariant media is to antenna problems. Let us assume the scenario shown in Fig.~\ref{fig_antenna} (top picture), where two antennas are placed in free space. In particular, we will consider a dipole antenna excited by a voltage source $V_{\rm A}$  and a receiving loop. The current excited in the dipole produces radiated fields which propagate in the background medium and induce a current in the loop antenna, denoted as $I_{\rm A}$.  In the reciprocal scenario, we consider the dipole as a receiving antenna and the loop becomes the transmitting antenna excited by the voltage source $V_{\rm B}$. In this case, the current excited in the loop emits propagating fields that induce a current $I_{\rm B}$ in the dipole (see the middle picture).  The Lorentz reciprocity theorem presented in Eq.~\r{ab3} can be simplified  as $V_{\rm A} I_{\rm B}=V_{\rm B} I_{\rm A}$ for this particular example. It is interesting to notice that the fields in the reciprocal scenario are not the time-reversed copy of the fields in the original example.   As it is shown in Fig.~\ref{fig_antenna} (bottom picture), in the time-reversal scenario, both loop and dipole resistors become active elements modeled by the negative resistors and the voltage source becomes a sink. Even more interesting, the impedance of the background medium will also become negative modeling energy deliver from infinity by the host medium. The negative resistors excite currents that produce exactly a time-reversed copy of the field excited in the original example. Under these considerations, the time-reversal scenario seems to be physically unrealistic. 
To understand the relation between the reciprocity theorem and the time-reversibility of Maxwell equations, one must consider that in both original and reciprocal scenarios the interaction between receiving and transmitting antennas is produced by “direct” rays that emanate from one antenna and induce a current in the second antenna. These rays are identical in the reciprocal and time-reversed scenario and linking the reciprocal scenario with the time-reversal problem.

\subsubsection{Time-varying media}
Let us rewrite  Maxwell's equations for two systems. In the original system, the time argument is~$t$, while in the time-reversed and  shifted by $\tau$ seconds system, the argument is $\tau-t$. Thus, we have
\begin{equation}
\begin{split}
&\nabla\times \time{\_E}(t)=-{\partial\over\partial t}\time{\_B}(t),\cr
&\nabla\times \time{\_H}(t)=\time{\_J}(t)+{\partial\over\partial t}\time{\_D}(t),\cr
&\nabla\times \time{\_E}(\tau-t)={\partial\over\partial t} \time{\_B}(\tau-t),\cr
&\nabla\times \time{\_H}(\tau-t)=\time{\_J}(\tau-t)-{\partial\over\partial t} \time{\_D}(\tau-t).
\end{split}
\end{equation}
Using the following identity: $\nabla\cdot(\_C\times\_D)=\_D\cdot(\nabla\times\_C)-\_C\cdot(\nabla\times\_D)$, and employing the above expressions based on Maxwell's equations, we can readily conclude that   
\begin{equation}
\begin{split}
&\nabla\cdot\Big(\time{\_E}_A(\tau-t)\times\time{\_H}_B(t)-\time{\_E}_B(t)\times\time{\_H}_A(\tau-t)\Big)=\cr
&\time{\_H}_A(\tau-t)\cdot{\partial\over\partial t}\time{\_B}_B(t)+\time{\_H}_B(t)\cdot{\partial\over\partial t}\time{\_B}_A(\tau-t)\cr
&-\time{\_E}_A(\tau-t)\cdot{\partial\over\partial t}\time{\_D}_B(t)-\time{\_E}_B(t)\cdot{\partial\over\partial t}\time{\_D}_A(\tau-t)\cr
&-\time{\_E}_A(\tau-t)\cdot\time{\_J}_B(t)+\time{\_E}_B(t)\cdot\time{\_J}_A(\tau-t).
\end{split}
\l{eq:llgctv}
\end{equation}
Here, for simplicity we assume that there are only electric current sources. If we integrate over a volume~$V$ which contains both sources and over time~$t$, we express the most general form of the Lorentz lemma (compare to \r{max22}). 

For a \textit{time-invariant} medium, two different forms of the Lorentz reciprocity can be introduced: Convolution-type and correlation-type reciprocity~\cite{AmirReciprocity,cheo1965}. Probably the most studied type is the convolution type which is given by
\begin{equation}
\begin{split}
\int_{-\infty}^{+\infty} {\rm d}t\int_V&\time{\_J}_A(\tau-t)\cdot\time{\_E}_B(t){\rm d}V=\cr
&\int_{-\infty}^{+\infty} {\rm d}t\int_V\time{\_J}_B(t)\cdot\time{\_E}_A(\tau-t){\rm d}V.
\end{split}
\l{eq:convtyrecip}
\end{equation}
This is a general definition in the time domain which is exactly equivalent to  the Lorentz reciprocity relation   in the frequency domain (see Eq.~\r{ab3}). This is due to the fact that in Eq.~\r{eq:convtyrecip}, as mentioned in the above, the convolution operation on the electric current density and the electric field is applied.
% Therefore, by simply taking the Fourier transform from both sides of the equation, we obtain the known Eq.~\r{ab3} in the frequency domain.
Let us develop Eq.~\r{eq:llgctv}. To do that, we can also include the material relations corresponding to a time-invariant medium. Remember that in the time domain, such relations are given by
\begin{equation}
\begin{split}
&\ve{\time{D}}(t)=\int_0^\infty\dya{\time{\varepsilon}}(\tau)\cdot\ve{\time{E}}(t-\tau){\rm d}\tau+\int_0^\infty\dya{\time{\xi}}(\tau)\cdot\ve{\time{H}}(t-\tau){\rm d}\tau,\cr
&\ve{\time{B}}(t)=\int_0^\infty\dya{\time{\zeta}}(\tau)\cdot\ve{\time{E}}(t-\tau){\rm d}\tau+\int_0^\infty\dya{\time{\mu}}(\tau)\cdot\ve{\time{H}}(t-\tau){\rm d}\tau.
\end{split} \l{dtbt}
\end{equation}
Now, by considering the Lorentz lemma~\r{eq:llgctv}, employing the convolution-type reciprocity~\r{eq:convtyrecip}, and substituting the material relations~\r{dtbt}, after doing some algebraic manipulations, we obtain the following expressions in time domain:
\begin{equation}
\dya{\time{\varepsilon}}(t)=\dya{\time{\varepsilon}}^T(t),\,\,\,\,\,\,
\dya{\time{\mu}}(t)=\dya{\time{\mu}}^T(t),\,\,\,\,\,\,
\dya{\time{\xi}}(t)=-\dya{\time{\zeta}}^T(t).
\end{equation}
These relations are equivalent to those for frequency-domain material tensors given by Eqs.~\r{onscas1}.

Regarding a linear \textit{time-varying} medium, whose macroscopic material parameters change in time, developing Eq.~\r{eq:llgctv} needs that we replace the material relations which take into account the general integral transform,  and this is not a simple task. According to that general form, the electric and magnetic flux densities are expressed as 
\begin{equation}
\begin{split}
&\ve{\time{D}}(t)=\int_0^\infty\hat{\time{\varepsilon}}(\tau,t)\cdot\ve{\time{E}}(t-\tau){\rm d}\tau+\int_0^\infty\hat{\time{\xi}}(\tau,t)\cdot\ve{\time{H}}(t-\tau){\rm d}\tau,\cr
&\ve{\time{B}}(t)=\int_0^\infty\hat{\time{\zeta}}(\tau,t)\cdot\ve{\time{E}}(t-\tau){\rm d}\tau+\int_0^\infty\hat{\time{\mu}}(\tau,t)\cdot\ve{\time{H}}(t-\tau){\rm d}\tau,
\end{split}
\l{eq:tvmdee}
\end{equation}
where $\hat{\time{\varepsilon}}(\tau,t)$, $\hat{\time{\mu}}(\tau,t)$, $\hat{\time{\xi}}(\tau,t)$, and $\hat{\time{\zeta}}(\tau,t)$ are operators depending on each moment of time $t$~\cite{pitaevskii_electric_1961}. In this case,  the response at any moment of time is associated strongly with these operators expressed at that moment. However, it is not the case for a time-invariant medium, in which Eq.~\r{eq:tvmdee} is simplified to Eq.~\r{dtbt}. This equation involves the conventional convolution integrals, while Eq.~\r{eq:tvmdee} takes into account the general integral transform  and as a consequence, developing the Lorentz lemma is not easy due to the dependency on $t$.

It should be mentioned that the Lorentz reciprocity relation~\r{eq:convtyrecip} can be applied to linear time-varying systems since all the requirements for the Onsager reciprocal relations (linearity, causality, microscopic reversibility, and thermodynamic quasi-equilibrium) are satisfied for such systems.

\subsection{Reciprocity applied to scattering parameters}

In many scenarios of solving an electromagnetic problem, it is not necessary to obtain exact wave solution at all points in space. Sometimes it is sufficient to determine the fields or voltages and currents only at specific boundaries (terminals). In this case, we model  a set of various electromagnetic components    of  arbitrary complexity as a ``black box'', to be exact, an electrical network. When an external electromagnetic signal or wave interact  with this network, we need to study only what output it produces for a given input, without solving   the fields inside the network. An example of an electrical network is a transmitting antenna. Fed  with an AC signal at its terminals, an antenna radiates electromagnetic waves in surrounding space. To improve radiation efficiency, one needs to decrease the parasitic reflections at the antenna terminals due to the impedance mismatch by adding a matching circuit. Full-wave solution  of this problem   (using the Maxwell equations) would be a resource-demanding task. Instead, we model the antenna as a ``black box'' with one input channel through  the cable (a one-port network) and the matching circuit as a two-port network. Subsequently, we determine the required properties of the matching circuit (its response to input) and design it using basic circuit elements. 

\begin{figure}[tb]
	\centering
\includegraphics[width=0.6\linewidth]{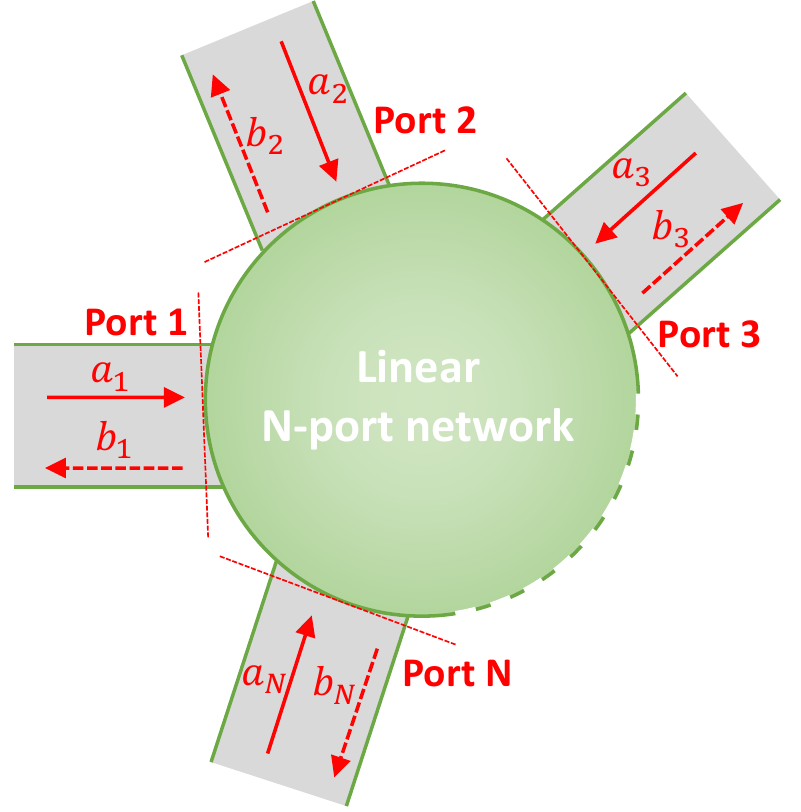} 	\caption{Schematic representation of a linear time-invariant $N$-port network and the scattering coefficients.}
\label{fig9schematic}
\end{figure}

There is plenty of different parameters for description of electrical networks~\cite[Ch.~4]{pozar_microwave_2012},\cite[Ch.~3]{marcuvitz_waveguide_1951}. Here, we will discuss only the scattering parameters (S-parameters) since they relate the input (the generalized forces) to the output signals or waves (response functions). As a consequence, we can directly apply to them the Onsager reciprocal relations. Other parameters, such as impedance, admittance, and transmission (ABCD) matrices, relate quantities which include both input and output signals. Reciprocity relations for such parameters can be derived from those of the scattering parameters (all these parameters can be expressed in terms of one another~\cite[p.~192]{pozar_microwave_2012}). 
% The scattering properties of linear   time-invariant (LTI) systems can be characterized using the scattering parameters for both reciprocal and nonreciprocal scenarios. 
According to the conventional notations of   $N$-port networks, the scattering coefficients $S_{ij}$ relate the normalized amplitudes of an incoming  and an outgoing signals. Scattering coefficients give information about the reflected or transmitted power and the phase shift produced by the system.  In the matrix form, the scattering parameters can be expressed as
\begin{equation}
   \left[ \begin{array}{cc}
b_1\\b_2\\\vdots\\b_{\rm N} \end{array} \right]=  \left[ \begin{array}{cccc}
S_{11} &
S_{12} & \cdots & S_{1 N} \\
S_{21} & & & \vdots\\
\vdots & & & \\
S_{ N1} &
\cdots &  & S_{N N} \end{array} \right] \left[ \begin{array}{cc}
a_1\\a_2\\\vdots\\a_{ N} \end{array} \right],
\l{eh2}
\end{equation} 
where $a_i$ and $b_i$ with $i=1,2, ...N$ represent the incoming  and outgoing signals,  respectively (see Fig.~\ref{fig9schematic}). With this definition, the tangential components of the fields in each port can be expressed as:
\begin{eqnarray}
 \ve{\freq{E}}_{t,i}= (a_i+b_i) \ve{\freq{e}}_i, \qquad \ve{\freq{H}}_{t,i}= (a_i-b_i) \ve{\freq{h}}_i
 \l{eh3}
\end{eqnarray}
with $i=1..N$ denoting the port number.
If scattering matrix is applied for circuits, the fields in \r{eh3} should be replaced by  voltages and currents.
The vectors  $\ve{\freq{e}}_i$ and $\ve{\freq{h}}_i$ represent the electric and magnetic modal fields in port~$i$. All the ports must be linearly independent, i.e. the fields should satisfy the orthogonality condition $\iint \freq{\_e}_i \cdot \freq{\_e}_j {\rm d}S= \delta_{ij}$~\cite[p.~5]{marcuvitz_waveguide_1951}, where the integration is extended over the cross section of the port and $\delta_{ij}$ denotes the Kronecker delta.   

Although usually scattering parameters are used for circuits in microwave engineering, they can be successfully applied for plane-wave propagation through different media. Consider   normal  incidence  of a wave of a given polarization on an interface of two materials.  This system can be characterized by two ports corresponding to the two sides of the interface. Now assume that the polarization of incident waves can be partially rotated by~$90^\circ$. In this case, it is convenient to model the interface by a four-port network: Two ports for the waves with original polarization and two other for the waves with rotated polarization. The ports are independent since the  two polarizations are orthogonal. The scattering matrix concept can be extended to  diffraction gratings with multiple orders~\cite{asadchy2017flat}. They provide a simple way to determine the power balance between different diffraction orders, the reciprocity conditions, etc.

As it was mentioned, the scattering parameters satisfy the conditions of the generalized susceptibilities~\r{ons3} and, additionally, all the requirements imposed on  the Onsager reciprocal relations~\r{ons6}. Since all the response functions in \r{eh2} have the same time-reversal symmetry (e.g., electric fields, magnetic fields, or currents), parameter $\sigma$ in~\r{ons6} must be taken $+1$ for all $i, k$ indices. As a result, the reciprocity condition  for scattering matrix is given by
\e \=S = \=S^T.
\l{eh4}
\f
In the general  case, \textit{all systems} (both reciprocal and nonreciprocal) must satisfy the Onsager-Casimir relation~\r{ons7} which is written for S-parameters as 
\e \=S (\_H_0)= \=S^T (-\_H_0).
\l{eh5}
\f
Interestingly, condition~\r{eh4} determines only overal reciprocity of a network.  A network can consist of multiple nonreciprocal components which compensate each other (pointwise nonreciprocity), while appear as reciprocal when probed at its ports. An example is a combination of two ferrite slabs magnetized in the opposite directions. Considering the system as a black box, one can only conclude  that it is  overal reciprocal.

Next, we will discuss how the characteristic of the system can be fathomed from the properties of the scattering matrix and give classical examples of reciprocal and nonreciprocal devices. 

\underline{\textit{Reciprocal  lossless systems}}: The scattering matrix is symmetric for reciprocal systems and unitary   for lossless systems~\cite[\textsection~4.3]{pozar_microwave_2012}. The latter condition is expressed as \mbox{$(\=S^{\ast})^T\cdot \=S =\=I$}. For example, if we consider a simultaneously lossless  and reciprocal 2-port system, the scattering matrix can be expressed as 
\begin{equation}\begin{array}{c}
\left[ \begin{array}{cc}
S_{11} &
S_{12}\\S_{21} &
S_{22} \end{array} \right]=
\left[ \begin{array}{cc}
{\rm e}^{j\theta}\sin{\beta} &
{\rm e}^{j\phi}\cos{\beta}\\{\rm e}^{j\phi}\cos{\beta} &
{\rm e}^{j\zeta}\sin{\beta} \end{array} \right],
\end{array} \label{eq78} \end{equation}
where $\beta \in [0;\pi/2]$ and the arguments satisfy the condition $2\phi - \theta-\zeta=\pi +2\pi m$ with $m$ being integer. 
An example of such a system is an isotropic non-dissipative dielectric slab (two interfaces define two ports), where we find symmetric transmissions and reflection. When multiple orthogonal modes are supported by each port (as in the case of plane waves with orthogonal polarizations), one can replace scalar elements in~(\ref{eq78}) by tensors $\=S_{11}$, $\=S_{22}$, $\=S_{12}$, and $\=S_{21}$.  
% a $4\times 4$ matrix can be written as a $2\times 2$ matrix whose each element is a  tensor: $\=S_{11}$, $\=S_{22}$, $\=S_{12}$, and $\=S_{21}$. This notation is useful when multiple orthogonal modes are supported in each port, for example in the case of plane waves with orthogonal polarization.
In this case, the reciprocal conditions are defined as $\=S_{11}=\=S_{11}^T$, $\=S_{22}=\=S_{22}^T$, $\=S_{12}=\=S_{21}^T$.

\underline{\textit{Nonreciprocal and lossless systems}}: It is evident that lossless and reciprocal systems provide a reduced number of degrees of freedom for the design. There are applications where it is necessary to break the strong condition imposed by reciprocity.
For example, one can think of phase shifters with different phase shifts depending on the direction. A canonical example of such devices is the \textit{gyrator}, a two-port network that introduces asymmetric phases in transmission with $\pi$ difference between them
\begin{equation}\begin{array}{c}
\=S_{\rm gyrator}=
\left[ \begin{array}{cc}
0 &
1\\-1 &
0 \end{array} \right].
\end{array} \end{equation}
A gyrator is considered as a fundamental non-reciprocal element that
in combination with four other     reciprocal elements, that is a resistor, capacitor, inductor, and ideal transformer, completes the set of building blocks needed for constructing an arbitrarily complex linear passive network~\cite{tellegen_gyrator_1948}.

For example, another
nonreciprocal and lossless device is a circulator, a three-port device where the signal can flow between ports 1$\rightarrow$2$\rightarrow$3, but not in the opposite direction. The scattering matrix of an ideal circulator can be expressed as 
\begin{equation}\begin{array}{c}
\=S_{\rm circulator}=
\left[ \begin{array}{ccc}
0 & 0 & 1\\
1 & 0 & 0\\
0 & 1 & 0  \end{array} \right].
\end{array} \end{equation}
A three-port circulator can be constructed using the basic nonreciprocal building block, the gyrator, and two quaterwave transmission lines. 

These scattering matrices that characterize these two examples are unitary, meaning that they are lossless systems. 
%Another example of nonreciprocal and lossless device is a circulator, a three-port device where the signal can flow from along ports 1$\rightarrow$2$\rightarrow$3, but not in the opposite direction. The scattering matrix of an ideal circulator can be expressed as 
%\begin{equation}\begin{array}{c}
%\=S_{\rm circulator}=
%\left[ \begin{array}{ccc}
%0 & 0 & 1\\
%1 & 0 & 0\\
%0 & 1 & 0  \end{array} \right].
%\end{array} \end{equation}
%These scattering matrices that characterize these two examples are unitary, meaning that they are lossless systems. 

\underline{\textit{Nonreciprocal and lossy systems}}: Finally, there are devices whose matrices are not symmetric nor unitary. One of the most important devices fulfilling these properties is the isolator:
\begin{equation}\begin{array}{c}
\=S_{\rm isolator}=
\left[ \begin{array}{cc}
0 &
0\\1 &
0 \end{array} \right].
\end{array} \end{equation}
This two-port device allows transmission in one direction, but both transmission or reflection are forbidden in the opposite direction. Importantly, lossless isolators cannot exist: A two-port network described by the above scattering matrix is matched at both ports, meaning that the wave falling on the isolated port cannot be reflected back and must be absorbed inside the isolator.

%\pink{From the previous examples, one can easily see that nonreciprocal behaviour is manifested as asymmetry in the scattering coefficients, $S_{ij}\neq S_{ji}$.}

Scattering parameters provide a most useful tool for the analysis of linear time-invariant systems that has been used in the  microwave engineering since the 1960´s. This formulation  has been extended to time-variant  systems~\cite{anderson_reciprocity_1965},\cite[\textsection~XIV]{caloz_electromagnetic_2018}, although these generalized parameters have restricted use. In the most general case, each terminal of the network is characterized by $M$-modes and $P$-frequencies. Considering that the system has $N$ different terminals, the characterization will be done using $N\times M \times P$ ports. For linear time-variant systems, the expression for each scattering parameter will be similar to the LTI case, $S_{ij}=b_i/a_j$. Lorentz reciprocity for time-variant systems was considered in~\cite{anderson_reciprocity_1965}.
%However, for nonlinear systems it will depend on the amplitude of all the input signals $S_{ij}=S_{ij}(a_1,a_2,...)$. 

%\begin{itemize}

%\item A lossless 1-port system can be only totally
%reflective, because $|S_{11}|^2=1$ even if it includes nonreciprocal materials.

%\item A lossless 2-port system is ``magnitude-wise reciprocal'': $|S_{12}|=|S_{21}|$. 
%General lossless 2-port: $$\overline{\overline{S}}=
%\begin{array}{c}
%\left[ \begin{array}{cc}
%S_{11} &
%S_{12}\\S_{21} &
%S_{22} \end{array} \right]=
%\left[ \begin{array}{cc}
%{\rm e}^{j\theta}\sin{\beta} &
%{\rm e}^{j{\color{red}\phi}}\cos{\beta}\\{\rm e}^{j{\color{red}%\psi}}\cos{\beta} &
%{\rm e}^{j\zeta}\sin{\beta} \end{array} \right]
%\end{array} $$

%\end{itemize}

\subsection{Different routes for breaking reciprocity } 

Here, we delve into the necessary physical conditions that warrant reciprocity in a system, as well as the possible ways to break it.
In the derivation of the Onsager reciprocal relations   presented in Section~\ref{onsrec}, the following physical assumptions were used:

\begin{enumerate}
    \item time-reversal symmetry of microscopic equations,
    \item linear response,
    \item causal response,
    \item thermodynamic quasi-equilibrium.
\end{enumerate}

The last condition should be discussed separately. All the previous formulations were supported by the assumption of  thermodynamic quasi-equilibrium or, in other words, assumption that the system is  in a stable and stationary state reached  after interactions with its surroundings for enough long time (so-called   linear or Onsager region~\cite{demirel_3_2007}). In this state, there are no net macroscopic flows of thermal energy.
%By contrast, non-equilibrium systems are characterized by net flows of energy. 
Particularly, in electromagnetic theory, this regime is achieved when the perturbations produced by the applied fields are slow enough to ensure that  the particles equilibrate to the surrounding particles. 

In order to achieve   nonreciprocity in a system, at least one of the mentioned conditions must be made invalid (however, it is not a sufficient condition). 
Thus, we can list several {\it possible} routes towards breaking reciprocity. 
The first condition of  time-reversal symmetry of microscopic equations can be violated by introducing to the system a time-odd external   force/parameter~$\_H_0$.   In this case, relation~\r{ons6} does not hold anymore $  \freq{\alpha}_{k i}(\omega, \_H_0) \neq \sigma \freq{\alpha}_{i k}(\omega, \_H_0)$, and the system may exhibit nonreciprocal response. Possible time-odd external parameters include but not limited to:
\begin{itemize}
\item external magnetic fields, e.g., applied to plasma or ferrite (see detailed discussion in Section~\ref{noneffects}),
\item exchange interaction force, e.g. in antiferromagnets,
\item linear velocity using linearly moving structures or linear space-time modulation (see detailed discussion in Section~\ref{nontv}),
\item angular velocity (rotating objects or  space-time modulation emulating rotation).
\end{itemize}
A separate discussion on the external time-odd parameters for breaking reciprocity will be given in Section~\ref{genclass}. Analogous routes towards electromagnetic nonreciprocity were reported in review paper~\cite[Table~I]{caloz_electromagnetic_2018}.

The second  condition of linear response can be naturally broken using nonlinear systems. However, as it will be discussed in Section~\ref{nonnon}, the nonlinearity route for breaking reciprocity is not universal and has its own limitations~\cite{shi_limitations_2015,sounas_nonreciprocity_2018}. 
The causality assumption does not apply to active systems\footnote{In active systems, the output may appear before input due to the  source external to the considered system (causality appears broken ``locally''). Naturally, in the global sense, all processes are causal.}, meaning that reciprocity can be broken in systems comprising amplifiers or parametric amplifiers~\cite{kodera_artificial_2011,wang_gyrotropic_2012,taravati_nonreciprocal_2017}.

The use of   systems far from equilibrium also appears possible for achieving nonreciprocity. It is  known that the fluctuation--dissipation theorem is violated  in non-equilibrium glassy systems (systems which slowly approach their equilibrium state)~\cite{crisanti_violation_2003}. In such systems the Onsager reciprocal relations do not necessarily hold.  

% \section{Examples of nonreciprocal meta-atoms and metasurfaces}

% Tellegen, ``moving'', time-modulated (discussing what is required for nonreciprocity)

\section{Nonreciprocity in linear time-invariant media}
In this section, phenomenological description of two nonreciprocal effects, namely Faraday rotation and Kerr ellipticity, is given. We list LTI materials in which these effects can occur. Furthermore, we introduce a general classification of nonreciprocal  LTI media based on their time- and space-reversal symmetries.

\subsection{Nonreciprocal effects using LTI materials } \label{noneffects}
Probably, the first known nonreciprocal effect dates back to the discovery by Faraday, made in 1845. He observed polarization rotation of linearly polarized light  propagating through a rod of lead borate glass placed in an external  static magnetic field~\cite{faraday_magnetization_1846}. By changing the direction of the magnetic field or the direction of the incident light beam, the sense of rotation is reversed. This property makes the Faraday rotation effect distinct from natural optical rotation in chiral materials. Linearly polarized light encounters double polarization rotation upon travelling through magnetically biased material forth and back. Whereas, the effect of natural optical rotation (chirality) vanishes in this case.
% Indeed, optical rotation effect in a chiral slab backed by a metal plate will be vanished since light experiences the same polarization rotation while travelling   in the forward and backward directions (note that clockwise rotation in the forward direction corresponds to counter-clockwise rotation in the backward direction). 

Let us present a phenomenological description of the Faraday effect in the framework of classical electrodynamics. Such description provides an intuitive route for understanding physics of the effect. However, its microscopic origin  is based on the spin-orbit interaction (Zeeman effect)   and relativistic effects~\cite[Ch.~5]{zvezdin_modern_1997}. 

\begin{figure}[tb!]
	\centering
	\subfigure[]{\includegraphics[width=0.44\linewidth]{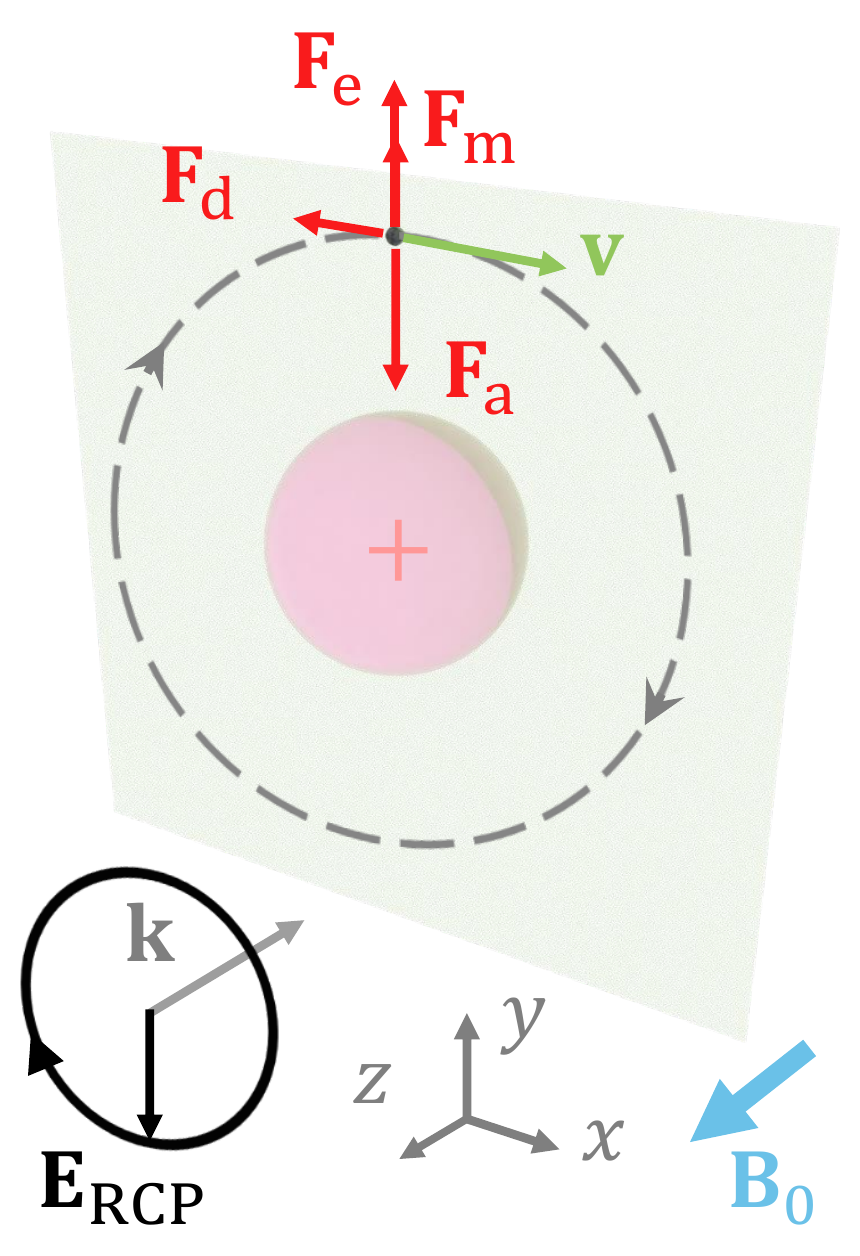} \label{fig9a}} \hspace{7mm}
	\subfigure[]{\includegraphics[width=0.44\linewidth]{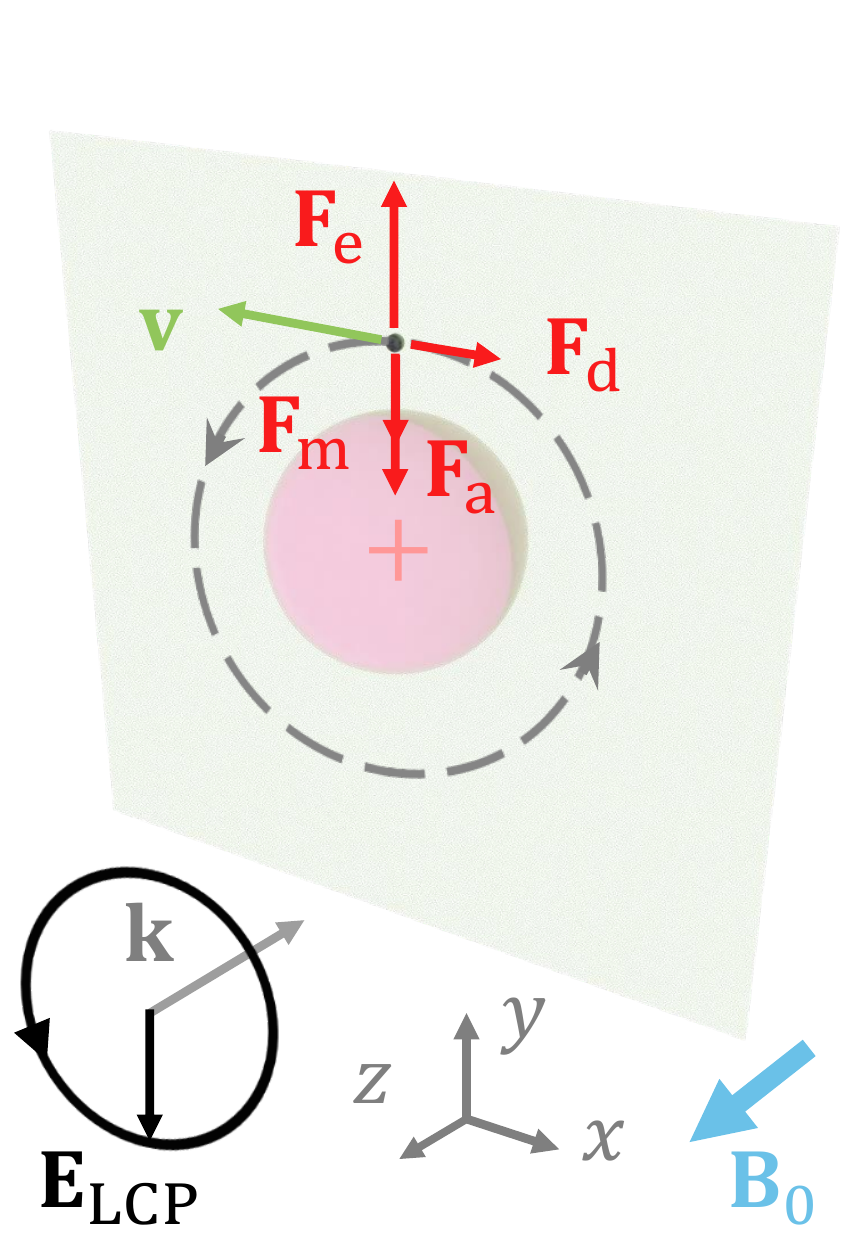}\label{fig9b}}
	\caption{Phenomenological description of circular birefringence of a magneto-optical material in external  magnetic field. The illustration depicts a free electron in the vicinity of positively charged ion.   Due to the oppositely directed Lorentz force, incident right and left polarized light cause electron's circulation at different orbit (cyclotron) radii, resulting in different refractive indices of the material. }
	\label{fig9}
\end{figure}

Consider a free electron located in the lattice of positively charged ions of a magneto-optical material. An incident light applies an external force on the electron, resulting in its displacement from the center of the atomic orbital. Here, we neglect the Lorentz force imposed on the electron by the \emph{alternating} magnetic field since it is typically much weaker than that by the electric field (see Ref.~\cite{oughstun_magnetic_2006} where the Lorentz force contribution is taken into account). Nevertheless, if the considered material is biased by a strong static magnetic field, the Lorentz force acting on the electron by this field must be included in the analysis. 
Let us assume the incident light with right  circular polarization (RCP) propagating in the material biased by an external magnetic flux density $B_0 \_z_0$ (with the direction towards the source of light).  It will cause the electron circulation in the polarization $x_0 y_0$ plane, as depicted in Fig.~\ref{fig9a}.  Note that the rotation of the electron occurs in the \textit{same} direction as the rotation of the incident electric field vector, despite the fact that the electron has negative electric charge. It is not the instanteneous   electric field that affects the electron motion direction, but its dynamics. 
 Here, we use the definition of the handedness sense as in electrical engineering literature~\cite[p.~24]{pozar_microwave_2012}, which is opposite to that in the optics literature~\cite[\textsection~2.4]{zvezdin_modern_1997}.

There are four forces acting on the electron in this configuration: The attractive Coulomb force pointing towards the center of the atomic orbital $\time{\_F}_{\rm a}=-k_{\rm a} \time{\_r}$, the  friction force $\time{\_F}_{\rm d}=-\Gamma {\rm d}\time{\_r}/{\rm d}t$ due to energy dissipation and directed opposite to electron's velocity, the force applied by the alternating electric field $\time{\_F}_{\rm e}=e \time{\_E}_{\rm i}$, and the Lorentz force acting on the circulating electron by the static magnetic field $\time{\_F}_{\rm m}= e B_0 {\rm d}\time{\_r}/{\rm d}t  \times  \_z_0$. Here $ k_{\rm a}$ is the effective stiffness coefficient, $\Gamma$ is the dissipation factor, $e$ is the elementary charge, and $\time{\_r}$ is the electron's position vector. The same electron under illumination by    incident light with left circular polarization (LCP) will circulate in the opposite direction and, therefore, will experience the oppositely directed  Lorentz force, as shown in Fig.~\ref{fig9b}. Thus, the total force acting on the electron towards the atomic orbital is different for the cases of light excitation with different circular polarizations. This results in different effective cyclotron radii of the electron orbit and,  subsequently, in different microscopic polarizabilities of the ion-electron pair and macroscopic refractive indices. Therefore, incident linearly polarized light (combination of right and left circularly polarized components) after propagating through the magnetized material acquires rotation of the polarization plane. 
Quantitative description of the mentioned effect can be made based on the Lorentzian model for the electron written in the form of  Newton's law of motion:
\e m \frac{{\rm d}^2\time{\_r}}{{\rm d}t^2}= -k_{\rm a} \time{\_r} - \Gamma \frac{{\rm d}\time{\_r}}{{\rm d}t}+ e \time{\_E}_{\rm i} + e B_0 \frac{{\rm d}\time{\_r}}{{\rm d}t}  \times  \_z_0,  \l{lorentz1}\f
where $m$ is the electon mass. By performing the direct Fourier transform to~\r{lorentz1}, one obtains the equation of motion in the frequency domain~\cite{kleemann_magneto-optical_2007,haider_review_2017}:
\e 
( m \omega_0^2 -m \omega^2 + j \omega \Gamma ) \, \freq{\_r}= e \freq{\_E}_{\rm i}  + j e B_0 \omega  \freq{\_r}  \times  \_z_0,  \l{lorentz2}\f
where $\omega$ is the angular frequency of electron motion, 
$\freq{\_r}$ and $\freq{\_E}_{\rm i}$ are the functions of $\omega$, $\omega_0=\sqrt{k_{\rm a}/m}$ is the material dependent constant with the dimensions of angular frequency. Writing the incident electric field  in the form  $\freq{\_E}_{\rm i}=\freq{E}_{{\rm i}x} \_x_0+ \freq{E}_{{\rm i}y} \_y_0$
and the position vector as $\freq{\_r}=x \freq{\_x}_0+y \freq{\_y}_0$, one obtains the following system of equations with respect to coordinates of the electron $\freq{x}(\omega)$ and $\freq{y}(\omega)$:
\e \begin{array}{c}\displaystyle
( m \omega_0^2 -m \omega^2 + j \omega \Gamma ) \freq{x}(\omega) - j \omega e B_0 \freq{y}(\omega)= e \freq{E}_{{\rm i}x},  \vspace{1mm} \\\displaystyle
( m \omega_0^2 -m \omega^2 + j \omega \Gamma ) \freq{y}(\omega) + j \omega e B_0 \freq{x}(\omega)= e \freq{E}_{{\rm i}y}.
\end{array}\l{lorentz3}\f

Assuming circular polarization of incident light $\freq{E}_{{\rm i}y}=\pm j \freq{E}_{{\rm i}x}= \freq{E}_{\rm i}/\sqrt{2}$ (upper and bottom signs stand for right and left circular polarizations, respectively), we conclude that   electron's motion is indeed circular $\freq{y}(\omega)=\pm j \freq{x}(\omega)=\freq{r}(\omega)/\sqrt{2}$ and the electric polarizability of the ion-electron pair $\freq{\alpha}_{\rm \frac{RCP}{LCP}}(\omega)= e \freq{r}(\omega)/\freq{E}_{\rm i}(\omega)/\varepsilon_0$ is given by\footnote{There is another mathematical solution when $ m \omega_0^2 -m \omega^2 + j \omega \Gamma = \mp e \omega B_0$. }
\e 
\freq{\alpha}_{\rm \frac{RCP}{LCP}}(\omega)=  \frac{e^2/ \varepsilon_0}{( m \omega_0^2 -m \omega^2 + j \omega \Gamma ) \pm e \omega B_0}.  \l{lorentz4}\f
The macroscopic refractive indices of the magnetized magneto-optical material for RCP and LCP read
\e 
\freq{n}_{\rm \frac{RCP}{LCP}}(\omega)=  \sqrt{1+N_{\rm e} \freq{\alpha}_{\rm \frac{RCP}{LCP}}(\omega)}=\freq{n}'_{\rm \frac{RCP}{LCP}}(\omega)- j \freq{\kappa}_{\rm \frac{RCP}{LCP}}(\omega), \l{lorentz5}\f
where $N_{\rm e}$ is the free electron concentration, $\freq{n}'$ and $\freq{\kappa}$ are the real part of refractive index and extinction coefficients. The Faraday rotation angle for linearly polarized light is readily calculated~\cite{kleemann_magneto-optical_2007,haider_review_2017} using
\e 
\theta_{\rm F}(\omega)=  \frac{\omega L}{2 c} [\freq{n}'_{\rm RCP}(\omega) -\freq{n}'_{\rm LCP}(\omega)   ],   \l{lorentz6}\f
where $L$ is the thickness of the magneto-optical slab and $c$ stands for speed of light in vacuum. The Faraday rotation angle is counted in such a way that it is related to the direction of $\_B_0$ by the right hand rule.
When $B_0=0$, the Faraday rotation angle becomes zero.

It is easy to demonstrate that the Faraday rotation angle flips sign for the opposite light propagation direction. Indeed, for the opposite illumination, direction of the magnetic static field with respect to the light wavevector reverses. This means that refractive indices for the opposite illumination $\freq{n}_{\rm \frac{RCP}{LCP}}^{\rm op}=\freq{n}_{\rm \frac{LCP}{RCP}}$ and the sense of rotation is flipped, i.e. $\theta_{\rm F}^{\rm op}=-\theta_{\rm F}$. This conclusion is in full agreement with Figs.~\ref{fig6c} and \ref{fig6d} and, as it was shown in Section~\ref{onscas}, such material response corresponds to a nonreciprocal effect. 
The Faraday ellipticity which determines how elliptical   linearly polarized light becomes after passing through the magneto-optical material is expressed as\cite{kleemann_magneto-optical_2007}
\e 
\tan \psi_{\rm F}(\omega)=  \tanh \left(\frac{\omega L}{2 c} [\freq{\kappa}_{\rm RCP}(\omega) -\freq{\kappa}_{\rm LCP}(\omega)   ] \right).   \l{lorentz7}\f
The ellipticity angle  $\psi_{\rm F}$ is defined as the angle between
the major axis of the ellipse and the diagonal of the rectangle that circumscribes the ellipse~\cite[\textsection~2.4]{zvezdin_modern_1997}. 

Next, solving Eqs.~\r{lorentz3} separately for    linearly polarized light beams along the $x$- and $y$-directions, one can determine the diagonal $\freq{\varepsilon}_{xx}=\freq{\varepsilon}_{yy}=\freq{\varepsilon}_{\rm s}$ and off-diagonal $\freq{\varepsilon}_{xy}=-\freq{\varepsilon}_{yx}=j\freq{\varepsilon}_{\rm a}$ components of permittivity tensor given by~\r{tensor}:
\e \begin{array}{l}\displaystyle
\freq{\varepsilon}_{xx}=\freq{\varepsilon}_{yy}=1+ N_{\rm e} e\frac{\freq{x}(\omega)}{\varepsilon_0 \freq{E}_{{\rm i}x}(\omega)}  
\vspace{1mm} \\\displaystyle
\qquad \qquad =1+ \Delta ( m \omega_0^2 -m \omega^2 + j \omega \Gamma ),  \vspace{1mm} \\\displaystyle
\freq{\varepsilon}_{yx}= -\freq{\varepsilon}_{xy}= N_{\rm e} e\frac{\freq{y}(\omega)}{\varepsilon_0 \freq{E}_{{\rm i}x}(\omega)}  = 
-j \Delta  \, \omega e B_0, 
\vspace{1mm} \\\displaystyle
\Delta= \frac{N_{\rm e} e^2 /\varepsilon_0 }{( m \omega_0^2 -m \omega^2 + j \omega \Gamma )^2- (\omega e B_0)^2 }.
\end{array}\l{lorentz8}\f
As is seen from \r{lorentz8}, the off-diagonal component of the permittivity tensor is  proportional  to the static magnetic flux density  $B_0$ (in the limit of small field, directly proportional).
It should be noted that in the accurate quantum mechanical description, permittivity tensor depends on the material magnetization $\mu_0 M$, rather than magnetic flux density $B_0$~\cite[\textsection~5.6.5]{coey_magnetism_2010},\cite[\textsection~5.2.2]{sugano_magneto-optics_2013}. 
 If dissipation is negligible ($\Gamma=0$), one can observe that the diagonal permittivity components become purely real, while the off-diagonal ones become purely imaginary. 
One can also see from \r{lorentz8} that the resonance of the cyclotron orbiting occurs at the frequency where the real part in the denominator of parameter $\Delta$ equals zero. This frequency is usually written as $\omega_{\rm c}= e B_0/m_{\rm eff}$~\cite[p.~571]{ashcroft_solid_1976}, where $m_{\rm eff}$ is the effective mass of the electron which takes into account the interaction with the ion. Since in  the majority of natural materials the cyclotron frequency $\omega_{\rm c}$ is relatively low (of the order of 10-1000~GHz for typical magnetic flux density $B_0 \sim 1$~T), in the optical range the magneto-optical effects are weak. Likewise, ferromagnetic properties of materials expressed by the antisymmetric part of the permeability  tensor are weak in the optical region, since the Larmor  resonance of the electron spin  $\omega_{\rm L}= e B_0/m$ (precession of the spin angular momentum) is located in the microwave range~\cite[p.~454]{pozar_microwave_2012}. 

By analogy with polarization rotation (conversion in general lossy case) of transmitted linearly polarized light through a sample of a magneto-optical material, one should also expect polarization conversion  for light reflected   from such material. Such an effect is nonreciprocal, as will be proved below, and is called    magneto-optical Kerr effect~\cite{ll.d_rotation_1877}. Depending on the mutual orientation of the magnetization direction of the material and the plane of incidence,   three basic configurations of the effect are distinguished: Polar, longitudinal, and transversal. Here, we consider only the former one, i.e. polar  magneto-optical Kerr effect.  A normally incident light beam is linearly polarized along the $x$-axis, and the external static magnetic flux density $B_0 \_z_0$ is along the $+z$-direction (antiparallel to the incident light direction). It is convenient to decompose the linearly polarized incident beam into two beams with left and right circular polarization, that is $\freq{E}_{\rm i} \_x_0=\freq{\_E}_{\rm RCP}+\freq{\_E}_{\rm LCP}=(\_x_0+j\_y_0) \freq{E}_{\rm i}/2+(\_x_0-j\_y_0) \freq{E}_{\rm i}/2$. Next, using the Fresnel formula for reflection coefficient, one can find the reflected field from the magneto-optical material:
\e 
\freq{\_E}_{\rm r}=  \frac{1-\freq{n}_{\rm RCP}}{1+\freq{n}_{\rm RCP}} \frac{\freq{E}_{\rm i}}{2} (\_x_0+j\_y_0) +
 \frac{1-\freq{n}_{\rm LCP}}{1+\freq{n}_{\rm LCP}} \frac{\freq{E}_{\rm i}}{2} (\_x_0-j\_y_0).   \l{lorentz9}\f
  \begin{figure}[tb]
	\centering
	\subfigure[]{\includegraphics[height=0.42\linewidth]{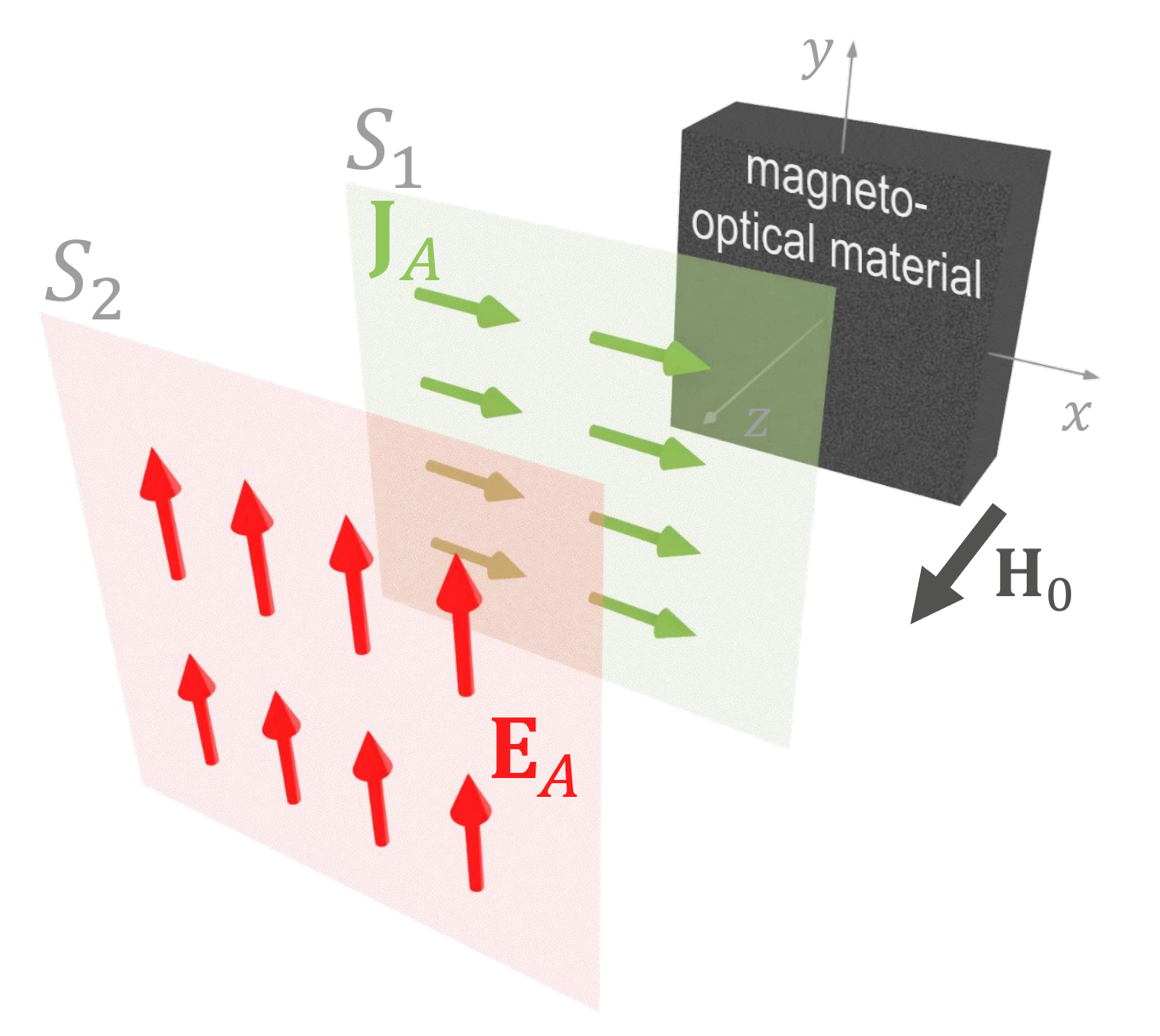} \label{fig10a}} 
	\subfigure[]{\includegraphics[height=0.42\linewidth]{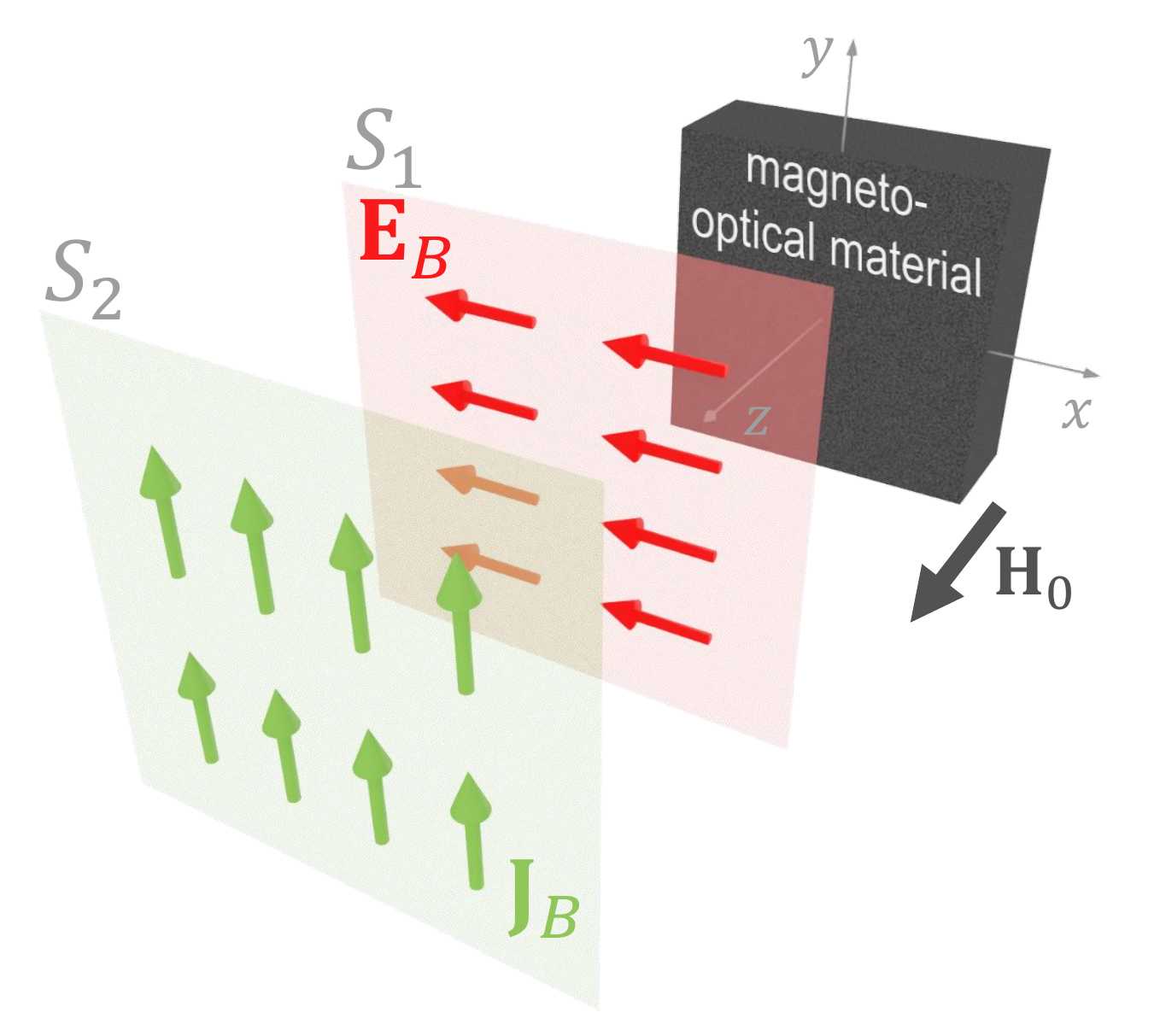}\label{fig10b}}
	\caption{ Application of the Lorentz reciprocity theorem for the magneto-optical Kerr effect. A biased slab of magneto-optical material reflects with additional polarization rotation an incident plane wave generated by current source (a) $\freq{\_J}_{A}$ and (b) $\freq{\_J}_{B}$.  The theorem does not hold since   $\int_{S_1} \freq{\_J}_A \cdot \freq{\_E}_B {\rm d}S  \neq \int_{S_2} \freq{\_J}_B \cdot \freq{\_E}_A {\rm d}S $. The illustrations do not depict the $x$-component of $\freq{\_E}_A$ and $y$-component of $\freq{\_E}_B$ (incident and reflected co-polarized  electric fields) since their contributions to the surface integrals are   zero.
	The two planes $S_1$ and $S_2$  were separated by one-wavelength distance for visual clarity.  }
	\label{fig10}
\end{figure}
Due to the magnetization by $B_0$, the reflected light acquired orthogonal polarization component along the $y$-direction. The ratio 
\e 
\frac{\freq{E}_{{\rm r}y}}{\freq{E}_{{\rm r}x}} =  j \frac{\freq{n}_{\rm RCP}-\freq{n}_{\rm LCP}}{\freq{n}_{\rm RCP}\freq{n}_{\rm LCP}-1}   
\l{lorentz10}\f
describes the Kerr rotation (real component of the ratio) and Kerr ellipticity (imaginary component)~\cite{kleemann_magneto-optical_2007}. It can be shown using \r{lorentz4}, \r{lorentz5}, and   \r{lorentz8} that~\footnote{This result is not accidental but can be alternatively derived based on the tensor conversion between the circular and linear bases. The generally correct formula $\freq{n}_{\rm \frac{RCP}{LCP}}^2=\freq{\varepsilon}_{\rm \frac{RCP}{LCP}}= (\freq{\varepsilon}_{xx}+\freq{\varepsilon}_{yy} \mp j \freq{\varepsilon}_{yx} \pm j \freq{\varepsilon}_{xy})/2$ is simplified to \r{lorentz11} due to the uniaxial symmetry of the tensor [see \r{lorentz8}].}
\e 
\freq{n}_{\rm \frac{RCP}{LCP}} =  \sqrt{\freq{\varepsilon}_{xx} \mp j \freq{\varepsilon}_{yx}}.
\l{lorentz11}\f
Note that in the lossless case both $\freq{\varepsilon}_{xx}$ and $j\freq{\varepsilon}_{yx}$ are purely real quantities. 
Since typically in natural magneto-optical materials the off-diagonal permittivity component is very small for realistic values of the bias field ($|\freq{\varepsilon}_{yx}| \ll |\freq{\varepsilon}_{xx}|$), we can rewrite \r{lorentz10} using \r{lorentz11} as
\e 
\frac{\freq{E}_{{\rm r}y}}{\freq{E}_{{\rm r}x}} =  
\frac{\freq{\varepsilon}_{yx}}{\sqrt{\freq{\varepsilon}_{xx} } (\freq{\varepsilon}_{xx}-1) }.
\l{lorentz12}\f
One can observe from \r{lorentz8} that for lossless magneto-optical materials ratio $\freq{E}_{{\rm r}y} /\freq{E}_{{\rm r}x}$ is purely imaginary, meaning that the reflected light has elliptical polarization. The axial ratio is proportional to the strength of the static magnetic  field.

\begin{figure*}[tb]
	\centering
\includegraphics[width=0.95\linewidth]{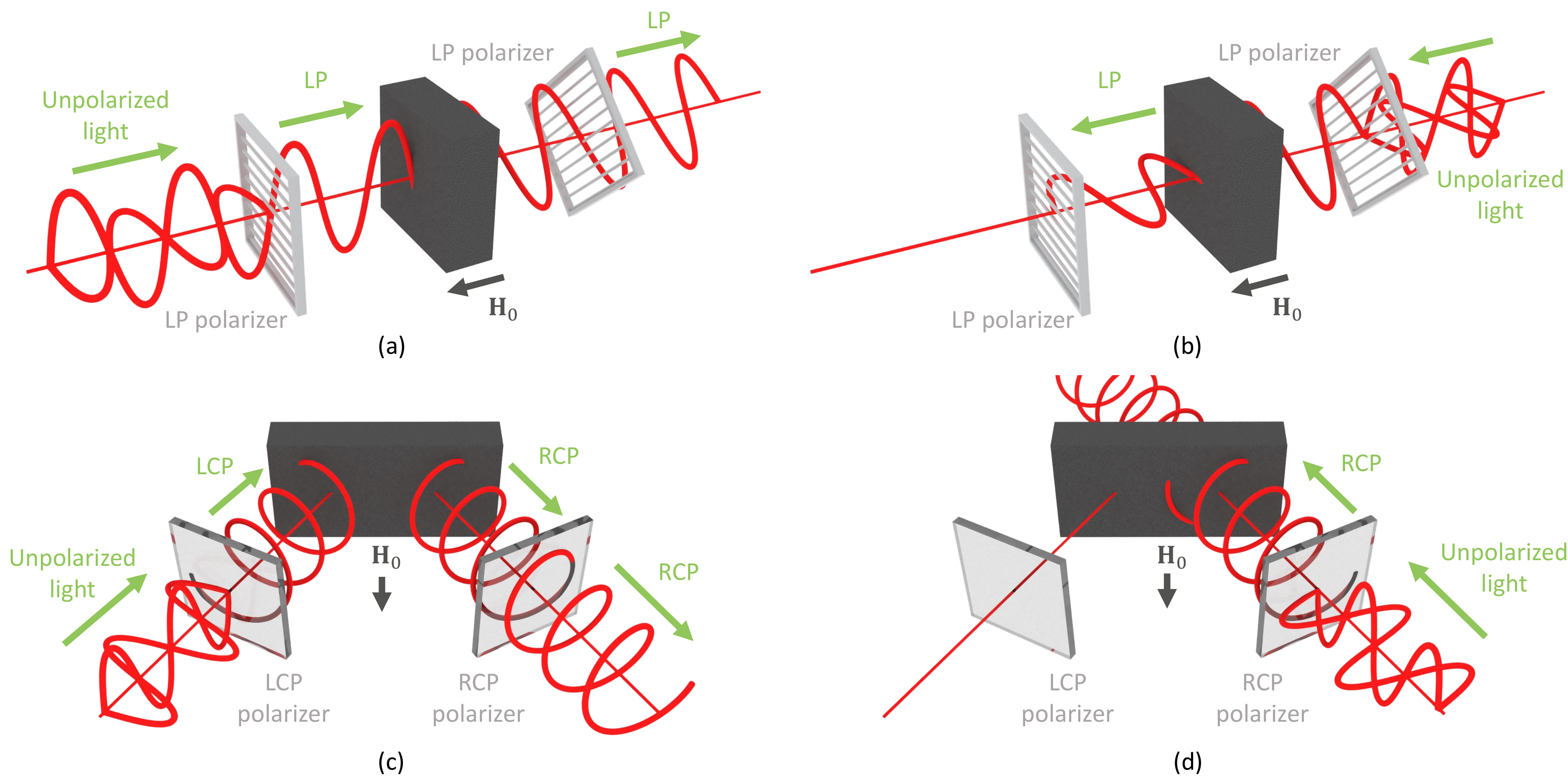} 
	\caption{ Electromagnetic isolation   based on (a--b) Faraday effect and (c--d) magneto-optical Kerr  effect. The first scenario requires presence of dissipation losses in the system, while the second scenario can be achieved in lossless systems. }
	\label{fig11}
\end{figure*}

In order to demonstrate that magneto-optical Kerr effect is a  manifestation of nonreciprocity, we need to find the reflected light for the case when the incident light is polarized along the $y$-direction. Carrying out analogous manipulations, we obtain for this scenario the axial ratio given by 
\e 
\frac{\freq{E}_{{\rm r}x}}{\freq{E}_{{\rm r}y}} =  
\frac{-\freq{\varepsilon}_{yx}}
{\sqrt{\freq{\varepsilon}_{xx} } (\freq{\varepsilon}_{xx}-1) },
\l{lorentz13}\f
which is precisely negative of \r{lorentz12}. In other words, some part (dictated by the right-hand side of \r{lorentz12}) of linearly $x$-polarized light  is reflected  in the $y$-polarization. However, when the incident light is $y$-polarized the same part of it is reflected in the $x$-polarization but with an additional phase flip of $180^\circ$ (reciprocal response would imply no phase change). Figures~\ref{fig10a} and \ref{fig10b} illustrate this  effect (note that for clarity 
 the figures   do not depict the incident and reflected co-polarized  electric fields, as explained in the caption). Using the Lorentz reciprocity theorem \r{ab3}, we conclude that   $\int_{S_1} \freq{\_J}_A \cdot \freq{\_E}_B {\rm d}S  \neq \int_{S_2} \freq{\_J}_B \cdot \freq{\_E}_A {\rm d}S $, and the polar magneto-optical  Kerr effect is nonreciprocal.

Thus, due to the Faraday and Kerr effects, light interacting with magnetized medium acquires polarization rotation. An important question is whether the opposite effects can exist, meaning that external light interacting with a medium leads to its magnetization. Indeed, such phenomena were theoretically predicted and are called inverse Faraday~\cite[eq. (36a)]{pitaevskii_electric_1961},\cite{pershan_nonlinear_1963} and inverse Kerr~\cite{belotelov_inverse_2012} effects. Their phenomenological description, analogous to the one in the present section, can be found, e.g. in~\cite{hertel_theory_2006,zhang_simple_2009}. The first experimental observation of the inverse Faraday effect was reported in~\cite{van_der_ziel_optically-induced_1965}. The magnetization is induced by a \emph{circularly polarized} optical pump of high energy (the circular  polarization of light beam generates  a  solenoid trajectories of the electrons which in turn result in net magnetic moments).  Therefore, this scheme can be exploited to obtain conventional Faraday polarization rotation of a probe signal without external magnetic field using only responsive optical pumping. 
Both these inverse effects are used for ultrafast (sub-picosecond) control of a medium magnetization  for modern
magnetic storage systems requiring very large operation rates~\cite{kimel_femtosecond_2007}.

Faraday and Kerr magneto-optical effects are not the only phenomena of nonreciprocal nature occuring in LTI materials. When a magnetic field is applied to a vapour or liquid 
through which light is passing perpendicularly to the field, magnetic linear birefringence takes place, resulting in so-called
Cotton-Mouton~\cite{cotton_sur_1907} or Voigt effect~\cite{smoluchowski_handbuch_1921}. This effect is typically very weak, depending quadratically on magnetization.

It should be mentioned that at microwaves, the magneto-optical effects are generally stronger than that at higher frequencies. The reason lies in the fact
that microwaves   are ``slow enough'' to excite resonant precession of electron's     magnetic moments (spin)  in   ferromagnetic materials.  In contrast to optical frequencies,   nonreciprocity in magneto-optical materials at microwaves is manifested by the off-diagonal components of the permeability tensor~\cite{polder_theory_1949}. Due to the duality of permittivity and permeability, the Faraday and Kerr effects can be observed also at microwaves~\cite{hogan_ferromagnetic_1952}.

\subsection{Applications of nonreciprocal effects and examples of LTI nonreciprocal materials}
Nonreciprocal effects in magneto-optical materials have found many important industrial applications (see a review in~\cite[Secs. 13,14]{zvezdin_modern_1997}). First of all, the Faraday effect enables control of polarization and amplitude of the transmitted light beam. This functionality is important for optical switches~\cite{shirasaki_bistable_1981} and light
modulators~\cite{scott_magnetooptic_1976} where fast electronic control  is required.  The latter is achieved via modulation of the electric current in the external electromagnet. Likewise, electronically tunable light deflection is possible using magnetic films with stripe   domain structure which behave as a diffraction grating~\cite{johansen_variation_1971}.   Another interesting application is   related to magneto-optical drives and requires ferromagnetic materials~\cite{chang_magnetooptical_1965}. The digital information is read from these drives based on the polar Kerr effect, discussed in the previous section. Nevertheless, after the success of the technology of magneto-optical drives in the 80-s and 90-s, it was completely surpassed by hard disk and solid-state drives. 

Probably,  the most eminent application of Faraday and Kerr effects is for electromagnetic  isolators~\cite{padula_optical_1967} and circulators~\cite{kobayashi_microoptic_1980}. One-way propagation based on Faraday effect is conventionally achieved in the configuration shown in Fig.~\ref{fig11}(a) where additional two linear polarizers are used from both sides of the magneto-optical material slab. The polarizers are rotated at $45^\circ$ with respect to one another, and the magneto-optical slab rotates polarization of the incident light by the same angle. Thus, for illumination from the left,  half of the unpolarized light is transmitted through both polarizers and the slab (see Fig.~\ref{fig11}(a)). On the contrary, light incident from the right side does not pass through the second polarizer being reflected (see Fig.~\ref{fig11}(b)). Importantly, in order to achieve perfect isolation effect, either polarizers or the material slab must possess certain level of dissipation losses (otherwise, isolation in the system would violate the second law of thermodynamics). Alternatively, such dissipation  can be mimicked by introducing additional channel to the system, e.g. by reflecting the light  from the second polarizer in Fig.~\ref{fig11}(b) towards the direction different from the isolator symmetry axis (in principle, it can be done by tilting the second polarizer).

\begin{table*}[tb]
\centering
 \begin{tabularx}{0.86\linewidth}{p{0.21\linewidth}p{0.15\linewidth}p{0.15\linewidth}p{0.15\linewidth}p{0.08\linewidth}}
\toprule
    \centering\textbf{Material \\ and Ref.}
    &\centering\textbf{Frequency range / Wavelength  range}  &\centering\textbf{Verdet constant, $V_{\rm c}$ $[{\rm rad} / ({\rm T} \cdot  {\rm m})]$}
    &\centering\textbf{Figure of merit, $V_{\rm c}/\alpha$  $[{\rm rad} / {\rm T}] $}
    &\centering\textbf{Temperature, \\$T$  $[K]$} \cr
    \midrule
    $\mathrm{Tb_3 Ga_5 O_{12}}$ (TGG) \cite{villora_faraday_2011,chen_preparation_2015} 
    & \centering 333--750~THz  / \\ 900--400~nm
    & \centering 60--475
    & \centering 175--10
    & \centering 300
    \cr
    \midrule
    Bi-doped $\mathrm{ YIG}$ \cite{bai_magneto-optical_2003} 
    & \centering 385--553~THz  / \\ 780--543~nm
    & \centering 384--5760
    & \centering 2.51--0.69
    & \centering 300
    \cr
\midrule
    $\mathrm{Cd_{0.39} Mn_{0.39} Hg_{0.22} Te}$ \cite{website} 
    & \centering 306--392~THz  / \\ 980--765~nm
    & \centering 262--1658
    & \centering 22.5--21.0
    & \centering 300
    \cr
\midrule
    $\mathrm{Rb}$ vapor \cite{weller_optical_2012} 
    & \centering 385~THz  / \\ 780~nm
    & \centering 1400
    & \centering 100
    & \centering 333
    \cr 
        \midrule
     $\mathrm{Cd_{0.95} Mn_{0.05} Te}$ \cite{gaj_giant_1978}
    & \centering 320--370~THz  / \\ 938--811~nm
    & \centering 140 -- 1396
    & \centering --
    & \centering 300
    \cr
    \midrule
    YbBi:YIG \cite{zhao_magneto-optic_2001}
    & \centering 170--300~THz  / \\ 1.76--1~$\mu$m
    & \centering  $(4.99 - 31.17)\cdot 10^3$
    & \centering  4.99--5.2
    & \centering 300
    \cr
    \midrule
    $\mathrm{In Sb}$ \cite{arikawa_giant_2012,tsakmakidis_breaking_2017} 
    & \centering 0.5--1.27~THz  / \\ 0.6--0.24~mm
    & \centering  $ (1.13 - 2.04)\cdot 10^4$
    & \centering  0.7--2.7
    & \centering 300
    \cr
\midrule
    $\mathrm{Sr Fe_{12} O_{19}}$ \cite{shalaby_magnetic_2013} 
    & \centering 80--800~GHz  / \\ 3.75--0.375~mm
    & \centering  1530
    & \centering  5
    & \centering 300
    \cr
\midrule
    ``Ferramic'' ferrite \cite{allen_microwave_1953}
    & \centering 9.3~GHz  / \\ 32.3~mm
    & \centering  $ 2.2 \cdot 10^4$
    & \centering  1420
    & \centering 300
    \cr
\bottomrule
\end{tabularx}
\caption{Properties  of some  magneto-optical materials in different frequency ranges (sorted in the order of decreasing frequency). }
\label{tab:2}
\end{table*}

Isolation based on the Kerr effect can be achieved using configuration illustrated in Fig.~\ref{fig11}(c) with two circular polarizers (the incident angle should be small). It should be mentioned that in this scenario the system can be \emph{lossless} and still respect   the second law of thermodynamics.     As it can be shown using \r{lorentz12} and \r{lorentz13}, in the lossless case, the magneto-optical material reflects incident LCP light into RCP (with complex amplitude $R_1$) and incident RCP light into LCP (with complex amplitude $R_2$). The same conclusion stems from  \r{lorentz9}. The reflection coefficients $R_1$ and $R_2$ become different if $\freq{n}_{\rm RCP} \neq \freq{n}_{\rm LCP}$, which is the case for magneto-optical materials  biased by an external field. 
In order to achieve perfect isolation, one must ensure enough strong gyromagnetic response from the material (magnitudes of the diagonal and off-diagonal permittivity or permeability components should be equal). Such strong response can be achieved at microwaves at the ferromagnetic resonance of some materials. In this case, $R_1$ can be designed to equal unity, while $R_2=0$. Thus, as shown in Fig.~\ref{fig11}(c), half of unpolarized light incident obliquely  from the left will be reflected as RCP light. At the same time, no light will be reflected for illumination from the right (Fig.~\ref{fig11}(d)), resulting in nonreciprocal isolation in reflection regime. It can be shown that if the system is lossless,   RCP light impinging on the magneto-optical slab from the right will be completely transmitted though it. In fact, the considered system   exhibits nonreciprocal wave propagation (isolation) also in the transmission regime. It fully transmits  RCP light incident opposite to the direction of $\_H_0$ but blocks it in the opposite direction. Likewise, it fully transmits LCP light incident along $\_H_0$ and blocks it in the opposite direction~\cite{silveirinha_optical_2016}. Being lossless, the system does not violate the unitarity condition of the scattering matrix because it includes four ports: With right and left circular polarizations on both sides of the slab.

Next, it is important to mention what materials with linear time-invariant response exhibit strong nonreciprocal effects. One of the ways to quantify the effect strength is based on the angle of Faraday rotation that  electromagnetic waves experience, passing through the material slab. Substituting Eq.~\r{lorentz11} into \r{lorentz6}, one can see that for the cases of small values of gyrotropy ($|\freq{\varepsilon}_{a}| \ll |\freq{\varepsilon}_{s}|$) the rotation angle is directly proportional to the slab thickness~$L$ and the bias magnetic flux density~$B_0$ ($\freq{\varepsilon}_{a} \propto B_0$).\footnote{As was mentioned above, the magneto-optical effects depend on the induced magnetization in the material $M$ rather than $B_0$. Therefore, $B_0$ in (93) should be replaced by $\mu_0 M$. } Therefore, it is convenient to write the expression for the rotation angle in the form:
\e 
\theta_{\rm F}=  - \frac{\omega L \freq{\varepsilon}_{a}}{2c \sqrt{\freq{\varepsilon}_{s}}}= V_{\rm c} B_0 L,
\l{lorentz14}\f
Here, $V_{\rm c}=V_{\rm c}(\omega,T)$ is the Verdet constant which describes the strength of Faraday rotation by the given material and depends only on the frequency and material temperature. For most nonreciprocal applications, materials with a high Verdet constant are preferable since in this case the required effect can be achieved with weaker bias field and thinner material slab, resulting in more compact and inexpensive design solutions.  An alternative commonly used constant to characterize the strength of the Faraday rotation is the Voigt constant, defined as $Q=\freq{\varepsilon}_{a}/\freq{\varepsilon}_{s}$. In the limit of small gyrotropy, it is directly proportional to the Verdet constant, as is seen from  \r{lorentz14}. It is important to note that Eq.~\r{lorentz14} is applicable only for the case when the off-diagonal permittivity component $\freq{\varepsilon}_{a}$ is much weaker than the diagonal one $\freq{\varepsilon}_{s}$. When  $\freq{\varepsilon}_{s}$ tends to zero (so-called ``epsilon-near-zero'' regime), the Faraday rotation angle $\theta_{\rm F}$ and the Verdet constant $V_{\rm c}$ in fact do not diverge, as it is predicted by~\r{lorentz14}. As it can be deduced from \r{lorentz11} and \r{lorentz6}, the largest Verdet constant for material with given $\freq{\varepsilon}_{a}$ is achieved when its off-diagonal permittivity reaches value $|\freq{\varepsilon}_{s}|=|\freq{\varepsilon}_{a}|$.
Another important characteristic of magneto-optical materials is the attenuation coefficient $\alpha_{\rm c}$ which determines the decay rate  of electromagnetic waves propagating through   the slab, i.e. ${\rm e}^{-\alpha_{\rm c} L}$. In most materials, a high Verdet constant is measured near the frequencies of optical transitions in atoms where strong absorption occurs. Therefore, the magneto-optical figure of merit $V_{\rm c}/\alpha_{\rm c}$ is usually adopted to characterize the suitability  of a given material for various applications (note that our figure of merit depends on  $B_0$ in contrast to that defined in~\cite[\textsection~9.6.5]{zvezdin_modern_1997}). Table~\ref{tab:2} contains the Verdet constant and the figure of merit for some magneto-optical materials in different frequency ranges (sorted in the order of decreasing frequency).

Most of magneto-optical materials exploited for electromagnetic wave processing could be divided into three groups: Magnetic insulators, diluted magnetic semiconductors, ferromagnetic semiconductors~\cite[Sec.~4]{kleemann_magneto-optical_2007}. The first group comprises magnetic ionic crystals such as magnetic garnets. The most common of them is yttrium iron garnet ($\mathrm{Y_3 Fe_5 O_{12}}$ or YIG). YIG  and its various doped versions, e.g.  co-doped Bi-substituted rare-earth garnet YbBi:YIG~\cite{zhao_magneto-optic_2001} and bi-doped YIG~\cite{bai_magneto-optical_2003}, possess high values of the Verdet constant  at microwaves, near infrared and visible frequencies (see Table~\ref{tab:2}). Although at microwaves losses are relatively weak in YIGs~\cite{allen_microwave_1953}, in the optical range they are relatively high, resulting in limited figure of merit (usually not exceeding five). Iron garnets (ferrites) found many applications in waveguide isolators and circulators at microwaves and even terahertz wavelengths (e.g. using strontium iron garnet $\mathrm{Sr Fe_{12} O_{19}}$ ~\cite{shalaby_magnetic_2013}).     
Another representative of the first group of magneto-optical materials are magnetoelectric and multiferroics which enable the manipulation of magnetic properties by an electric field~\cite{spaldin_advances_2019}.

The second group of magneto-optical materials is represented by  dilute magnetic (paramagnetic or semimagnetic)  semiconductors  that   are based on traditional semiconductors, but are doped with transition metals instead of, or in addition to, electronically active elements. The characteristic examples of such semiconductors are terbium gallium garnet ($\mathrm{Tb_3 Ga_5 O_{12}}$)~\cite{villora_faraday_2011,chen_preparation_2015} and cadmium manganese telluride ($\mathrm{Cd_{1-x} Mn_x Te}$)~\cite{gaj_giant_1978}. Both of them exhibit giant Faraday rotation with relatively small dissipation losses, resulting in a very high figure of merit (see Table~\ref{tab:2}). Such properties make these two materials best candidates for  commercial optical isolators. The high Verdet constant allows to design isolators of a millimeter scale.  Although paramagnetic semiconductors have a very  high Verdet constant  at room temperature, it can be further enhanced at cryogenic temperatures.

The third group is represented by ferromagnetic semiconductors which exhibit hysteretic magnetization behavior. Typical example is gallium manganese arsenides ($\mathrm{Ga_{1 - x} Mn_x As}$)~\cite{kimel_picosecond_2004}. The applications of ferromagnetic semiconductors include gateable ferromagnetism (where an electric field is used to control the ferromagnetic properties) and creation of  spintronic materials~\cite{neal_room-temperature_2006,gould_magnetic_2007}.

It should be mentioned that known magneto-optical materials are not limited to the mentioned three groups. For example, Faraday rotation of radio waves occurs in ionosphere plasma which consists of free electrons. Giant Faraday rotation  with low attenuation was observed in rubidium vapor~\cite{weller_optical_2012} (see Table~\ref{tab:2}) and in organic molecules with Verdet constant $V_{\rm c}=4300$~${\rm rad} / ({\rm T} \cdot  {\rm m})$~\cite{vandendriessche_giant_2013}. Furthermore, magneto-optics based on graphene  attracted considerable attention~\cite{gusynin_magneto-optical_2006,crassee_giant_2011}. Very recently, magneto-optical properties were demonstrated in semiconductors, such as   indium antimonide (InSb)~\cite{arikawa_giant_2012,tsakmakidis_breaking_2017}, which have response of  topological insulators~\cite{shuvaev_universal_2016,wu_quantized_2016,dziom_observation_2017}.
%~\footnote{Data is accurate for magnetic bias field $B_0=1$~T.}

Finally, it is important to mention  the emerging class of materials, magnetic Weyl semimetals~\cite{yan_topological_2017,kar_weyl_2019}, that exhibit   anomalous Hall effect and chiral magnetic effect~\cite{hofmann_surface_2016}. Due to their large Berry curvature, it was theoretically predicted that  Weyl semimetals may possess giant gyrotropic properties with or without external magnetic field (solely due to the internal magnetic ordering of the crystal structure). For example at the temperature $T=150$~K, Weyl semimetal  ${\rm Co}_3{\rm S}_2{\rm Se}_2$ can have 
the off-diagonal permittivity component $\freq{\varepsilon}_{a} \sim 1$   even in the optical range~\cite{kotov_giant_2018}. In Ref.~\cite{shuvaev_giant_2011} it was demonstrated that semimetal mercury telluride (HgTe) placed in an external magnetic field exhibits the Verdet constant of $V_{\rm c}=3 \cdot 10^6$~${\rm rad} / ({\rm T} \cdot  {\rm m})$.
Such strong gyrotropic properties   make Weyl semimetals best candidates for modern compact nonreciprocal optical devices with dimensions that are reduced by three orders of magnitude compared to conventional magneto-optical conﬁgurations~\cite{asadchy_sub-wavelength_nodate}.
% and in hybrid
% \cite{Surface Plasmon Resonance Enhanced
% Magneto-Optics (SuPREMO) Faraday
% Rotation Enhancement in Gold-Coated
% Iron Oxide Nanocrystals} with V=4.1e4 rad
% but no losses number was given

As modern photonics requires ultimate miniaturization of optical components, in the last decades, significant efforts have been devoted to the design of
compact optical nonreciprocal components. As was mentioned above, most magneto-optical materials, with a few exceptions such as indium antimonide and magnetic Weyl semimetals, exhibit low Verdet constant and require long propagation distances. Nevertheless, the dimensions of the nonreciprocal components based on such materials can be to some extent  reduced using several means. Characteristic examples include techniques based on   ring resonators or Mach--Zehnder interferometers~\cite{du_monolithic_2018,zhang_monolithic_2019}, 
  magnetic photonic crystals~\cite{figotin_nonreciprocal_2001,lyubchanskii_magnetic_2003,khanikaev_two-dimensional_2005,yu2007},
coupled surface plasmon polaritons~\cite{khanikaev_anomalous_2007,belotelov_extraordinary_2009,khanikaev_one-way_2010}, 
resonant metasurfaces~\cite{mousavi_gyromagnetically_2014,christofi_giant_2018},  and
Zeeman splitting in two-level systems~\cite{ying2018}.

\subsection{General classification of  LTI media based on space and time symmetries }  \label{genclass}
In the previous sections, we have discussed the concepts of time reversibility and reciprocity as well as phenomenology of nonreciprocal effects in linear time-invariant media. We mostly concentrated on particular effects in particular materials. On the contrary, in this section, we will make general observations that apply to all linear  time-invariant materials and can serve as an effective tool  for analyzing and  designing  new nonreciprocal systems. This study covers the general case of bianisotropic materials, i.e. materials where electric (magnetic) flux density can be induced by magnetic (electric) field. An overview of physics and applications of bianisotropic systems can be found in~\cite{Asadchy2018}.  

According to Noether's theorem, each  conservation law is associated with a specific symmetry property of a given system. For example, if a physical process exhibits the same outcomes regardless of  time, it leads to the fact that energy is conserved in this system. Since in this tutorial we consider only processes of electromagnetic nature\footnote{As an exception,  weak interaction processes do not conserve parity.}, they must obey both time- and space-reversal symmetries.   This property   does not imply that under space or time inversion the system remains unchanged (in fact, it can be even or odd with respect to these inversions). Instead, it implies that if space  and time inversions are applied to the entire process in the system, its result must also be a \emph{possible} physical process. Thus, space and time inversion symmetries provide an important constraint in addition to other constraints like energy conversation~\cite{rinard_faraday_1971,barron_parity_1972}. 
%In this section, for simplicity we assume the absence of energy dissipation in the considered media. The presented results can be extended to the lossy case via restricted time reversibility concept discussed in Section~\ref{restricted}. 
%In this section, we exploit the notion of time reversal in the way it was described in~Section~\ref{Loschmidt}, i.e. assuming that under reversal all the microscopic initial conditions are satisfied.

The electromagnetic quantities change signs under time reversal according to the list given in Table~\ref{tab:1}. Let us analyze time-inversion properties of material tensors in the frequency domain. We can rewrite material relations~\r{matrel1} in the general form as
\e \begin{array}{l}\displaystyle
\ve{\freq{D}}(\omega)=\dya{\freq{\varepsilon}}(\omega,\ve{Q})\cdot\ve{\freq{E}}(\omega)
+\dya{\freq{\xi}}(\omega,\ve{Q})\cdot\ve{\freq{H}}(\omega), \vspace{1mm} \\ \displaystyle
\ve{\freq{B}}(\omega)=\dya{\freq{\zeta}}(\omega,\ve{Q})\cdot\ve{\freq{E}}(\omega)
+\dya{\freq{\mu}}(\omega,\ve{Q})\cdot\ve{\freq{H}}(\omega),
\end{array}\l{class1}\f
where $\_Q$ is an arbitrary bias vector that  defines some external physical quantity (time-even or time-odd).  One can split each material tensor into two parts, representing separately \emph{linear} dependence on a time-even vector quantity in the form $\dya{\varepsilon}_{\rm TE}(\omega,\ve{Q}_{\rm TE})=\dya{\varepsilon}_1(\omega) + \dya{\varepsilon}_2(\omega)\ve{Q}_{\rm TE}$ and linear dependence on a time-odd quantity  $\ve{Q}_{\rm TO}$ in the form  $\dya{\freq{\varepsilon}}_{\rm TO}(\omega,\ve{Q}_{\rm TO})=\dya{\varepsilon}_3 \ve{Q}_{\rm TO}$, where $\dya{\varepsilon}_1$,
$\dya{\varepsilon}_2$, and $\dya{\varepsilon}_3$ are arbitrary tensors. The absence of dependence on  any external vector is modeled by assuming  $\ve{Q}_{\rm TE}=\ve{Q}_{\rm TO}=0$. Thus, we represent the permittivity tensor as  $\dya{\varepsilon}_{\rm TE}(\omega,\ve{Q})= \dya{\varepsilon}_{\rm TE}(\omega,\ve{Q}_{\rm TE})+\dya{\varepsilon}_{\rm TE}(\omega,\ve{Q}_{\rm TO})$ (likewise, for other three material tensors).   Applying time reversal to both sides of Eqs.~\r{class1}, we obtain 
\e \begin{array}{l}\displaystyle
\ve{\freq{D}}^\ast(\omega)=\left[\dya{\freq{\varepsilon}}_{\rm TE}' (\omega,\ve{Q}_{\rm TE}) + \dya{\freq{\varepsilon}}_{\rm TO}' (\omega,-\ve{Q}_{\rm TO}) \right]
\cdot\ve{\freq{E}}^\ast(\omega) \vspace{1mm}\\\displaystyle
\hspace{11mm}
-\left[\dya{\freq{\xi}}_{\rm TE}' (\omega,\ve{Q}_{\rm TE}) + \dya{\freq{\xi}}_{\rm TO}' (\omega,-\ve{Q}_{\rm TO}) \right] 
\cdot\ve{\freq{H}}^\ast(\omega), \vspace{2mm} \\
\displaystyle
-\ve{\freq{B}}^\ast(\omega)=\left[\dya{\freq{\zeta}}_{\rm TE}' (\omega,\ve{Q}_{\rm TE}) + \dya{\freq{\zeta}}_{\rm TO}' (\omega,-\ve{Q}_{\rm TO}) \right]
\cdot\ve{\freq{E}}^\ast(\omega) \vspace{1mm}\\\displaystyle
\hspace{15mm}
-\left[\dya{\freq{\mu}}_{\rm TE}' (\omega,\ve{Q}_{\rm TE}) + \dya{\freq{\mu}}_{\rm TO}' (\omega,-\ve{Q}_{\rm TO}) \right] 
\cdot\ve{\freq{H}}^\ast(\omega),
\end{array}\l{class2}\f 
where we have denoted time-reversed tensors with primes `` $'$ '' and exploited identities  $T\{ {\ve{\freq{H}}}(\omega)\}=-{\ve{\freq{H}}}^\ast(\omega)$ and $T \{ {\ve{\freq{E}}}(\omega) \}=+{\ve{\freq{E}}}^\ast(\omega)$ from~\r{tr} and \r{tr101} (similar identities hold for $\freq{\_B}$ and $\freq{\_D}$ vectors, respectively). 

Because of the time-reversal symmetry of the field equations (importantly, note that we have reversed also the \emph{external} fields), the system does not change its properties. Thus, the material relations of the time-reversed system must not change. This property allows us to find relations between the material parameters of the original and time-reversed system. To do that, we complex conjugate both sides of \r{class2} and compare the obtained equations with  Eqs.~\r{class1}. As a result, we come to the following expressions:
\e \begin{array}{l}\displaystyle
\dya{\freq{\varepsilon}}_{\rm TE}' (\omega,\ve{Q}_{\rm TE})=\dya{\freq{\varepsilon}}_{\rm TE}^\ast(\omega,\ve{Q}_{\rm TE}), \vspace{1mm}\\
\dya{\freq{\varepsilon}}_{\rm TO}' (\omega,\ve{Q}_{\rm TO})=-\dya{\freq{\varepsilon}}_{\rm TO}^\ast(\omega,\ve{Q}_{\rm TO}), 
\vspace{1mm}\\
\dya{\freq{\xi}}_{\rm TE}' (\omega,\ve{Q}_{\rm TE})=-\dya{\freq{\xi}}_{\rm TE}^\ast(\omega,\ve{Q}_{\rm TE}),
\vspace{1mm}\\
\dya{\freq{\xi}}_{\rm TO}' (\omega,\ve{Q}_{\rm TO})=\dya{\freq{\xi}}_{\rm TO}^\ast(\omega,\ve{Q}_{\rm TO}),
\end{array}\l{class3}\f
\e \begin{array}{l}\displaystyle
\dya{\freq{\zeta}}_{\rm TE}' (\omega,\ve{Q}_{\rm TE})=-\dya{\freq{\zeta}}_{\rm TE}^\ast(\omega,\ve{Q}_{\rm TE}),
\vspace{1mm}\\
\dya{\freq{\zeta}}_{\rm TO}' (\omega,\ve{Q}_{\rm TO})=\dya{\freq{\zeta}}_{\rm TO}^\ast(\omega,\ve{Q}_{\rm TO}),
\vspace{1mm}\\
\dya{\freq{\mu}}_{\rm TE}' (\omega,\ve{Q}_{\rm TE})=\dya{\freq{\mu}}_{\rm TE}^\ast(\omega,\ve{Q}_{\rm TE}), \vspace{1mm}\\
\dya{\freq{\mu}}_{\rm TO}' (\omega,\ve{Q}_{\rm TO})=-\dya{\freq{\mu}}_{\rm TO}^\ast(\omega,\ve{Q}_{\rm TO}), 
\end{array}\l{class4}\f
in which  we have used the fact that for arbitrary material tensor (let us denote it as $\dya{\freq{\rho}}$) relation
$\dya{\freq{\rho}} (\omega,-\ve{Q}_{\rm TO})=-\dya{\freq{\rho}}  (\omega,\ve{Q}_{\rm TO})$ 
holds due to the linear dependence on $\ve{Q}_{\rm TO}$. Although~\r{class3} and \r{class4} include eight tensors, only six of them are in fact independent. This conclusion can be deduced from the generalized Onsager-Casimir relations~\r{got}. Next, we rewrite~\r{got} with the present notations ($\_Q_{\rm TE}$ and $\_Q_{\rm TO}$ instead of $\_H_0$) and redefine tensors with notations commonly exploited in the literature~\cite[Eq.~(2.74)]{serdyukov_electromagnetics_2001}. Thus, we have
\e \begin{array}{l}\displaystyle
\dya{\freq{\varepsilon}}_{\rm TE} (\omega,\ve{Q}_{\rm TE})=\dya{\freq{\varepsilon}}_{\rm TE}^T (\omega,\ve{Q}_{\rm TE})=
\dya{\freq{\varepsilon}}_{\rm r} (\omega,\ve{Q}_{\rm TE}),
\vspace{2mm}\\
\dya{\freq{\varepsilon}}_{\rm TO} (\omega,\ve{Q}_{\rm TO})=-\dya{\freq{\varepsilon}}_{\rm TO}^T (\omega,\ve{Q}_{\rm TO})=
\dya{\freq{\varepsilon}}_{\rm n} (\omega,\ve{Q}_{\rm TO}),
\vspace{2mm}\\
\dya{\freq{\mu}}_{\rm TE} (\omega,\ve{Q}_{\rm TE})=\dya{\freq{\mu}}_{\rm TE}^T (\omega,\ve{Q}_{\rm TE})=
\dya{\freq{\mu}}_{\rm r} (\omega,\ve{Q}_{\rm TE}),
\vspace{2mm}\\
\dya{\freq{\mu}}_{\rm TO} (\omega,\ve{Q}_{\rm TO})=-\dya{\freq{\mu}}_{\rm TO}^T (\omega,\ve{Q}_{\rm TO})=
\dya{\freq{\mu}}_{\rm n} (\omega,\ve{Q}_{\rm TO}),
\vspace{1mm}\\ \displaystyle
\dya{\freq{\xi}}_{\rm TE} (\omega,\ve{Q}_{\rm TE})=-\dya{\freq{\zeta}}_{\rm TE}^T (\omega,\ve{Q}_{\rm TE})=- \frac{j}{c} \dya{\freq{\kappa}} (\omega,\ve{Q}_{\rm TE}),
\vspace{1mm}\\ \displaystyle
\dya{\freq{\xi}}_{\rm TO} (\omega,\ve{Q}_{\rm TO})=\dya{\freq{\zeta}}_{\rm TO}^T (\omega,\ve{Q}_{\rm TO})= \frac{1}{c} \dya{\freq{\chi}} (\omega,\ve{Q}_{\rm TO}).
\end{array}\l{class5}\f
Equations~\r{class5} determine the symmetry of the material tensors. For example,  tensors corresponding to reciprocal electromagnetic response $\dya{\freq{\varepsilon}}_{\rm r}$ and $\dya{\freq{\mu}}_{\rm r}$ are symmetric (as in usual dielectrics and magnetics). At the same time, those associated with nonreciprocal response $\dya{\freq{\varepsilon}}_{\rm n}$ and $\dya{\freq{\mu}}_{\rm n}$ are antisymmetric  and responsible for Faraday rotation in magneto-optical materials.  Reciprocal chirality tensor    $\dya{\freq{\kappa}}$ and  nonreciprocal Tellegen tensor $\dya{\freq{\chi}}$ (named after B.~Tellegen who introduced realization of such nonreciprocal material~\cite{tellegen_gyrator_1948})  have general form and can be further decomposed to symmetric and antisymmetric parts. It is important to notice that   in \r{class5} the imaginary unit $j$ appears only for the chirality tensor  $\dya{\freq{\kappa}}$. It is an important consequence of   the fact that chirality is an effect of spatial dispersion and, therefore, vanishes when $\omega \rightarrow 0$, while  Tellegen magnetoelectric coupling can exist even in locally uniform external fields (when sizes of the medium constituents  are
negligibly small compared to the wavelength). This difference can be easily observed in the time-domain form of material relations within the Condon model~\cite{condon_theories_1937,sihvola_bi-isotropic_1991}
\e \begin{array}{l}\displaystyle
\ve{\time{D}}(t)=\dya{\time{\varepsilon}}_{\rm c}(\ve{Q})\cdot\ve{\time{E}}(t)
+\frac{1}{c} \left[ \dya{\time{\chi}}_{\rm c}(\ve{Q})-
 \dya{\time{\kappa}}_{\rm c}(\ve{Q}) \cdot \frac{\partial}{\partial t} \right]  \ve{\time{H}}(t), \vspace{1mm} \\ \displaystyle
\ve{\time{B}}(t)=\dya{\time{\mu}}_{\rm c}(\ve{Q})\cdot\ve{\time{H}}(t)
+\frac{1}{c} \left[ \dya{\time{\chi}}_{\rm c}^T(\ve{Q})+
 \dya{\time{\kappa}}_{\rm c}^T(\ve{Q}) \cdot \frac{\partial}{\partial t} \right]  \ve{\time{E}}(t),
\end{array}\l{class6}\f
where for harmonic fields $\omega \dya{\time{\kappa}}_{\rm c}=\dya{\freq{\kappa}}$. Comparing the Condon model written above with~\r{class1}, we can conclude that the Condon model can be used in time domain assuming that  $\dya{\time{\varepsilon}}_{\rm c}$, $\dya{\time{\kappa}}_{\rm c}$, 
$\dya{\time{\mu}}_{\rm c}$, and $\dya{\time{\chi}}_{\rm c}$ are constant values\footnote{For artificial magnetics one should use $\dya{\time{\mu}}_{\rm c}=\mu_0$ in time-domain models.} (note that dimensions of some of these quantities are different from those in \r{dtbt}). In this case, the model neglects dispersion of the permittivity and permeability and correctly accounts for the frequency dispersion of chirality at low frequencies (well below the resonances of molecules or meta-atoms). However,  the Condon model can be modified so that in the above equation we have the convolution integrals instead of the direct multiplication. Using convolution integrals, one can fully take into account the frequency dispersion of all parameters and achieve a similar form mentioned in~\r{class1}. As a consequence, we still can define $\omega \dya{\freq{\kappa}}_{\rm c}=\dya{\freq{\kappa}}$ in the frequency domain and $\dya{\freq{\kappa}}$ is proportional to $\omega$:
\begin{equation}
\dya{\freq{\xi}}=-{j\over c}\omega\dya{\freq{\kappa}}_{\rm c}=-{j\over c}\dya{\freq{\kappa}}.
\l{eq93}
\end{equation}
It is worth noting that the Condon model is based on the assumption  that the time derivative of the magnetic field is the cause of electric polarization contributing the response function~$\mathbf{\time{D}}$. There are two consequences of this assumption. The first is that, similarly to permittivity and permeability, $\dya{\time{\kappa}}_{\rm c}$ can be considered as a generalized  susceptibility in Eq.~\r{ons3}\footnote{Equation~\r{ons3} implies that $\dya{\freq{\kappa}}_{\rm c}$ in the frequency domain satisfies relation $\dya{\freq{\kappa}}_{\rm c}^\ast (\omega)=\dya{\freq{\kappa}}_{\rm c}(-\omega)$~\cite[Eq.~(123.6)]{landau_course_1980}.}, where  the generalized force is the time derivative of magnetic field. %Notice that based on~\r{eq93}, one can simply deduce that in contrast to $\dya{\time{\kappa}}_{\rm c}$, the parameter $\dya{\time{\kappa}}$ does not satisfy the generalized susceptibility condition. 
The second consequence is about time-reversal transformation. Since the time derivative of the magnetic field and the electric flux density are invariant under time reversal,  $\dya{\time{\kappa}}_{\rm c}$ also does not change sign under this  transformation.
Basically, the modified Condon model accounting for frequency dispersion and model~\r{class1} are equivalent according to~\r{eq93}. The only difference is in the definition of chirality parameter.  In what follows, we use the model given by~\r{class1} since it is more common in the literature.
% Here, we do not introduce these convolution integrals and let us say that more accurate form with frequency dispersion can be achieved by direct Fourier transform of~\r{class1}.
% {\bf I do not quite understand.  My understanding is the following: 
% Since the chirality effects are due to dispersion (there is no chiral coupling in statics), the material relations in the Tellegen form (with the chirality parameter $\kappa$) can be used only in the frequency domain. In time domain, one has to write convolution integrals to account for dispersion. This difficulty can be partially overcome using the Condon form, where the dispersive nature of chirality is modelled by coupling terms which are proportional to the time derivatives of the fields. This formalism can be used for weak chirality (small frequency, far from resonances of the molecules or inclusions). In this case, equations can be written in time domain, assuming that the parameter $\kappa_c$ is a constant. This, of course, corresponds to the approximation of $\kappa$ in the Tellegen model as a linear function of $\omega$, which is also valid at low frequencies, far from resonances. About the notion of "true parameter", if we use it, we need to explain the meaning. }

\begin{figure}[tb]
\centering
   \includegraphics[width=0.95\columnwidth]{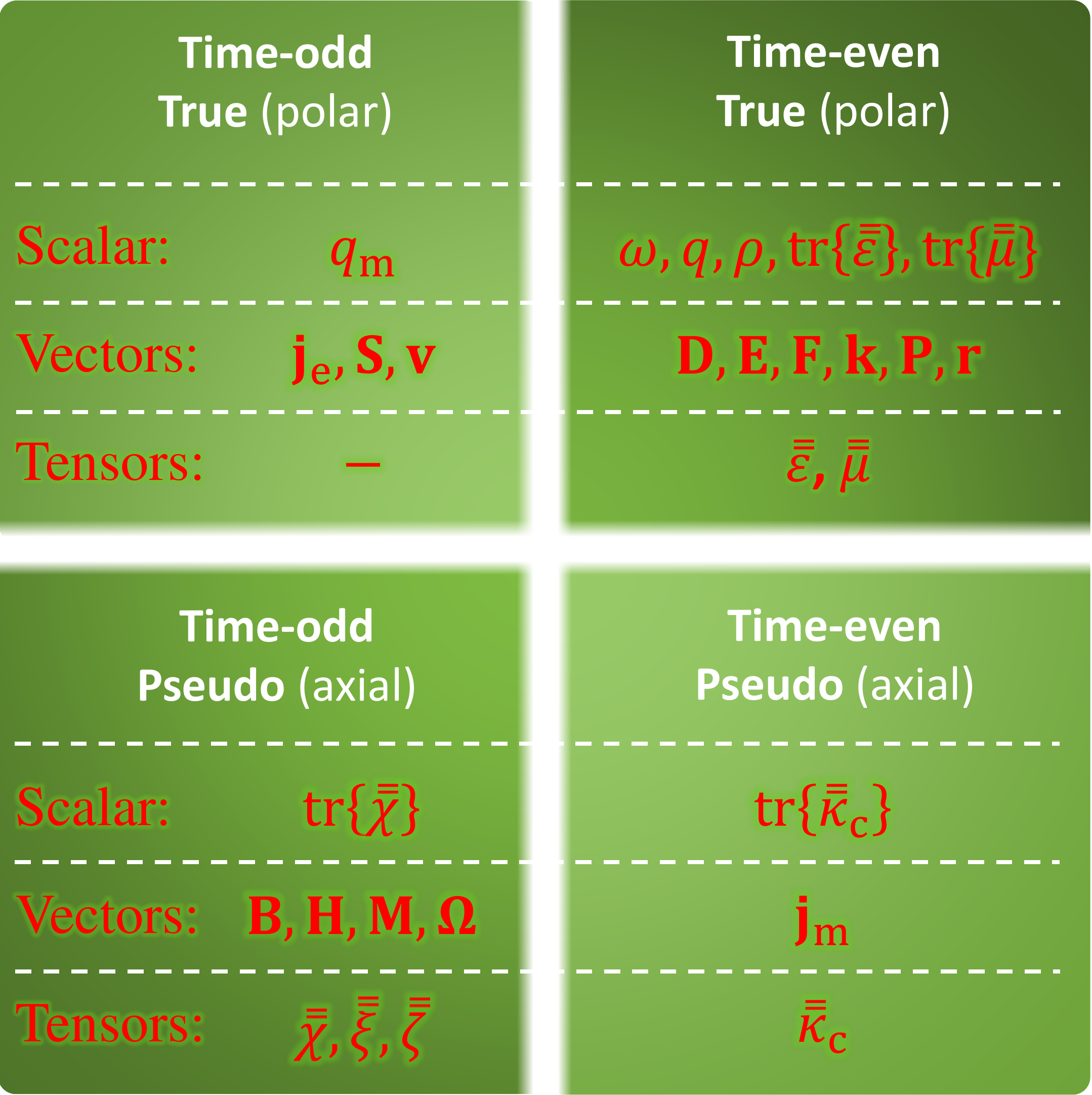}
\caption{Properties of some electromagnetic scalars, vectors, and tensors with respect to frequency and parity inversions. }
\label{fig12}
\end{figure}

From \r{dtbt}, one can deduce that the time-domain kernels $\dya{\time{\varepsilon}}(t)$ and $\dya{\time{\mu}}(t)$ are even under time reversal, while   $\dya{\time{\xi}}(t)$  and $\dya{\time{\zeta}}(t)$ are odd. The   chirality tensor  $\dya{\time{\kappa}}_{\rm c}(t)$ is time-even, as is seen from~\r{class6}. The Tellegen coupling coefficient is, in contrast, time-odd, i.e. $T\{ \dya{\time{\chi}}(t) \}=  \dya{\time{\chi}}'(t) = -\dya{\time{\chi}}(-t)$. These facts are reflected in Fig.~\ref{fig12}. 
Applying \r{class5} to \r{class3} and \r{class4},  we can get the time-reversal transformations of the material tensors in the frequency domain: 
\e \begin{array}{l}\displaystyle
 \hspace{-3mm} \dya{\freq{\varepsilon}}'_{\rm r}(\omega,\ve{Q}_{\rm TE})= \dya{\freq{\varepsilon}}_{\rm r}^\ast (\omega,\ve{Q}_{\rm TE}), \hspace{2mm} \dya{\freq{\varepsilon}}'_{\rm n}(\omega,\ve{Q}_{\rm TO})= -\dya{\freq{\varepsilon}}_{\rm n}^\ast(\omega,\ve{Q}_{\rm TO}),
 \vspace{1mm} \\ \displaystyle
 \hspace{-3mm} \dya{\freq{\mu}}'_{\rm r}(\omega,\ve{Q}_{\rm TE})= \dya{\freq{\mu}}_{\rm r}^\ast(\omega,\ve{Q}_{\rm TE}), \hspace{1mm} \dya{\freq{\mu}}'_{\rm n}(\omega,\ve{Q}_{\rm TO})= -\dya{\freq{\mu}}_{\rm n}^\ast (\omega,\ve{Q}_{\rm TO}),
  \vspace{1mm} \\ \displaystyle
 \hspace{-3mm} \dya{\freq{\kappa}}'(\omega,\ve{Q}_{\rm TE})=\dya{\freq{\kappa}}^\ast(\omega,\ve{Q}_{\rm TE}), \hspace{5mm} \dya{\freq{\chi}}'(\omega,\ve{Q}_{\rm TO})= \dya{\freq{\chi}}^\ast(\omega,\ve{Q}_{\rm TO}).
\end{array}\l{class7}\f
% From these relations one can observe that tensors $\dya{\freq{\varepsilon}}_{\rm r}$, $\dya{\freq{\mu}}_{\rm r}$ and  $\dya{\freq{\chi}}$ are even functions with respect to frequency, whereas $\dya{\freq{\varepsilon}}_{\rm n}$, $\dya{\mu}_{\rm n}$ and  $\dya{\kappa}$ are odd functions~\footnote{In  time domain, $\dya{\kappa}_\omega(t,\ve{Q}_{\rm TE})$ is an even function of time.}. Therefore, material tensors $\dya{\varepsilon}_{\rm n}$, $\dya{\mu}_{\rm n}$ and  $\dya{\kappa}$ tend to zero when $\omega \rightarrow 0$. 

Figure~\ref{fig12}   richly illustrates the contrast between different material tensors and their space-time symmetry. This information is essential for understanding   phenomena which might have similar character but drastically distinct origins. For example, optical rotation effect can be observed in   chiral materials and in biased magneto-optical materials, as seen in Fig.~\ref{fig6}, however, these effects have a different nature. This difference was   understood already by Faraday and commented by Kelvin in his lectures~\cite{kelvin_baltimore_1904}. 
\begin{figure*}[tb]
	\centering
\includegraphics[width=0.98\linewidth]{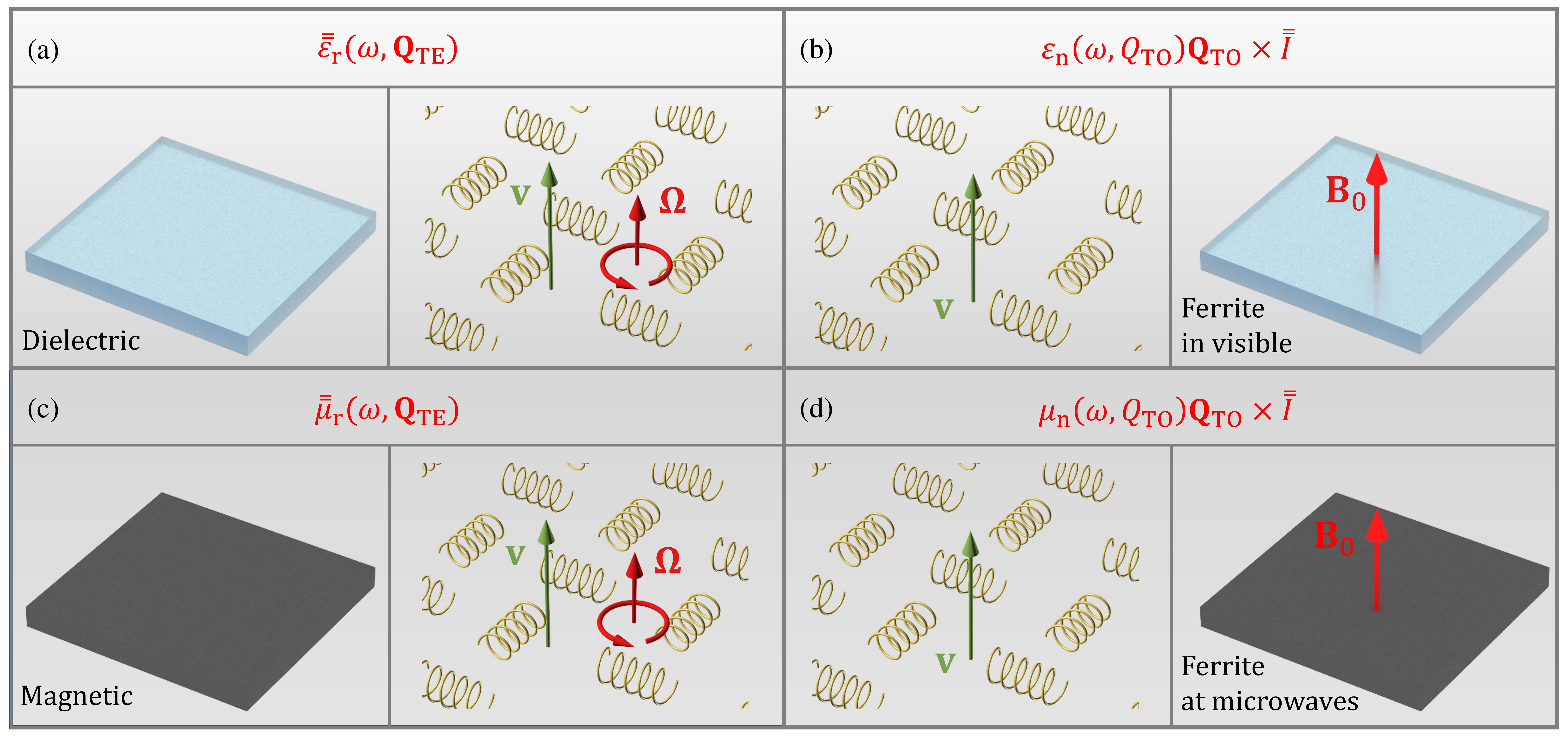}  
	\caption{ Conceptual implementation of different material tensors. Two characteristic examples are shown for each tensor.   }
	\label{fig13}
\end{figure*}

Next, we investigate the properties of material tensors with respect to parity inversion.  As it was discussed above, the necessary condition for the existence of  nonreciprocal effects in a material is that its response depends on an external   parameter  which has time-odd symmetry $\_Q_{\rm TO}$. However, in order to achieve nonreciprocal bianisotropic response of a specific type, one should also consider spatial symmetry properties of the material constituents.

Let us consider an arbitrary vector $\_a= a_x \_x +a_y\_y+a_z\_z$ in the Cartesian coordinate system  ($\_x$, $\_y$, and $\_z$ are the  basis  unit vectors). Under ``active''\footnote{Here ``active'' parity inversion applies to the object while the coordinate system is unchanged~\cite[p.~268]{Jackson1999}. Alternative definition of ``passive'' parity inversion implies that the object is unchanged while the coordinate system is reversed.} parity inversion (point reflection),  the coordinates of the vector change and  the vector in the new system will have form $\_a'= -a_x \_x -a_y\_y-a_z\_z=-\_a$. This transformation is equivalent to direction inversion of the vector. 
Vectors which obey this transformation rule are called \textit{true} (polar) vectors. Physical vectors such as linear speed $\_v$, force $\_F$, wavevector $\_k$,  differential operator $\nabla$, and position vector $\_r$ are true vectors.

Next, let us consider a vector which is a result of the vector product of two true vectors, i.e., $\_a=\_b\times\_c$. In the initial physical  system, this vector has the   form of
$\_a= (b_y c_z- b_z c_y)\_x+(b_z c_x- b_x c_z)\_y+(b_x c_y- b_y c_x)\_z$.
Under the ``active'' parity inversion, both true vectors $\_b$ and $\_c$ flip sign (the cross product operation does not change), while their vector product $\_a'=\_b'\times\_c'$ remains unchanged, i.e. $\_a'=\_a$.
Thus, vector $\_a$,  formed as   cross products of two true vectors, transforms differently under parity inversion compared to true vectors. Such vectors are called \textit{pseudovectors} (axial vectors).  Under parity inversion a pseudovector transforms as a true vector with an    additional sign flip. 
It should be noted that the difference in the properties of true vectors and pseudovectors occurs only under parity inversion and does not appear under rotational coordinate transformations. It is easy to check that scalar product of two true vectors gives a true scalar, i.e. a scalar which does not change under parity inversion.

Similarly, it can be shown that a cross product of a true vector and a pseudovector results in a true vector (cross product can be thought as a pseudo-operator itself).  An example of pseudovectors in electrodynamics is the magnetic induction vector $\_B$. Indeed, according to $\_F=q \_E+q\_v \times \_B$, the force which is  acting on a moving electric charge by the field $\_B$ can be a true vector only if $\_B$ is a pseudovector. Other examples of pseudovectors are magnetic field $\_H$, magnetization $\_M$, and orbital angular velocity vector $\Omega$ (see Fig.~\ref{fig12}). From the macroscopic Maxwell equations, it follows that  electric field $\_E$, displacement field  $\_D$, electric current density $\_j_{\rm e}$ and polarization $\_P$ are all true vectors. 
Classification to true and pseudo quantities can be extended to scalars and tensors of arbitrary rank assuming that pseudo quantity transforms under parity inversion like a true quantity but with an additional sign flip. True scalars include electric charge $q$, magnetic charge $q_{\rm m}$, frequency $\omega$, electric charge density $\rho$ and the traces (sums of diagonal components) of permittivity and permeability tensors. An example of a   pseudoscalar is the trace of chirality  tensor ${\rm tr}\{ \dya{\kappa}\}$.  Under parity inversion of  chiral isotropic material formed, for example, by helical inclusions, the sign of the chirality parameter of the material changes since the handedness of the helices flips. 

\begin{figure*}[tb]
	\centering
\includegraphics[width=0.98\linewidth]{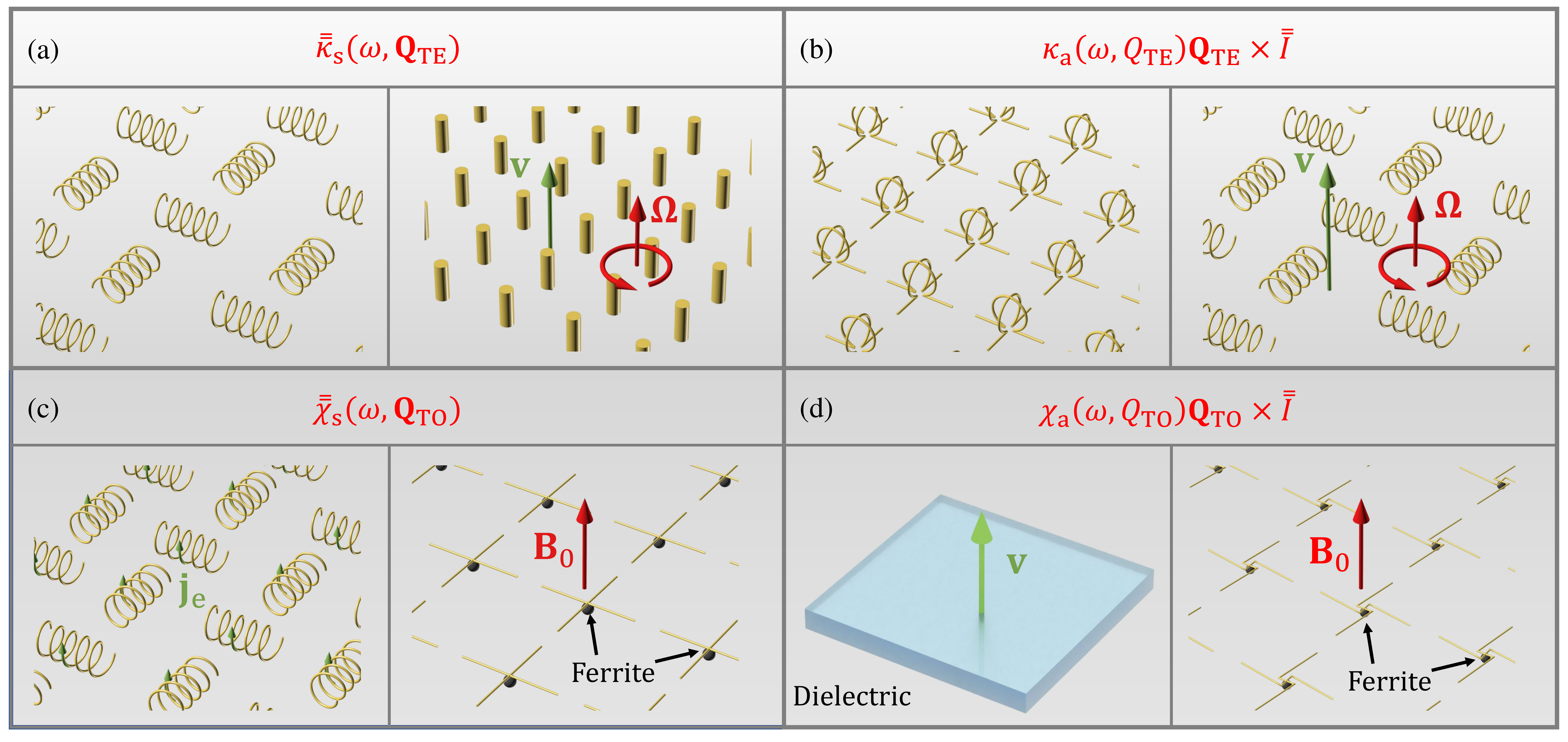}  
	\caption{ Conceptual implementation of different material tensors. Two characteristic examples are shown for each tensor.   }
	\label{fig14}
\end{figure*}
Taking into account the aforementioned parity symmetry properties of the electromagnetic field quantities, let us re-examine   constitutive relations~\r{class1}. One can use the fact that multiplication of a tensor with a vector results in a true vector only  if both of them are either true or pseudo quantities. Therefore, one can deduce that 
$\dya{\freq{\varepsilon}}_{\rm r}$, $\dya{\freq{\varepsilon}}_{\rm n}$, $\dya{\freq{\mu}}_{\rm r}$, and $\dya{\freq{\mu}}_{\rm n}$ are true tensors, while $\dya{\freq{\kappa}}$ and $\dya{\freq{\chi}}$ are pseudotensors. 
It is convenient to decompose the latter ones to symmetric and antisymmetric parts so that $\dya{\freq{\kappa}}=\dya{\freq{\kappa}}_{\rm s}+\dya{\freq{\kappa}}_{\rm a}$ and $\dya{\freq{\chi}}=\dya{\freq{\chi}}_{\rm s}+\dya{\freq{\chi}}_{\rm a}$.
By representing all antisymmetric tensors in the form 
$ \dya{\freq{\rho}} (\omega, {\_Q}) =  {\freq{\rho}} (\omega, {Q}) \ve{Q}  \times \dya{I}$ (here $\_Q$ is a vector dual to the antisymmetric tensor, $\dya{I}$ is a unit tensor and cross product denotes dyadic multiplication), we can summarize the parity symmetry of all the material tensors:
\e \begin{array}{lr}\displaystyle
  \dya{\freq{\varepsilon}}_{\rm r}(\omega,\ve{Q}_{\rm TE}) & {\rm - \,  true\,\,tensor},  \vspace{1mm} \\
  \dya{\freq{\varepsilon}}_{\rm n}(\omega,\ve{Q}_{\rm TO})=  {\freq{\varepsilon}}_{\rm n}(\omega, |{Q}_{\rm TO}|) \ve{Q}_{\rm TO} \times \dya{I} & 
  {\rm - \, true\,\,tensor},
 \vspace{1mm} \\ \displaystyle
   \dya{\freq{\mu}}_{\rm r}(\omega,\ve{Q}_{\rm TE}) & {\rm - \,  true\,\,tensor}, \vspace{1mm} \\
  \dya{\freq{\mu}}_{\rm n}(\omega,\ve{Q}_{\rm TO})=  {\freq{\mu}}_{\rm n}(\omega, |{Q}_{\rm TO}|) \ve{Q}_{\rm TO} \times \dya{I} & 
  {\rm - \, true\,\,tensor},  
\end{array}\l{class8}\f
\e \begin{array}{lr}\displaystyle
      \dya{\freq{\kappa}}_{\rm s}(\omega,\ve{Q}_{\rm TE}) & {\rm - \,  pseudotensor}, \vspace{1mm} \\
  \dya{\freq{\kappa}}_{\rm a}(\omega,\ve{Q}_{\rm TE})=  {\freq{\kappa}}_{\rm a}(\omega, |{Q}_{\rm TE}|) \ve{Q}_{\rm TE} \times \dya{I} & 
  {\rm - \, pseudotensor}, 
   \vspace{1mm} \\ \displaystyle
        \dya{\freq{\chi}}_{\rm s}(\omega,\ve{Q}_{\rm TO}) & {\rm - \,  pseudotensor}, \vspace{1mm} \\
  \dya{\freq{\chi}}_{\rm a}(\omega,\ve{Q}_{\rm TO})=  {\freq{\chi}}_{\rm a}(\omega, |{Q}_{\rm TO}|) \ve{Q}_{\rm TO} \times \dya{I} & 
  {\rm - \, pseudotensor}. 
\end{array}\l{class9}\f
As was mentioned above, the case when there is no external field $\_Q$ acting on the material can be easily taken into account by assuming $Q_{\rm TE}=1$ and $\_Q_{\rm TO}=0$.

Now we have eight tensors each corresponding to specific material response. Each symmetric tensor  can be diagonalized in a new basis. Let us note that a symmetric tensor $ \dya{\freq{\rho}} (\omega, {\_Q})  $ is a \textbf{pseudotensor} if \textit{one and only one} of the two following conditions is correct: 
\begin{itemize}
    \item $\_Q$ is a pseudovector,
    \item the material has chiral topology (in a sense of broken inversion symmetry).
\end{itemize}
If \textit{one and only one} of the two above conditions is correct, an antisymmetric tensor $ \dya{\freq{\rho}} (\omega, {Q})  \ve{Q}  \times \dya{I} $ is, on contrary, a \textbf{true tensor}.  Thus, material response attributed to each tensor in~\r{class8} and \r{class9} can be achieved by two opposite scenarios: 1) when the external vector $\_Q$ is a true vector; 2) when it is a pseudovector.  Figures~\ref{fig13} and \ref{fig14} summarize all these scenarios (two scenarios for each of the eight material tensors) with some specific known or possible conceptual implementations. 
It is important to mention that each scenario can be implemented in practice in many different ways, while the figures depict only one of them.  
Moreover, satisfying the space and time symmetry conditions  is a \emph{necessary} but \emph{not sufficient} condition to achieve the desired response.

Figure~\ref{fig13}a corresponds to realization of material response described by symmetric tensor $\dya{\freq{\varepsilon}}_{\rm r} (\omega, \ve{Q}_{\rm TE})$. Naturally, such response is realized with simple dielectrics and magnetics without an external bias vector ($Q_{\rm TE}=1$ and $\_Q_{\rm TO}=0$). Another alternative is a medium with broken inversion symmetry on which an external time-even pseudovector acts (simplifying, one can say that the ``pseudo symmetry'' of the medium in this case is compensated by the pseudovector symmetry, and the permittivity tensor becomes  a true tensor). Looking at Fig.~\ref{fig12}, one can find that the magnetic current density $\_j_{\rm m}$ is the only example of vector quantity which is time-even and at the same time a pseudovector.  However, conceptually,
time-even pseudoquantity can be  also synthesized by combining two external vectors: Linear speed $\_v$ and orbital angular velocity $\Omega$. In other words, the medium with broken inversion symmetry moving linearly and, simultaneously, rotating around the same direction  may
%\footnote{Remember that these symmetry considerations constitute only necessary conditions for the desired response.} 
also have material tensor $\dya{\freq{\varepsilon}}_{\rm r} (\omega, \ve{Q}_{\rm TE})$ (see right panel of Fig.~\ref{fig13}a). 

The antisymmetric permittivity tensor $\dya{\freq{\varepsilon}}_{\rm n}(\omega, \_Q_{\rm TO})$ can be achieved with a true time-odd external vector such as linear speed~$\_v$ (see Fig.~\ref{fig13}b). Note that from~\r{class8} it follows that scalar quantity $ {\freq{\varepsilon}}_{\rm n}(\omega, |{Q}_{\rm TO}|) $ is time-odd since the entire tensor 
$ {\dya{\freq{\varepsilon}}}_{\rm n}  $ is time-even. This fact implies that the material response expressed by $ {\freq{\varepsilon}}_{\rm n} $ must include some time-odd parameter or time derivative,  in addition to the external vector $\_Q_{\rm TO}$. Figure~\ref{fig13}b (left side) depicts a conceptual example of a composite which is characterized by  tensor~$\dya{\freq{\varepsilon}}_{\rm n}(\omega, \_Q_{\rm TO})$. The constitutive relations of an isotropic chiral medium (here chirality is required to ensure that $\dya{\freq{\varepsilon}}_{\rm n}$  is a true tensor) linearly moving along a given direction contain (in addition to symmetric) an antisymmetric permittivity tensor ${\freq{\varepsilon}}_{\rm n}(v) \ve{v} \times \dya{I}$, where 
\e {\freq{\varepsilon}}_{\rm n}(v)= \frac{2j \freq{\kappa}_{\rm i} \freq{\varepsilon}_{\rm i} \left(1-\frac{\displaystyle v^2}{\displaystyle c^2} \right)/c}{ 2 \freq{\kappa}_{\rm i}^2 \frac{\displaystyle v^2}{\displaystyle c^2} \left(1+\frac{\displaystyle v^2}{\displaystyle c^2} \freq{\varepsilon}_{\rm i} \freq{\mu}_{\rm i} \right) - \freq{\kappa}_{\rm i}^4 \frac{\displaystyle v^4}{\displaystyle c^4} - \left(\frac{\displaystyle v^2}{\displaystyle c^2} \freq{\varepsilon}_{\rm i} \freq{\mu}_{\rm i} -1 \right)^2  }
\f
and $\freq{\varepsilon}_{\rm i}$, $\freq{\mu}_{\rm i}$, and $\freq{\kappa}_{\rm i}$ stand for isotropic permittivity, permeability and chirality parameter of the medium when it is at rest.  We have derived this nontrivial result   from~\cite[Sec.~3.4.2]{serdyukov_electromagnetics_2001}. It is easy to see that scalar function ${\freq{\varepsilon}}_{\rm n}(v)$ in fact includes a time derivative ($j \kappa_{\rm i}$ is equivalent to $j \omega \kappa_{\rm c}$ according to~\r{eq93}), as required by the time-reversal symmetry.
Another  and more traditional way to achieve the antisymmetric permittivity is based on biasing magneto-optical materials with a time-odd pseudovector field, namely magnetic flux density $\_B_0$ (see the right panel of Fig.~\ref{fig13}b). To satisfy the time-reversal symmetry, the scalar function  $ {\freq{\varepsilon}}_{\rm n} $ includes a time derivative ($j\omega$ in the frequency domain), as is seen in~\r{lorentz8} for the off-diagonal permittivity components. Alternatively, similar effect was reported using another pseudovector bias field, orbital angular velocity  $\Omega$, and was coined as ``rotatory ether drag''  in a rapidly
rotating rod made of Pockels glass~\cite{jones_reginald_victor_rotary_1976}. This is an analogue of Faraday rotation without external magnetic field.

The symmetric and antisymmetric parts of the permeability tensor can be synthesized using the same techniques as those for the  permittivity tensor (see Figs.~\ref{fig13}c and~d). The only difference is that the antisymmetric part resulted from magnetic biasing occurs due to the ferromagnetic resonance at microwaves, rather than electron cyclotron orbiting at optical frequencies.

Figure~\ref{fig14}a (left panel) depicts  a typical implementation of the symmetric chirality tensor $\dya{\freq{\kappa}}_{\rm s} (\omega, \_Q_{\rm TE})$ via a random mixture of metal helices (so-called Pasteur medium). This realization does not require external bias ($Q_{\rm TE}=0$). Another interesting route is based on a time-even pseudovector quantity originated via linear velocity $\_v$ and simultaneous angular velocity $\Omega$. Interestingly, in this case the material constituents can be achiral (such as \emph{conceptual} cylinders   shown in right panel of Fig.~\ref{fig14}a), chirality is generated solely through the external bias quantities~\cite[Sec.~1.9.5, Fig.~1.21]{barron_molecular_2009}. Another route to achieve chirality is a gyrotropic (biased magneto-optical) medium moving linearly with speed $\_v$ in the direction of the bias~\cite[\textsection~7.4c]{kong_electromagnetic_1986},\cite[\textsection~5]{dmitriev_group_1999}.

The antisymmetric tensor $\dya{\freq{\kappa}}_{\rm a}$ can be gained without an external field in an achiral so-called omega medium in which $\_Q_{\rm TE}=\_r$ (description of the constituent geometry and scattering physics behind can be found e.g. in~\cite{Asadchy2018}) or, potentially, in a chiral medium moving linearly and rotating around the same axis (see Fig.~\ref{fig14}b).

Nonreciprocal symmetric material tensor $\dya{\freq{\chi}}_{\rm s} (\omega, \_Q_{\rm TO})$ might be achieved in a  medium  with broken inversion symmetry biased by some external time-odd true vector field. It is worth mentioning that a chiral medium moving with linear speed $\_v$, although is satisfies the symmetry requirements~\r{class9}, in fact does not possess  Tellegen coupling. This confirms our previous remark that the symmetry conditions are  necessary but not sufficient. Probably, a chiral medium biased by external electric current $\_j_{\rm e}$ could generate response in form of $\dya{\freq{\chi}}_{\rm s} (\omega, \_Q_{\rm TO})$ (see Fig.~\ref{fig14}c). The second and feasible route is based on achiral media with ferrite-based constituents biased by pseudovector external magnetic flux density  $\_B_0$~\cite{tretyakov_nonreciprocal_1998}, as shown in the right panel of Fig.~\ref{fig14}c. It is so far, probably, the most practical implementation of artificial Tellegen medium proposed in~\cite{tellegen_gyrator_1948}. Magnetoelectric properties of Tellegen type appear also in natural materials, such as topological insulators~\cite{laforge_optical_2010} and multi-ferroic media~\cite{pyatakov_magnetoelectric_2012}.

Finally, the antisymmetric tensor $\dya{\freq{\chi}}_{\rm a}$ can be attained in a moving isotropic dielectric medium with linear speed~$\_v$~\cite[Sec.~7.4a]{kong_electromagnetic_1986},\cite{mazor_nonreciprocal_2019} or in a chiral medium biased by an external pseudovector such as magnetic flux density $\_B_0$ (see Fig.~\ref{fig14}d). Interesting alternatives to the first approach were proposed based on dielectric  scatterers rotating with angular speed $\_v=\Omega \times \_r$~\cite{shiozawa_phenomenological_1973,steinberg_two-dimensional_2006,mazor_rest_2019} ($\_v$ is a true vector), based on synthetic motion~\cite{mazor_one-way_2020}, and based on materials biased by static  electric and magnetic fields which  are orthogonal to one another and to the light wavevector, resulting in $\dya{\freq{\chi}}_{\rm a}$  being proportional to $\_E \times \_B$~\cite{baranova_new_1977,ross_selection_1989,rikken_observation_2002} ($\_E \times \_B$ is a true vector). 
Structures with antisymmetric tensor $\dya{\freq{\chi}}_{\rm a}$ were proposed in optical~\cite{baranova_theory_1979,mazor_metaweaves:_2014} and microwave~\cite{tretyakov_nonreciprocal_1998,mirmoosa_polarizabilities_2014,radi_one_way_2014} regimes and experimentally demonstrated  in works~\cite{rikken_observation_1997,rikken_observation_2002,vehmas_transmission_2014}. Recently, the effect of similar symmetry type was proposed for phonons~\cite{hamada_phonon_2020}.

Thus, space and time inversion symmetries are powerful tools for analyzing and designing materials with arbitrary  electromagnetic properties, which can be required by various applications. One of the early examples of exploiting symmetry arguments for obtaining new results can be found in~\cite{zocher_about_1953}.
Space and time inversion symmetries provide a simple but fundamental classification of all possible linear effects in matter. Moreover, this approach allows us to draw analogies between seemingly distinct effects which  have the same physical origin.

\section{Nonreciprocity in nonlinear  systems} \label{nonnon}

\begin{figure}[tb]
\centering
\epsfig{file=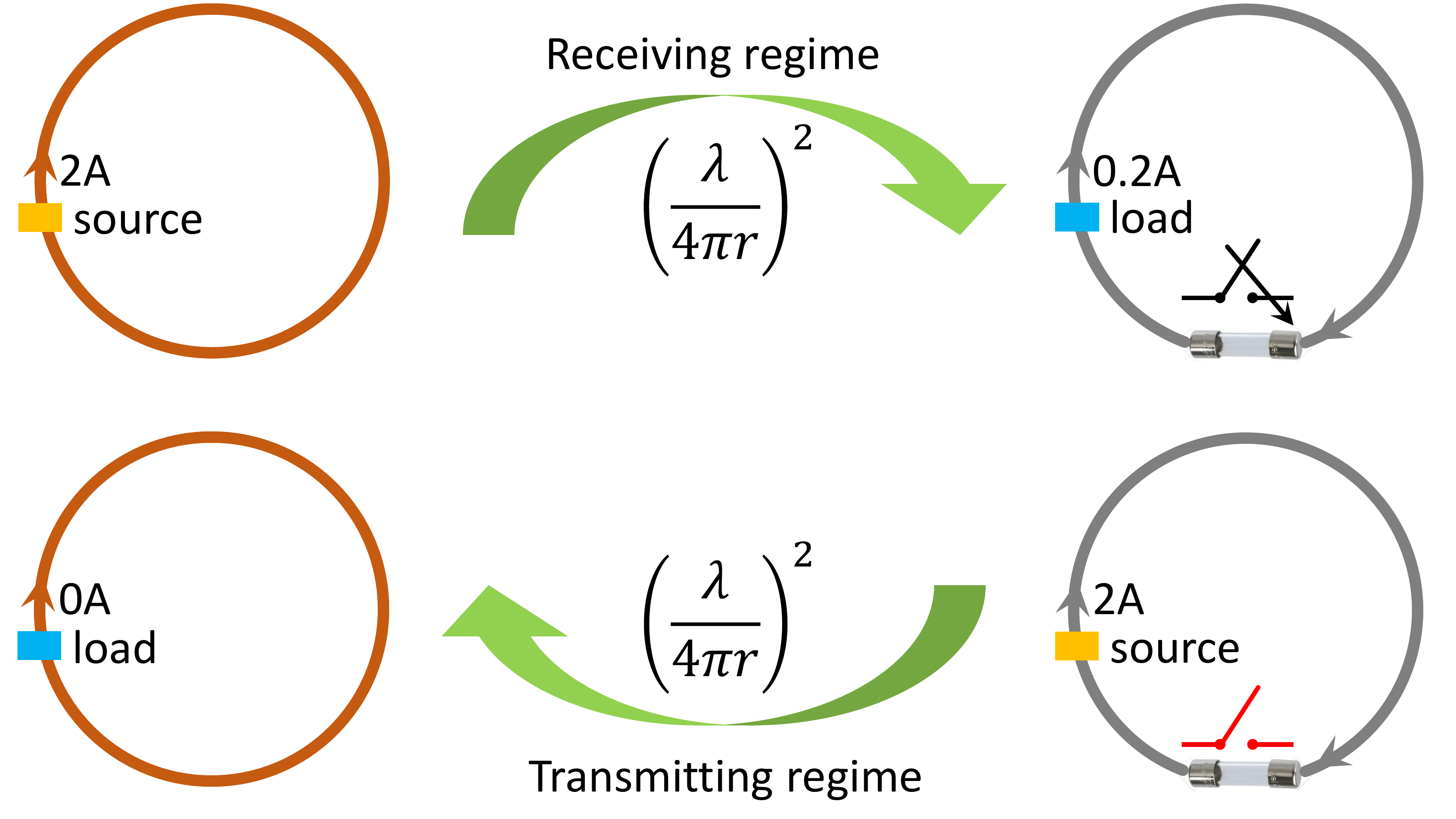, width=0.99\linewidth} 
\caption{Nonreciprocal communication between two small loop antennas. The loop antenna on the right side includes a nonlinear device (electrical fuse) for controlling the magnitude of the electric current flowing through the loop. Here, $r$ is the distance between the two antennas and $\lambda$ denotes the operating wavelength in free space.}
\label{NrecipNlinear}
\end{figure} 

\begin{figure}[tb]
\centering
\epsfig{file=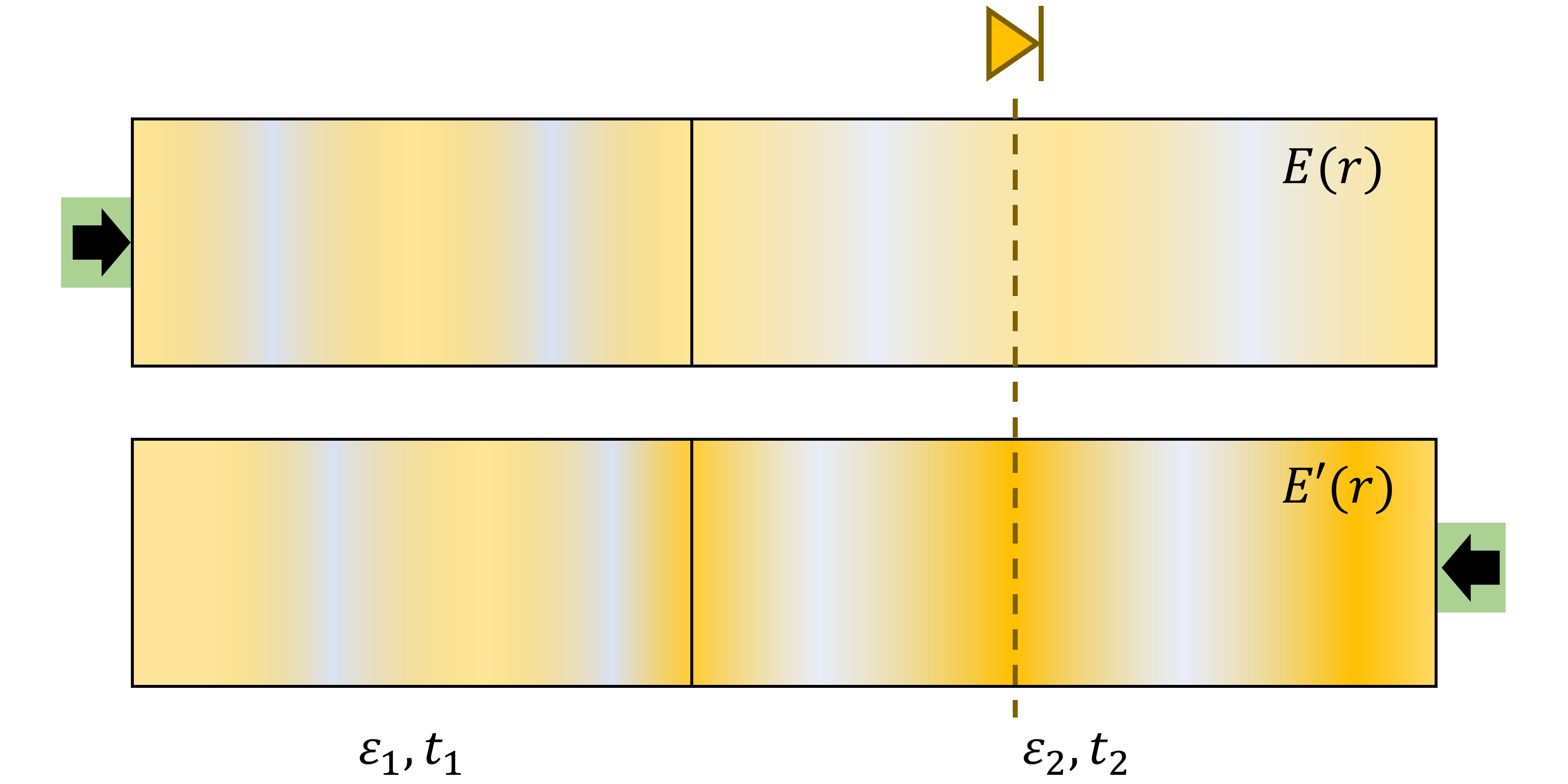, width=0.99\linewidth} 
\caption{Forward (up) and backward (bottom) wave propagation through an infinitely-wide slab consisting of two layers of thicknesses $t_1$ and $t_2$. The regions outside the slab are assumed to be vacuum.  The gradient color shows the magnitude of the total electric field  $E(r)$ inside the slab. The dashed line shows the location where the ratio of the total fields for the forward and backward illuminations is the maximum. By positioning a nonlinear element at this location, one can achieve nonreciprocal wave transmission through the slab. }
\label{fig17}
\end{figure}

Nonlinearity can be used to break the Lorentz reciprocity, and therefore, gives an opportunity for manipulation of the transmission characteristics of systems in order to design nonreciprocal components~\cite{tocci_thinfilm_1995,gallo_all-optical_1999,Gallo2,Zhou1,Lin1,Manipatruni1,Roy1,miroshnichenko_reversible_2010,trendafilov_hamiltonian_2010,Zhukovsky1,lepri_asymmetric_2011,Shadrivov1,fan_subwavelength_2011,Grigoriev1,fan_all-silicon_2012,Ding22,Anand1,Fan22,roy_cascaded_2013,chang_paritytime_2014,Xu22,nazari_optical_2014,Peng1,shi_limitations_2015,Yu22,mahmoud_all-passive_2015,sounas_time-reversal_2017,roy_critical_2017,Aleahmad1,bino_microresonator_2018,
rosario_hamann_nonreciprocity_2018,sounas_broadband_2018,sounas_nonreciprocity_2018}. 
Nonlinear nonreciprocal devices exploit the fact that field distributions in direct and reciprocal processes can be very different. The main idea is to position nonlinear objects where the field strength is high for one propagation direction and weak for the opposite one. 

Perhaps the simplest example is coupling between two small antennas in free space as shown in Fig.~\ref{NrecipNlinear}. We consider two loop antennas whose radii are much smaller than the free-space wavelength, and they are separated by a sufficient distance. One of the antennas (the right one in Fig.~\ref{NrecipNlinear}) is connected to a nonlinear object (here, an electrical fuse as the simplest conceptual example) which is sensitive to the magnitude of the electric current flowing through the loop. If the current magnitude is large enough, the object works similarly to a switch and opens the circuit, while it is short-circuited if the current magnitude is not high. In the receiving regime, when the right  antenna   is a receiver, the induced electric current is small due to the free-space path loss $[\lambda/(4\pi r)]^2$ despite the magnitude of the transmitting current is considerably large. Therefore, the nonlinear object is short circuited and the load receives the electric power. However, in the transmitting regime (when the right antenna is transmitting), since the magnitude of the transmitting current is large, the nonlinear object is consequently open circuited, and no energy is transferred to the load in the left  loop antenna. Thus, one could simply design a nonreciprocal link between the two antennas by using the concept of nonlinearity.    

In the above example, we had the free-space path loss causing the attenuation of energy and ensuring the asymmtetry of the field distribution. However, it is also possible to design a nonreciprocal system  where there is no power spread, like a nonreciprocal waveguide. To understand the principle of nonreciprocal isolators in waveguides, let us discuss an explicit example~\cite{mahmoud_all-passive_2015}. Consider a double slab which is extended infinitely in the transverse plane and has a finite thickness in the longitudinal direction. It comprises two dielectric layers made of isotropic linear materials which have contrasting relative permittivities $\varepsilon_1$ and $\varepsilon_2$.  In addition, the thicknesses of those layers can be also different, namely $t_1$ and $t_2$ (see Fig.~\ref{fig17}). Due to such nonidentical permittivities and thicknesses, the whole system is spatially asymmetric. This asymmetry plays the key role for obtaining   nonreciprocity. Illuminating the double slab by an incident wave, we excite a standing wave within the slab (meaning that there are maxima and minima for the amplitude of the field). However, intriguingly, such standing waves are different for forward and backward illuminations, and the spatial distribution of the field  inside the slab is, indeed, dependent on the propagation direction of the input wave, as is shown with gradient color in Fig.~\ref{fig17}. 
While we certainly achieve the same transmission coefficient as the system is reciprocal by this point, the field distribution inside the slab is not the same. As a consequence, there exist locations at which the ratio between the local fields induced by two incident waves illuminated from opposite sides, is  high. Next, to complete the design of a nonlinear isolator, let us    place a \textit{nonlinear} resonant element at the location  where the  ratio is maximum (shown by the dashed line in the figure). Since the performance of the element depends on the field amplitude,  its interaction  with the field in the two cases (illuminations from the right and from the left side) will be  dramatically different. For that reason, by proper design of the system, noticeable transmission for one direction and low transmission for the opposite direction can be attained. Importantly, nonlinearity must be accompanied by the spatial asymmetry to break the Lorentz reciprocity.

\begin{figure}[t!]
\centering
\epsfig{file=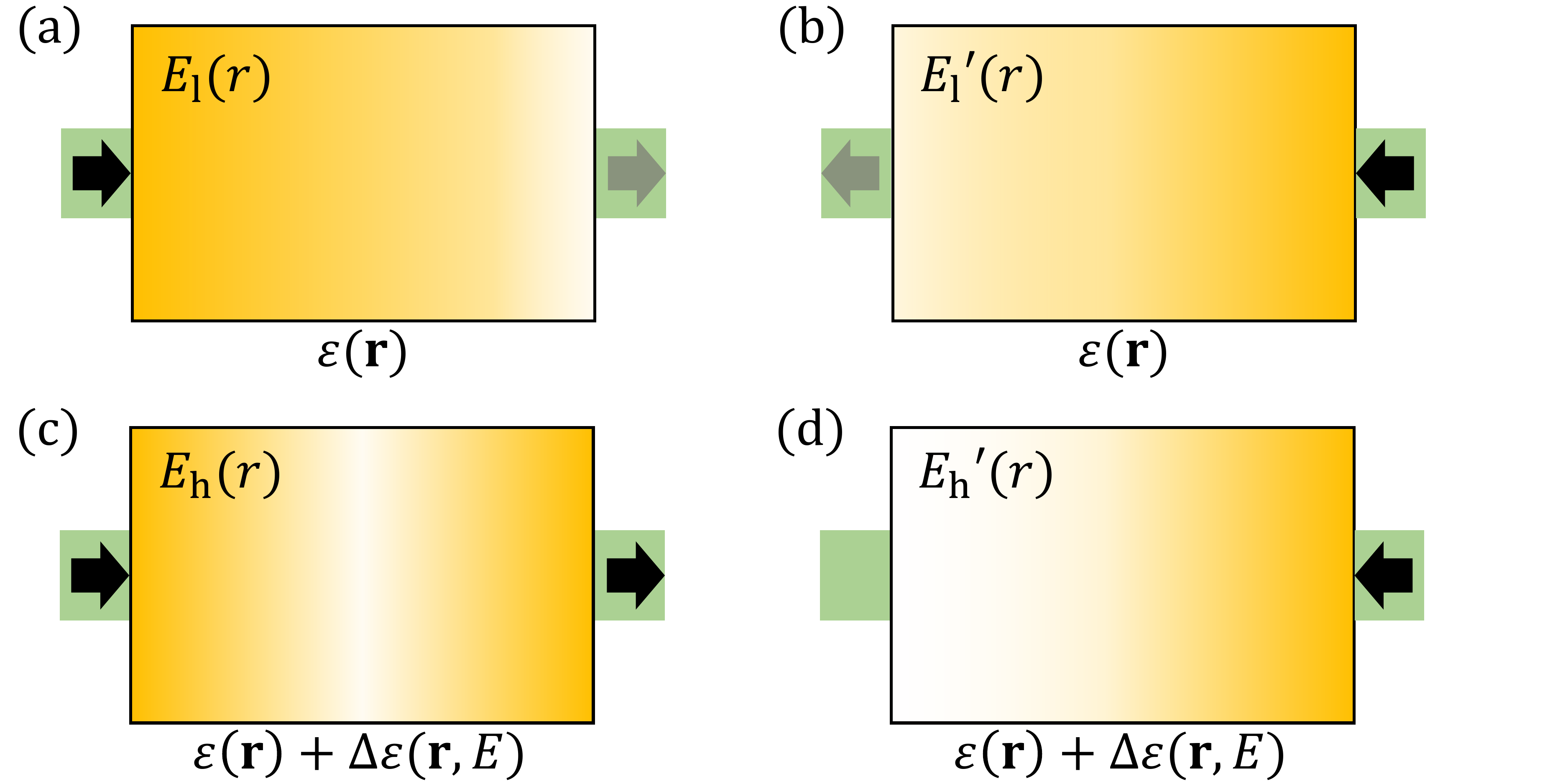, width=0.99\linewidth} 
\caption{
The general schematic view of a nonreciprocal device based on a nonlinear dielectric slab. The slab is spatially asymmetric and is described by linear permittivity $\freq{\varepsilon}(\_r)$ for weak excitations and by nonlinear permittivity $\freq{\varepsilon}(\_r)+ \Delta \freq{\varepsilon}(\_r, \freq{E})$ for strong excitations.
The gradient color shows the magnitude of the total electric field $\freq{E}(r)$ inside the slab. (a--b) Illuminations from different sides by a low-power wave (reciprocal  transmission). (c--d) Same for a high-power wave  (nonreciprocal  transmission).   }
\label{example2}
\end{figure}

However, these conditions cannot provide us with an ideal nonreciprocity, and in practice nonlinear isolators   face limitations. The above explanations were based on assumption that  the forward and backward incident waves are not \textit{concurrently} present. Let us give another example to understand the problem. Now, the system consists of a single dielectric layer. The dielectric is nonhomogeneous (to provide necessary spatial asymmetry) and described by permittivity $\freq{\varepsilon}(\_r)$ for weak excitations and by $\freq{\varepsilon}(\_r)+\Delta\freq{\varepsilon}(\_r,E)$ for incident waves with strong fields (in the nonlinear regime). Naturally, for weak excitations, the layer will exhibit reciprocal transmission for forward and backward illuminations, as shown in Fig.~\ref{example2}(a--b). The field distribution will be different in the two cases, similarly to Fig.~\ref{fig17}. For strong excitations, nonlinear dielectric polarization will occur, causing   different transmission levels for opposite illuminations (see Fig.~\ref{example2}(c--d)). High isolation can be achieved by designing specific distribution of $\freq{\varepsilon}(\_r)$. 
% This single layer is similar to the above considered double-layer slab
Consider that the dielectric layer is illuminated by  a low-power incident wave  together with the a high-power wave (the two waves propagate in the same forward direction). In this case, the low-power wave will have the same high transmission, like in Fig.~\ref{example2}(c). 

Now, let us assume that an input high-power wave  illuminates the dielectric layer in the forward direction, while, at the same time, a low-power  wave is incident on it from the backward direction. 
% In this case, the small-amplitude   wave will experience the same high transmission as the main wave.
% Since we get high transmission in the forward direction, the small-amplitude incident wave also experiences high transmission. The question is that what the scenario is if we illuminate the main input wave in the forward direction but the the small-amplitude signal is sent from the backward direction. 
The question is whether the low-power wave will pass through (like in Fig.~\ref{example2}(b)) or be blocked by the layer (like in Fig.~\ref{example2}(d)).  In fact, it will pass through since in the presence of the high-power wave, the   polarization induced by the low-power wave will linearly depend on its electric field. This effect was known in the microwave community for long time in electronic diodes, which are essentially nonlinear systems, as ``small-signal approximation''~\cite[\textsection~3.3.8]{sedra_microelectronic_2003}. In   photonics, the effect was proposed recently and coined as  ``dynamic reciprocity''~\cite{shi_limitations_2015}. 
% The nonlinear Maxwell equations working for high-power are linear around the solution corresponding to the high-power forward signal. This is in analogy with the performance of a diode. Reminding that diode is a nonlinear component.
Next, we will provide derivations explaining the origin of the effect. According to nonlinear optics, the wave equation inside  nonlinear nonhomogeneous dielectric can be written in the form~\cite[eq.~(2.1.21)]{boyd_nonlinear_2008},\cite{shi_limitations_2015}:
\begin{equation}
\nabla\times\nabla\times\mathbf{\time{E}}+\mu_0\varepsilon_0\freq{\varepsilon}(\mathbf{r}){\partial^2\mathbf{\time{E}}\over\partial t^2}+\mu_0{\partial^2\mathbf{\time{P}}_{\rm NL}\over\partial t^2}=0,
\label{eq:kerrnon}
\end{equation}
where $\mathbf{\time{E}}$ is the time-harmonic electric field, $\freq{\varepsilon}$ is the frequency dependent permittivity in  the linear regime, and 
$\mathbf{\time{P}}_{\rm NL}$ denotes the nonlinear polarization density vector which in the general case depends on the electric field as~\cite[eq.~(1.1.2)]{boyd_nonlinear_2008}
\begin{equation}
\mathbf{\time{P}}_{\rm NL}=\varepsilon_0\freq{\chi}^{(2)}(\mathbf{r})\mathbf{\time{E}}^2+\varepsilon_0\freq{\chi}^{(3)}(\mathbf{r})\mathbf{\time{E}}^3+ \dots
\end{equation}
Here, $\freq{\chi}^{(n)}(\mathbf{r})$ is the $n$-th order component of the electric susceptibility of the medium. Let us consider the optical (AC) Kerr nonlinearity. For media which have a significant Kerr effect, the third-order component $\freq{\chi}^{(3)}$  is dominant and the other components are neglected, i.e. $\mathbf{\time{P}}_{\rm NL} \approx \varepsilon_0\freq{\chi}^{(3)}(\mathbf{r})\mathbf{\time{E}}^3$. 

Since in our example, the dielectric layer is illuminated by the high-power and low-power waves simultaneously, the total  electric field can be written as  $\mathbf{\time{E}}=\mathbf{\freq{E}}_{\rm{h}}\exp(j\omega_{\rm{h}} t)+\mathbf{\freq{E}}_{\rm{l}}\exp(j\omega_{\rm{l}} t)$, in which $\vert\mathbf{\freq{E}}_{\rm{h}}\vert\gg\vert\mathbf{\freq{E}}_{\rm{l}}\vert$ are the complex amplitudes of the time-harmonic waves. The frequencies of the two waves $\omega_{\rm{h}}$ and $\omega_{\rm{l}}$ are different. Note that the dynamic reciprocity restriction is not applied when $\omega_{\rm{h}}$ and $\omega_{\rm{l}}$ are equal since the resulting linearized equation is nonreciprocal~\cite[Suppl. Inf.]{shi_limitations_2015},\cite{roy_critical_2017}. 
After substituting $\mathbf{\time{E}}$ in the expression for $\mathbf{\time{P}}_{\rm NL}$ and
doing some algebraic manipulations, one obtains~\cite[eq.~(4.1.12)]{boyd_nonlinear_2008},\cite{shi_limitations_2015}
\begin{equation}
\mathbf{\time{P}}_{\rm NL} \approx \mathbf{\time{P}}_0+\mathbf{\time{P}}_{\omega_{\rm{l}}}=\mathbf{\time{P}}_0+6\varepsilon_0\freq{\chi}^{(3)}(\mathbf{r})\vert\mathbf{\freq{E}}_{\rm{h}}\vert^2\mathbf{\freq{E}}_{\rm{l}} {\rm e}^{j\omega_{\rm l} t},
\l{sedreq}
\end{equation}
where $\mathbf{\time{P}}_{\omega_{\rm{l}}}$ is the polarization density due to the low-power incident wave and 
$\mathbf{\time{P}}_0$ is the rest of the polarization terms oscillating at different frequencies.
It is clear that the above equation is a linear equation for the low-power incident wave with frequency~$\omega_{\rm{l}}$. It provides a \textit{linear} susceptibility which is proportional to the third-order   susceptibility and to the square of the magnitude of the high-power signal. Put another way, one can introduce the effective susceptibility as $ \mathbf{\time{P}}_{\omega_{\rm{l}}}= \varepsilon_0 \freq{\chi}_{\omega_{\rm{l}}} \time{\_E}_{\rm l} $, where
\begin{equation}
\freq{\chi}_{\omega_{\rm{l}}}\approx6\freq{\chi}^{(3)}(\mathbf{r})\vert\mathbf{\freq{E}}_{\rm{h}}\vert^2.
\end{equation}
Note that~\r{sedreq} is physically equivalent to Eq.~(3.15) in~\cite{sedra_microelectronic_2003} written in terms of the current and voltage for electronic diodes. That equation implies that a low-amplitude current flowing through the diode in the presence of a high-amplitude current will  ``sense'' the diode as a simple linear resistance. 

Having~\r{sedreq} in mind, in the frequency domain, we can rewrite (\ref{eq:kerrnon}) as a linearized equation for the low-power incident wave  
\begin{equation}
\nabla\times\nabla\times\mathbf{\freq{E}}_{\rm{l}}-k^2_{\rm{l}}\Big[\freq{\varepsilon}(\mathbf{r})+6\freq{\chi}^{(3)}(\mathbf{r})\vert\mathbf{\freq{E}}_{\rm{h}}\vert^2\Big]\mathbf{\freq{E}}_{\rm{l}}=0,
\end{equation}
where $k_{\rm{l}}=\omega_{\rm{l}}\sqrt{\mu_0\varepsilon_0}$ is the free-space wave number. We conclude that the effective relative permittivity corresponding to this \textit{linear} equation is indeed equal to $\freq{\varepsilon}_{\rm eff}(\mathbf{r})= \freq{\varepsilon}(\mathbf{r})+6\freq{\chi}^{(3)}(\mathbf{r})\vert\mathbf{\freq{E}}_{\rm{h}}\vert^2$.
Since  the   dielectric function $\freq{\varepsilon}_{\rm eff}$ is  scalar and time-independent,  propagation of the low-power incident wave will be  reciprocal. Therefore, it will not be blocked  by the layer during the backward illumination. Such functionality precludes the described  nonlinear system  from operating as an ideal isolator. 
% According to this emergence of dynamic reciprocity, the device cannot be an ideal isolator. If we send the small-amplitude signal from the opposite direction, the transmission will be reciprocal.
% (unitary in the considered scenario)

In addition to the considered constraint, the nonreciprocity via nonlinearity has another fundamental limitation due to the second law of thermodynamics~\cite{RAYLEIGH1901,sounas_nonreciprocity_2018}. Accordingly, the nonlinear isolators cannot operate from both sides {\it simultaneously}. It can be readily expected that if such an isolator could operate, the radiative thermal power transferred between the ports would  not be zero while the two ports are kept in the same temperature. This violates the second law of thermodynamics which stresses that the total radiative power transferred must be zero. For linear isolators, for example, we do not have this limitation because the power illuminated from one of the isolated port is absorbed~\cite{RAYLEIGH1901}. Finally,  in~\cite{Sounasfundamental,sounas_nonreciprocity_2018}   the third limitation was pointed out. It states that there is a tradeoff between transmission in the forward direction and the level of input intensity for which large isolation (in an isolator) can be obtained. This limitation comes from the fact that the field asymmetry reduces when the transmission through the structure increases~\cite{sounas_time-reversal_2017}.
It is worth mentioning that similar conclusions were obtained    in~\cite[\textsection~XXI]{caloz_electromagnetic_2018}.

\section{Nonreciprocity in linear time-variant systems }\label{nontv}

%time-modulation + spatial asymmetry. Three ways: oblique incidence, bulk metamaterial, spatial dispersion. Some examples...

An alternative approach for breaking reciprocity, without using magnetic or nonlinear materials, is to break the time-invariance of the system. The possibility of using time-variant systems (also called time-modualted), has  been known for many
years~\cite{cullen_travelling-wave_1958,slater_interaction_1958,fettweis_steady-state_1959,kamal_parametric_1960,currie_coupled-cavity_1960,simon_action_1960,macdonald_exact_1961,oliner_wave_1961,
anderson_reciprocity_1965,holberg_parametric_1966,Time_Modulation_Energy2019,mirmoosa_instantaneous_2020,ptitcyn_time-modulated_2019}. 
However, due to the  technological advances on the implementation of time-variant systems, it has been in the recent years when these solutions have been proposed for the design of compact nonreciprocal devices~\cite{yu_complete_2009,yu_optical_2009,kang_reconfigurable_2011,
lira_electrically_2012,fang_photonic_2012,fang_realizing_2012,
wang_optical_2013,horsley_optical_2013,sounas_giant_2013,fang_controlling_2013,
qin_nonreciprocal_2014,estep_magnetic-free_2014,tzuang_non-reciprocal_2014,lin_light_2014,shaltout_time-varying_2015,hadad2015space,hadad_breaking_2016,correas-serrano_nonreciprocal_2016,taravati_mixer-duplexer-antenna_2017,guo_nonreciprocal_2019,zhang_breaking_2019}.

Before discussing these methods we need to clarify the definition of reciprocal and nonreciprocal processes for time-varying systems. Processes in time-invariant systems are called  reciprocal  if the two reaction integrals are equal, see \r{green10}. In the frequency domain, for a two-port device, we simply have the symmetry relation for the S-parameters $S_{12}=S_{21}$. For time-varying systems, the definition of reciprocity   should be given in time domain. It is usually formulated in terms of equality of convolution integrals \r{eq:convtyrecip} or correlation integrals~\cite{cheo1965,anderson_reciprocity_1965,AmirReciprocity}. However, especially for harmonic pumping, the use of this general definition is not convenient.

Next, we derive reciprocity relations in systems whose material properties vary harmonically with time. We consider  a special case when the permittivity of the background material   is \textit{globally} modulated according to   the following \textit{symmetric} (with respect to time) modulation function
\e
{\varepsilon}(\omega, \_r, t)={\varepsilon}_{\rm st}(\omega, \_r) + M(\_r) \cos(\Omega t+\phi),
\l{time1}\f 
where  $\time{\varepsilon}_{\rm st}$ is the static permittivity, $M(\_r)$ is the modulation strength function, $\Omega$ is the modulation frequency, and $\phi$ is an arbitrary phase. 
Here, for simplicity, we use the adiabatic
model for temporal modulations, assuming that the operational frequency  $\omega$ is very low compared to the lowest resonance frequency of the material. 
Note that this model can be used for arbitrary modulation frequency $\Omega$ (see ~\cite{mirmoosa_dipole_2020},\cite[Suppl. Inf.]{wang_nonreciprocity_2020} for details). In the general
non-adiabatic case, the following derivations could still
be performed, writing the material parameters using integrals over past time.
Here, for simplicity, we assume that $M(\_r)$  has no frequency dispersion. This assumption is realistic for gaseous plasmas with properly modulated in time  charge concentration $N_{\rm e}(t)$~\cite[Eq.~(53)]{mirmoosa_dipole_2020}.
 The reciprocity relation which we discuss below, however, can be derived also for other   \mbox{models} of the modulation function. 
%Here we use the adiabatic model, assuming that the external modulation is very slow as compared to the relaxation processes in the medium. 
%Equation~\r{time1} is accurate only for modulation frequencies that are  lower  than the frequency scale of  all polarization processes in the material~\cite{ptitcyn_time-modulated_2019}.
% Also, we neglect dissipation losses in permittivity and assume instantaneous response. 
The permeability is assumed to be static in time. Global modulation implies that at all points in space the permittivity alternates with the same phase.  Frequency-domain wave equation  for the time-varying material reads~\cite{shi_multi-frequency_2016}
\e
\nabla\times \frac{1}{\freq{\mu}(\omega)} \nabla\times \freq{\_E}
- \omega^2 \left[ \freq{\varepsilon}_{\rm st}(\omega) \freq{\_E} +\freq{\_P}_{\rm M} \right] = -j \omega \freq{\_J}, 
\l{time2}\f
where  the spatial dependency is assumed implicitly and  $\freq{\_P}_{\rm M}(\omega)$ is the additional polarization density of the material due to dynamic modulation which can be found from~\r{time1}:
\e
\freq{\_P}_{\rm M}(\omega)= \frac{M}{2} \left[ {\rm e}^{j\phi} \freq{\_E}(\omega -\Omega) +{\rm e}^{-j\phi} \freq{\_E}(\omega +\Omega)
\right]. 
\l{time3}\f
Modulation of the permittivity at a frequency $\Omega$ induces a number of sideband field harmonics $ \freq{\_E}_n$ of  frequencies $\omega_n = \omega_0 +n\Omega$, where $n$ is an integer. This is due to the  periodicity of the electric field which is the solution of wave equation~\r{time2} and can be written in terms of the Fourier series.
Substituting \r{time3} in \r{time2} and matching specific frequency components with temporal ${\rm e}^{j\omega t}$ variations, one can obtain wave equation in the  form
\e
\hat{\Theta}_n   \freq{\_E}_n -\omega_n^2 \frac{M}{2}
\left[ {\rm e}^{j\phi} \freq{\_E}_{n-1} +{\rm e}^{-j\phi} \freq{\_E}_{n+1} \right] 
= -j  \omega_n \freq{\_J}_n.
\l{time4}\f
Here, $\hat{\Theta}_n$ is an operator defined through its action:  $\hat{\Theta}_n  \freq{\_E}_n=    \nabla \times \frac{1}{\freq{\mu}_{n}} \nabla \times  \freq{\_E}_n -\freq{\varepsilon}_{{\rm st},n} \omega_n^2 \freq{\_E}_n  $. For generality, we assume a current source with multiple side bands, with $\freq{\_J}_n$ denoting the source at frequencies $\omega_n = \omega_0 + n \Omega$. 
% Note that   $\freq{\_E}_n$ and    $\freq{\_J}_n$ are represented as vectors of length $n$ whose each element is a spatial vector. 
It can be shown that writing~\r{time4} in the matrix form and expressing electric field vector as $\freq{\_E}_n= \=f_{nm}   \freq{\_J}_m$, we always obtain asymmetric matrix $\=f$. However, by   dividing both sides of~\r{time4}   by $\omega_n^2$ and  using replacement $\freq{\_J}'_m =  \freq{\_J}_m /\omega_m $, we obtain relation
\e
\freq{\_E}_n(\_r) = \=F_{nm}(\_r)   \freq{\_J}'_m(\_r),
\l{time45}\f
where $\=F$ is a symmetric matrix if $\phi=0$ is chosen. Note that phase $\phi$ can be chosen arbitrarily by time translation $t \rightarrow t +\Delta t$~\cite{fang_photonic_2012}. Formally solving~\r{time45},  
the radiated field harmonics by given current source read as
\e
 \freq{\_E}_n(\_r) = \int_V \dya{\freq{G}}_{nm}(\_r,\_r')  \freq{\_J}'_m(\_r') {\rm d}V'.
\l{time5}\f
Since $\=F$ is symmetric,  dyadic Green's function of time-varying material is also symmetric 
\e
\=G(\_r,\_r')=\=G^T(\_r,\_r').
\l{time6}\f
Here $\_r'$ denotes coordinates of   points inside source volume~$V$. Comparing~\r{time5} and \r{time6} to  \r{green1} and \r{green4}, we see that electromagnetic radiation in time-varying material with global modulation~\r{time1} satisfies the Onsager reciprocity conditions, i.e. is always reciprocal. The important difference between~\r{green1} and \r{time5} is that in the latter case the current    harmonics are normalized by their frequencies. Repeating derivations analogous to~\r{green103} and \r{green110}, we obtain equation
\e  
\int\limits_{V_A} \sum_{n}   \frac{  \freq{\_E}_n^{(B)}   \cdot \freq{\_J}_n^{(A)}}{\omega_n} {\rm d}V_A  = 
\int\limits_{V_B} \sum_{n}   \frac{  \freq{\_E}_n^{(A)}   \cdot \freq{\_J}_n^{(B)}}{\omega_n} {\rm d}V_B,
\l{time7} 
\f
which relates the fields and current harmonics generated via interaction between a pair of two sources~$A$ and $B$ (similarly to the scenario shown in 
Fig.~\ref{fig3_8}). Relation \r{time7} is the electromagnetic reciprocity theorem for time-varying systems. Let us write~\r{time7} for the special case when each current source is represented by one frequency harmonic, i.e. $\freq{\_J}_n^{(A)}=\freq{\_J}_p^{(A)} \delta_{pn}$ and $\freq{\_J}_n^{(B)}=\freq{\_J}_q^{(B)} \delta_{qn}$:
\e  
\int\limits_{V_A}     \frac{  \freq{\_E}_p^{(B)}   \cdot \freq{\_J}_p^{(A)}}{\omega_p} {\rm d}V_A  = 
\int\limits_{V_B}    \frac{  \freq{\_E}_q^{(A)}   \cdot \freq{\_J}_q^{(B)}}{\omega_q} {\rm d}V_B.
\l{time8} 
\f
When $p=q$ (sources~$A$ and $B$ are at the same frequency), equation~\r{time8} simplifies to the conventional form of reciprocity theorem~\r{green10}. In the case when $p=1$ and $q=2$ and $\freq{\_J}^{(A)}(\omega_1)=\freq{\_J}^{(B)}(\omega_2)$, the radiated fields harmonics are not equal $\freq{\_E}^{(A)}(\omega_2) \neq \freq{\_E}^{(B)}(\omega_1) $, differing by the ratio of frequencies $\omega_1/\omega_2$ (for uniform current sources). Reciprocity relation~\r{time8} applies restriction on how a two-port device converts waves of $\omega_1$ into $\omega_2$ in the forward direction and $\omega_2$ into $\omega_1$ in the backward direction: The conversion efficiencies of these two processes are not equal and must be related through $\omega_1$ and $\omega_2$. It is important to note that~\r{time7} does not impose direct  constraints on the identical conversions from $\omega_1$ into $\omega_2$ (in the forward direction) and from $\omega_1$ into $\omega_2$ (in the backward direction).
% , despite of some attempts of using such a constraint in the literature~\cite{kruk_nonlinear_2019}. 

Reciprocity relation~\r{time7} connects reaction functions normalized by frequencies or, alternatively, energies of photons if we divide both sides of the relation by the reduced Planck constant~$\hbar$. Thus, the fractions inside the volume integrals have dimensions of photon number flux (number of photons per second per unit area). In other words, reciprocity implies restriction on the evolution of number of photons in the direct and inverse processes, rather than intensities of waves. Analogous observation can be applied to the   Manley-Rowe relations for nonlinear processes~\cite[\textsection~2.5]{boyd_nonlinear_2008}.

Thus, we have concluded that systems with \textit{global symmetric time-harmonic} modulation of permittivity~\r{time1} are always reciprocal in the sense that relation~\r{time7}  is always satisfied. Interestingly, although such systems break time-reversal symmetry of Maxwell equations (since $\time{\varepsilon}(-t)\neq \time{\varepsilon}(t)$ for given $\phi$), they obey the so-called generalized time-reversal symmetry expressed as $\time{\varepsilon}(-t+t_0)= \time{\varepsilon}(t+t_0)$~\cite{williamson_breaking_2020}. For any given $\phi$, it is always possible to find such value of $t_0$ that  the above equality holds.

If modulation is not global and can be represented by two modulation functions~\cite{fang_photonic_2012}
\e
\time{\varepsilon}(\omega,\_r,t)=\time{\varepsilon}_{\rm st}(\omega, \_r) + M_1(\_r) \cos(\Omega t+\phi_1) + M_2(\_r) \cos(\Omega t+\phi_2)
\l{time9}\f
with $\phi_1 \neq \phi_2$ and $M_1(\_r) \neq \alpha M_2(\_r)$ ($\alpha   $ is an arbitrary real constant), then equation~\r{time4} cannot be written in the form $\freq{\_E}_n(\_r) = \=F_{nm}(\_r)   \freq{\_J}'_m(\_r)$ with $\=F$ being symmetric matrix (time translation affects both $\phi_1$ and $\phi_2$ and does not   make $\=F$ symmetric). Therefore, dyadic Green's function is asymmetric, and this time-modulated system will break reciprocity relation~\r{time7}. Nevertheless, one can write Onsager-Casimir relation  which imposes constraint even on time-modulated nonreciprocal systems:  
\e  
\int\limits_{V_A} \sum_{n}   \frac{  \freq{\_E}_n^{(B)} (\Omega)  \cdot \freq{\_J}_n^{(A)}}{\omega_n} {\rm d}V_A  = 
\int\limits_{V_B} \sum_{n}   \frac{  \freq{\_E}_n^{(A)} (-\Omega)  \cdot \freq{\_J}_n^{(B)}}{\omega_n} {\rm d}V_B.
\l{time10} 
\f
Here, similarly to~\r{eq104}, the argument $-\Omega$ implies that the corresponding quantity should be considered in the medium with time modulation reversed, i.e. assuming $t\rightarrow -t$ in \r{time9} or, equivalently, reversing signs of $\phi_1$ and $\phi_2$. This result is apparent from equality $\=F(\_r,\_r',\Omega)=\=F^T(\_r,\_r',-\Omega)$. 

\begin{figure}[tb!]
\centering
\epsfig{file=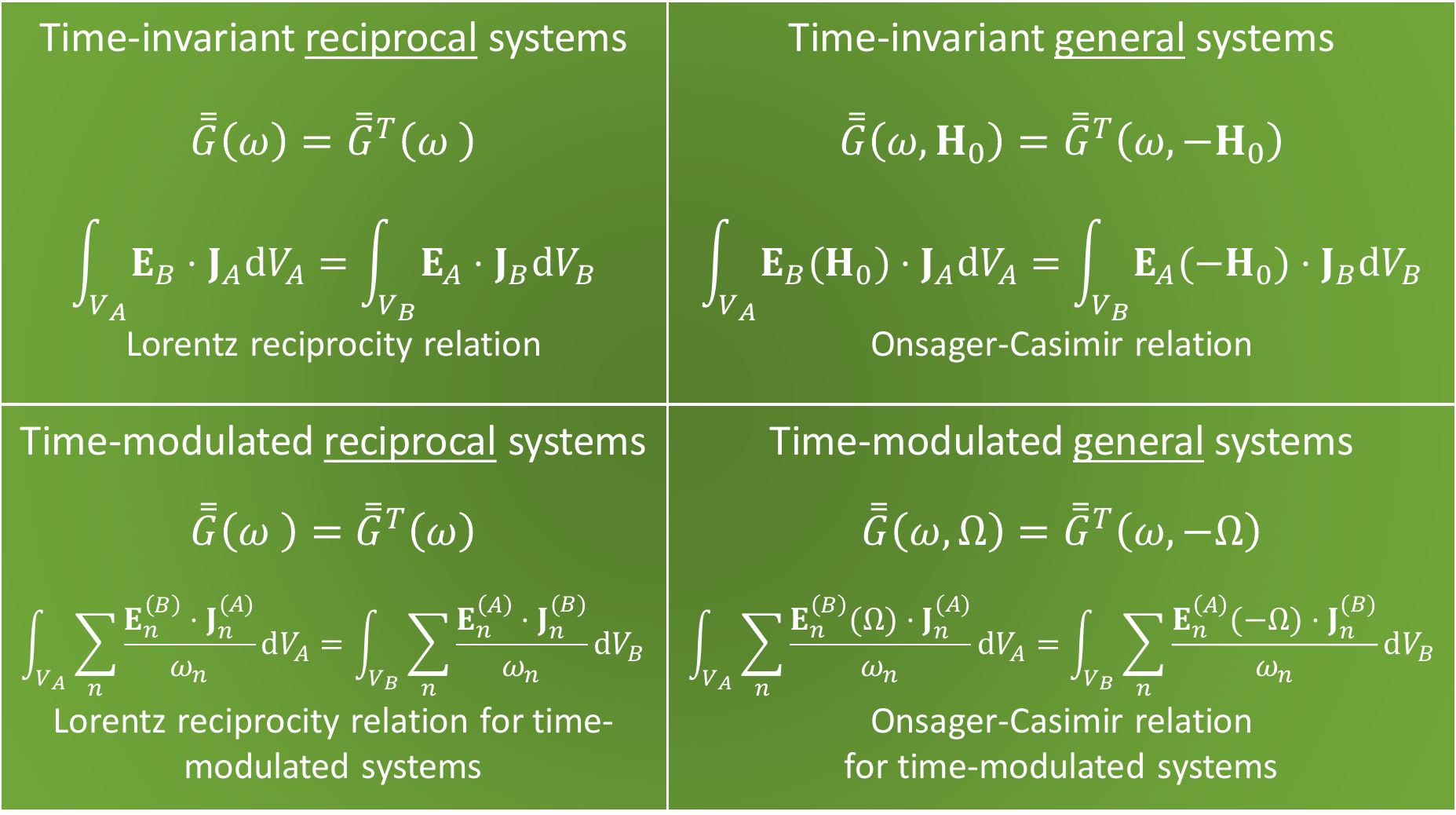, width=0.99\linewidth} 
\caption{Summary of the Lorentz reciprocity relations (for reciprocal electromagnetic systems) and Onsager-Casimir relations (for both reciprocal and nonreciprocal systems) for time-invariant and time-varying materials.   The time-modulated system is called reciprocal if the Lorentz reciprocity relation holds for some specific time translation $\Delta t$. }
\label{onsager_table}
\end{figure}

It is important to note that  reciprocity can be broken even in globally (uniformly) modulated systems which include materials whose properties vary according to an asymmetric function (e.g., with sawtooth profile)~\cite{williamson_breaking_2020} or
 bianisotropic materials with varying magnetoelectric coupling (even with symmetric variation profile). The latter scenario can be verified following derivations~\r{time2}--\r{time4} and using bianisotropic constitutive relations~\r{matrel1}. For example, in~\cite{wang_nonreciprocity_2020}, strong nonreciprocity was reported for a single bianisotropic metasurface of the ``omega'' type with uniform time modulation.  
Figure~\ref{onsager_table} summarizes the   Lorentz reciprocity relations   and Onsager-Casimir relations   for time-invariant and time-varying materials.

Next, we present  two main approaches for creation of nonreciprocal devices based on time modulation.

\begin{figure}[t!]
\centering
\epsfig{file=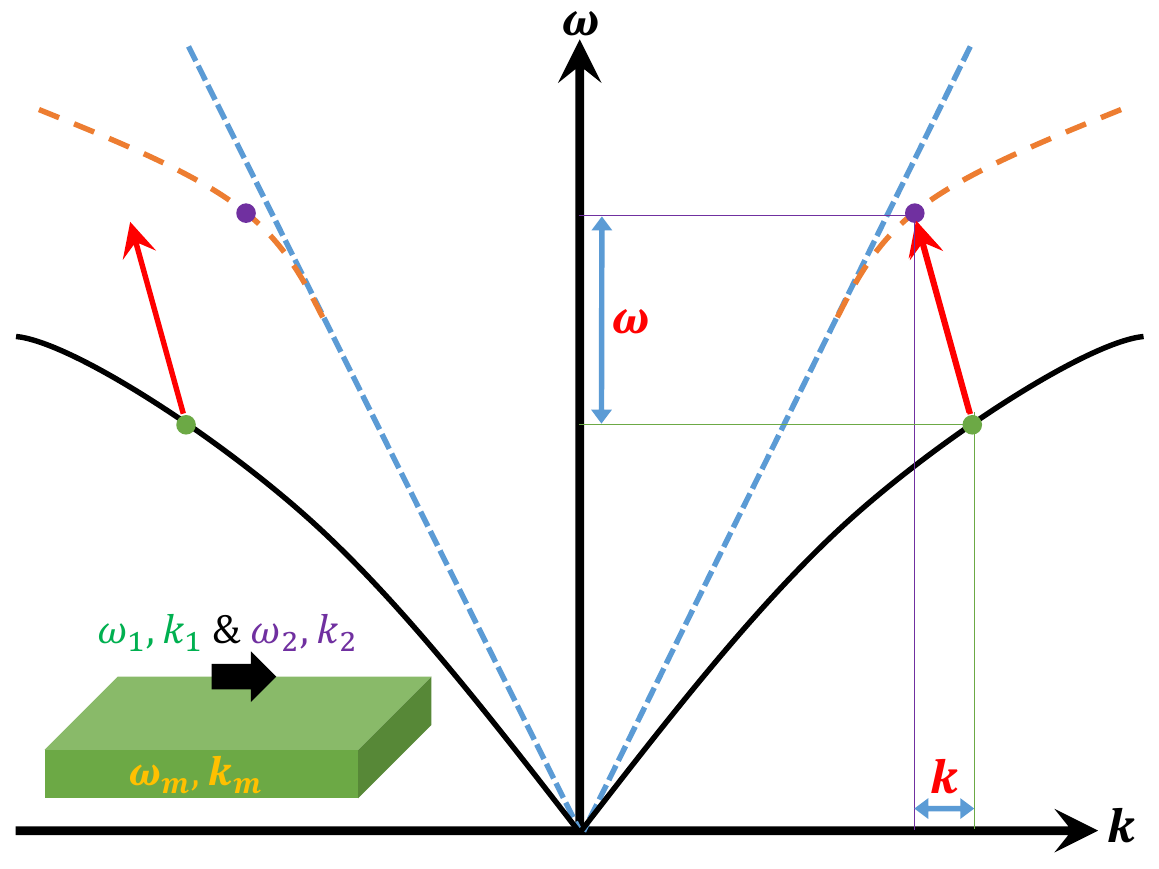, width=0.79\linewidth} 
\caption{The dispersion curve for a waveguide supporting two different modes. If   waveguide's dielectric constant is spatially and temporally modulated with the frequency $\omega_{\rm{m}}$ and the momentum $k_{\rm{m}}$, an indirect wave transition happens. This transition is the key point in achieving different transmission from opposite ports.}
\label{fig19}
\end{figure}

\subsubsection{Travelling-wave modulation}
It is clear that studies and applications of travelling-wave modulations have a long history at least in microwave engineering. It appears that this concept was developed in 1950s and 1960s for parametric amplifiers~\cite{cullen_travelling-wave_1958}. It is known that for a lossless transmission line with a distributed inductance per unit length   $L=L_0+M\sin2(\omega t-kz)$, the amplitude of the current wave is exponentially growing or decaying. Precise synchronization between  the oscillator (which produces the variation of the inductance in time) and the incoming signal gives rise in maximum possible amplification. A simple  pictorial example of such synchronization is a 
child at a playground swing. The child ``pumps'' the swing by periodically standing and squatting to increase the amplitude of the swing's oscillations.  The maximum efficiency is achieved when the "pump" motions is at twice the frequency of   swing's oscillations~\cite{roura_towards_2010}.

Travelling-wave modulation for achieving nonreciprocity (in contrast to amplification) has recently attracted strong attention of researchers. In analogy with the above description of parametric amplifiers, let us consider an inhomogeneous isotropic dielectric waveguide with permittivity~\cite{yu_complete_2009}
\begin{equation} 
\time{\varepsilon}(x,y,z,t)=\varepsilon_{\rm st}(x,y)+M(x,y)\cos(\omega_{\rm{m}}t-k_{\rm{m}}z).
\end{equation}
Here, $\varepsilon_{\rm st}(x,y)$ denotes the static permittivity, $M(x,y)$ is the modulation amplitude distribution, $\omega_{\rm{m}}$ represents the modulation angular frequency, $k_{\rm{m}}$ is the modulation wave number, and finally $z$ is the axis of propagation inside the waveguide. We suppose that the modulation frequency is small compared to the frequency of the wave signal. The dielectric waveguide supports two oppositely  propagating modes whose angular frequencies $\omega_1$ and $\omega_2$ correspond to the phase constants $k_1$ and $k_2$, respectively. Regarding the forward direction, if the difference between the two angular frequencies is equal to the modulation angular frequency, i.e.~$\omega_{\rm{m}}=\omega_2-\omega_1$, and if also the same condition holds for the phase constants $k_{\rm{m}}=k_1-k_2$, an indirect photonic transition happens (see Fig.~\ref{fig19}). Accordingly, the first mode fully transits to the second mode. Such mode conversion is found to be made after propagation over a distance (called coherence length~\cite{yu_complete_2009}).
It should be mentioned that  the modulation frequency can
be much smaller than the carrier frequency of the input signal.  
If the above strict conditions are not obeyed, the conversion will not happen and the modulation does not influence the incoming signal, as is shown in Fig.~\ref{fig19}. Indeed, this is the true scenario related to the backward direction. Remind that for the backward propagation the phase constants along the waveguide are $-k_1$ and $-k_2$. Therefore, due to the spatiotemporal modulation, both time-reversal symmetry and spatial-inversion symmetry are broken\footnote{Here, broken symmetries should be understood in the sense that backward wave propagation is different from the forward  one since the time modulation of permittivity is kept the same for both processes. Naturally, if time reversal is applied globally, including the modulator device ($\time{\varepsilon}_{\rm rev}(t)=\time{\varepsilon}_{\rm orig}(-t)$), the forward and backward propagations will be equivalent.}, and the system becomes nonreciprocal. From microwave engineering point of view, this is similar to inserting properly modulated capacitors in a transmission line. Such a line offers  nonreciprocal propagation~\cite{qin_nonreciprocal_2014}.

\begin{figure}[t!]
\centering
\epsfig{file=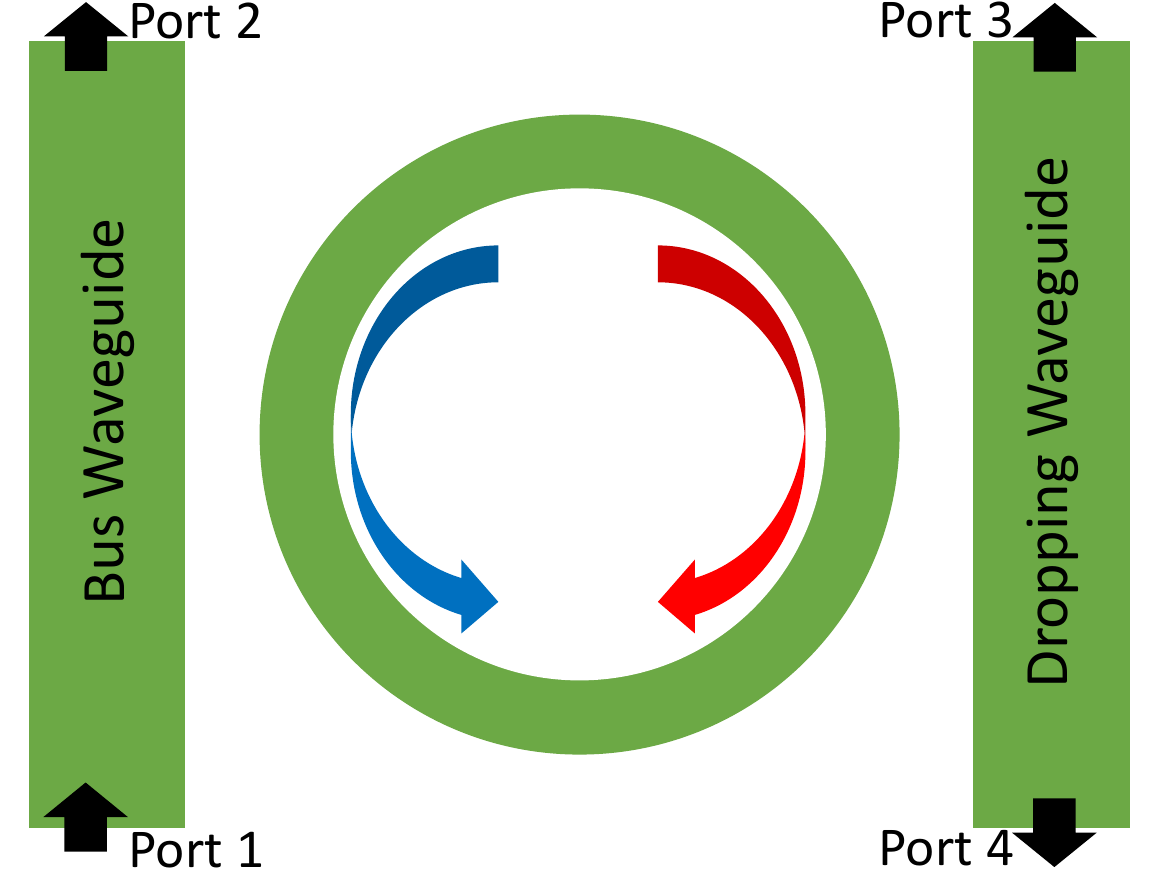, width=0.75\linewidth} 
\caption{An add-drop filter based on a ring resonator. Azimuthal spatiotemporal modulation of the ring resonator removes the degeneracy between the two modes of opposite handedness in the modulated ring. The input signal from port~1 is not transmitted to  port~2, while the signal is fully transmitted to  port~1 when it is incident from port~2. In the case of no modulation, the degeneracy exists and the transmission coefficients $S_{21}$ and $S_{12}$ are identical.}
\label{fig:angularmontum}
\end{figure}

\subsubsection{Angular momentum}
Let us consider a simple example of a circular cavity which is accompanied with a waveguide structure~\cite{Sounas_time_modulation_2017}, as shown in  Fig.~\ref{fig:angularmontum}. Here, the ring cavity is located between two parallel waveguides. This ring resonator simultaneously supports two  modes propagating in the opposite directions: Righ-handed and left-handed modes. The parallel waveguides are coupled to the resonator. Accordingly, if the frequency of the incident signal from the waveguide port is the same as the resonance frequency of the resonator, the transmission coefficient (from the input port 1 to the output port 2) will be zero. At other frequencies, full transmission occurs. The more important point is that the dip in the transmission for both modes is identical giving rise to a reciprocal structure since $S_{12}=S_{21}$ ($S$ denotes the scattering matrix). However, if the degeneracy of these two counterpropagating modes is revoked, we can create a nonreciprocal response. When the  degeneracy is lifted, the resonance frequency of the cavity is different for the two modes. Consequently, the transmission coefficients $S_{12}$ and $S_{21}$ are not equal. One way to obtain different resonance frequencies is to use magnetic (or magneto-optical) materials which will result in different guided wavelengths for opposite propagation directions. However, similarly to the previous approach for achieving nonreciprocity,  the alternative way is to apply the time modulation to the material filling the cavity such that~\cite{Sounas_time_modulation_2017}
\begin{equation}
\time{\varepsilon}(r,\phi,z,t)=\varepsilon_{\rm st}(r,z)+M(r,z)\cos(\omega_{\rm{m}}t-L_{\rm{t}}\phi),
\end{equation}
where $L_{\rm{t}}$ is the so-called   orbital angular momentum. The modulation frequency can be small and the angular momentum should be chosen such that the two dominant counterpropgating modes in the cavity are excited (in fact, $L_{\rm{t}}=2L$, where $L$ is a nonzero integer number). Note that the fields inside the cavity  are dependent on the exponential function $\exp(\pm jL\phi)$, where $\phi$ is the azimuthal component of the coordinate system. Due to the mixing, the resonance frequency of the modes shift and differ from each other. Therefore, degeneracy is cancelled and one-way transmission can be achieved.

The concept of ``temporal modulation" or more generally ``linear time-varying systems" for the purpose of nonreciprocity has recently attracted great attention of the engineers and physicists who work in both radio frequency   and photonics community. They have immensely published about various nonreciprocal  magnetless devices~\cite{estep_magnetic-free_2014,Reiskarimian10,Reiskarimian200,Dinc1,Nagulu1,Biedka1,qin_nonreciprocal_2014,Estep200,Kord1,Kord2,Kord3,Kord4,Kord5,Sounas1,Bhave1}.

%%%%%%%%%%%%%%%%%%%%%%%%%%%%%%%%%%%%%%%%%%%%%%%%%%%%%%%%%%%%%%%%%%%%%%%%%%%%%%%
%%%New Section 

\section{Asymmetric transmission in reciprocal systems } 

Unfortunately, in the modern literature, one can find many examples of misconceptions about nonreciprocity (see e.g., \cite{fan_comment_2012,mutlu_diodelike_2012,petrov_comment_2013}). The main confusion stems from the apparent similarity between \textit{asymmetric transmission in reciprocal} systems and \textit{isolation in nonreciprocal} systems. It is important to note that isolation \textit{cannot} be achieved without nonreciprocity since it requires   violation of the Lorentz reciprocity theorem~\cite{jalas_what_2013}. Nevertheless, in some works~\cite{mutlu_diodelike_2012}, the opposite erroneous statement was made. Moreover, the use of terms such as ``emulating nonreciprocity''~\cite{pfeiffer_emulating_2016,qiu_emulating_2019} may lead to further confusions  for  inexperienced readers. Another possible reason for confusions is the use of unorthodox terminology like \emph{optical diode} in the meaning of  \emph{optical isolator}. Diodes are nonlinear devices which strongly modify the signal spectrum, while isolators are linear nonreciprocal devices.

\begin{figure}[tb!]
	\centering
\includegraphics[width=0.65\linewidth]{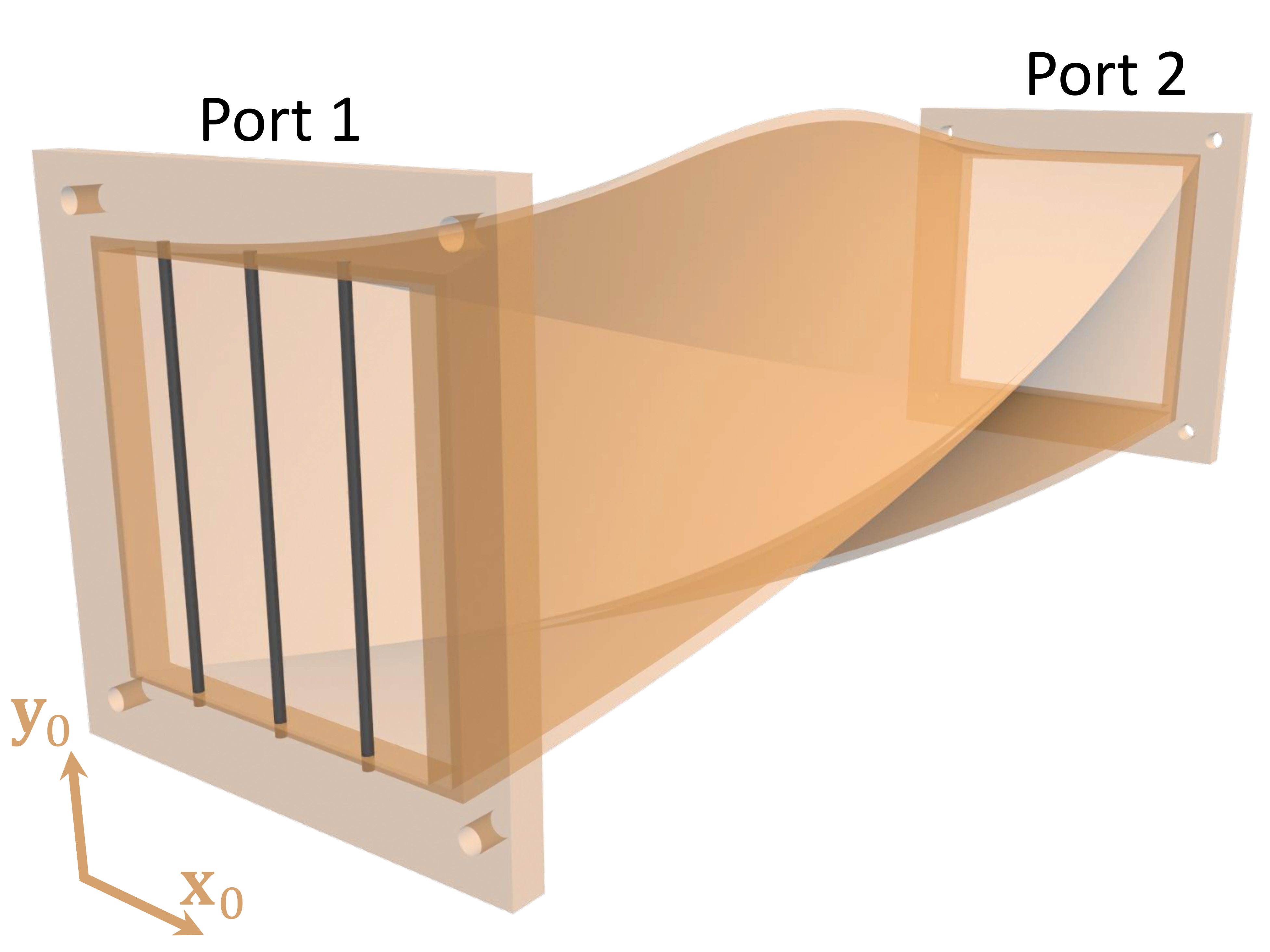}  
	\caption{ Twisted square-cross-section waveguide. The walls are depicted as semi-transparent to show that there is no polarization filter at the back end of the waveguide.  }
	\label{waveguide}
\end{figure}

As was said earlier, asymmetric transmission   can be achieved in reciprocal systems. To clarify, let us consider a simple example of a twisted waveguide section, equipped with a polarizer at one end, as shown in Fig.~\ref{waveguide}. The waveguide has a square cross section, which means that there are two degenerate fundamental modes propagating along  the waveguide. These modes, namely TE$_{10}$ and TE$_{01}$ modes, have the transverse electric polarization, and therefore the corresponding electric field does not possess longitudinal component along the propagation direction. In other words, the electric field has $\mathbf{x}_0$ component (TE$_{01}$ mode) or $\mathbf{y}_0$ component (TE$_{10}$ mode) depending on the mode. First, the waveguide is excited from port~1 with the $\_x_0$ polarization that passes through the polarizer, as shown in Fig.~\ref{waveguide}. Due to twisting of the waveguide, the excited mode during propagation acquires orthogonal $\_y_0$ polarization at the output port. Note that there is no polarization filter located at the output port. Now, if we excite the $\_x_0$-polarized  mode at the opposite port (port 2), the scenario is different due to the polarization filter placed in port 1. Since the output electric field is parallel to the wire grid (polarizer), the reflection is nearly perfect for illumination from port~2. Thus, we have strong asymmetry of  transmission for opposite illuminations. We stress that using polarization filter, we do not break the Lorentz reciprocity, which can be verified applying the Lorentz reciprocity theorem at the two ports of the waveguide.
Basically, this device is a reciprocal asymmetric polarization converter.

\begin{figure}[t!]
\centering
	\centering
	\subfigure[]{\includegraphics[height=0.43\linewidth]{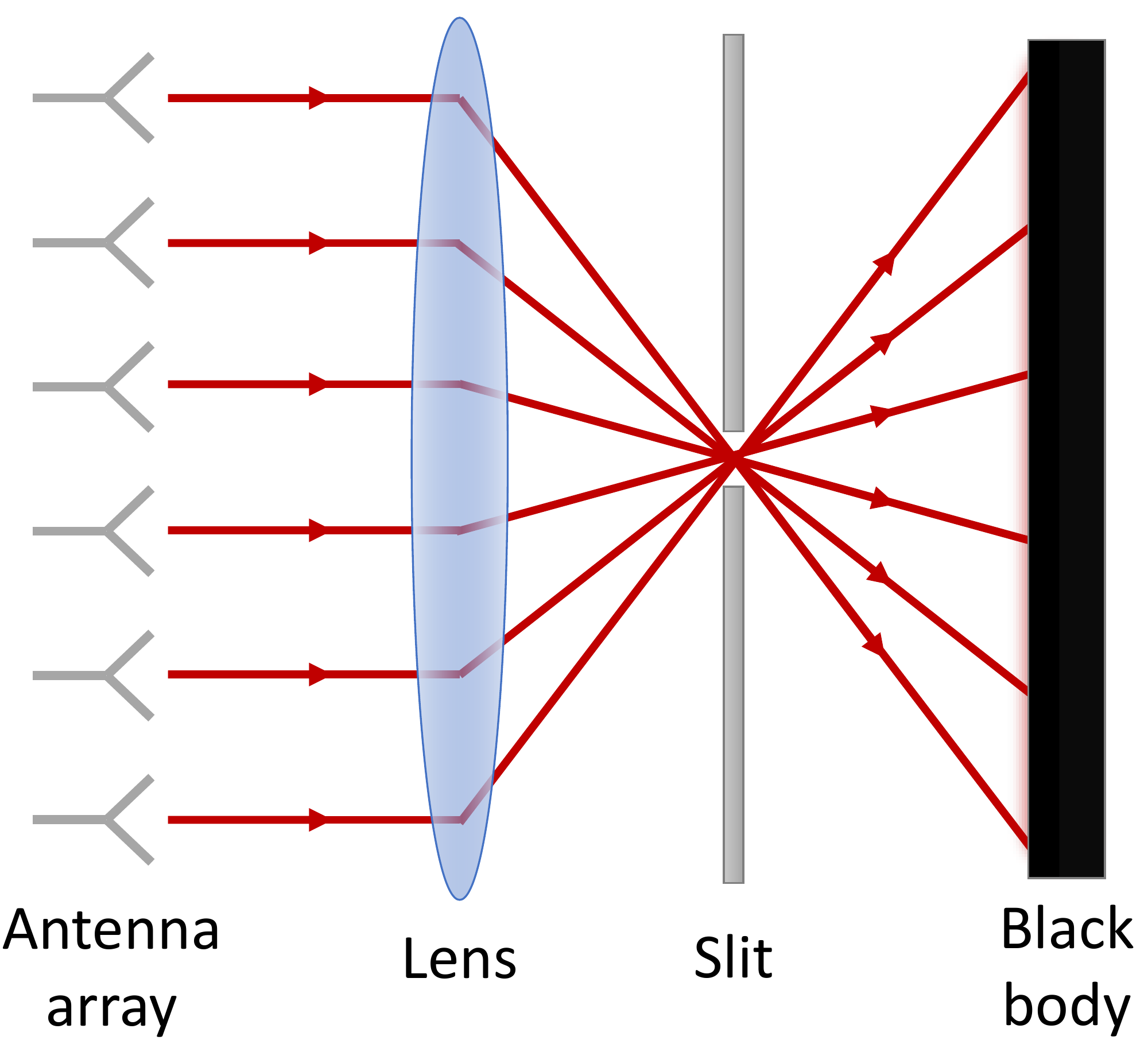} \label{hole_a}} 
	\subfigure[]{\includegraphics[height=0.43\linewidth]{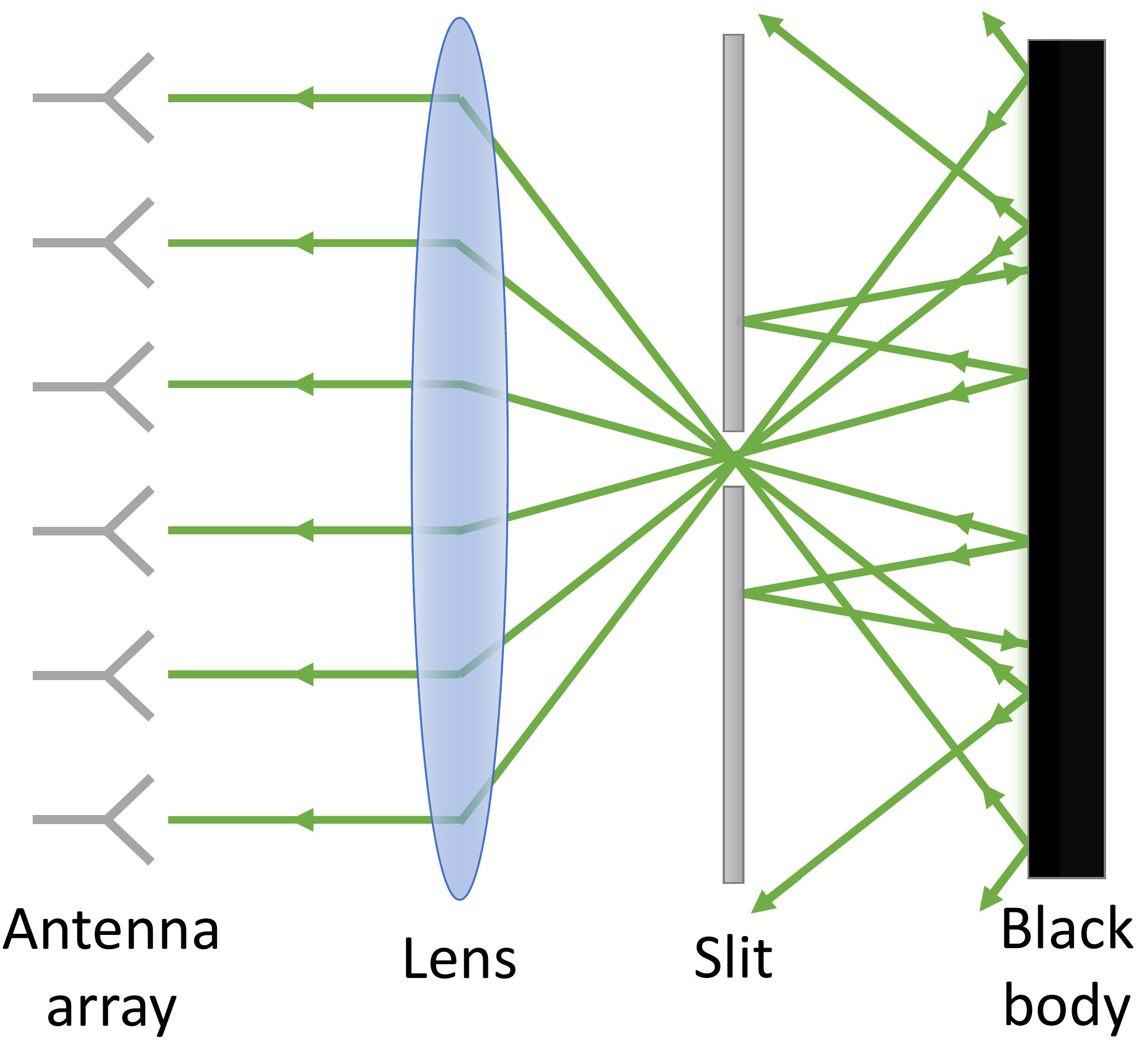}\label{hole_b}}
\caption{Asymmetric transmission in a reciprocal system formed by a perfectly conducting screen with a slit and a focusing lens. }
\label{hole}
\end{figure}

% Another simple example, illustrating the asymmetric transmission in a reciprocal system, is a lens in front of  a perfectly conducting plane screen which has a small hole at its center (see Fig.~\ref{hole}).  The lens can focus the electromagnetic energy at the location of the hole when an incident plane wave is illuminating the lens, and the power will be effectively transported across the screen.  If the incident plane wave is coming from the right side of the screen, due to the reflection from the conduction screen, most of the power will be reflected, and we realize ``unidirectional propagation''. Note, however, that if the two sources (for instance, antenna arrays, as illustrated in the picture) are the same, the power transmission between the two antennas is perfectly symmetric, as it should be in any reciprocal system. The left array creates a plane wave in the normal direction, and the right array can receive power only from the same direction. Obliquely incident waves will be reflected from the array (converted to other plane-wave modes) and dissipated elsewhere. 

Another simple example, illustrating the asymmetric transmission in a reciprocal system, is a lens in front of  a perfectly conducting plane screen which has a small hole at its center (see Fig.~\ref{hole}). The lens can focus the electromagnetic energy at the location of the hole when an incident plane wave is illuminating the lens from the left, and the power will be effectively transported across the screen. In Refs.~\cite{maznev_reciprocity_2013,caloz_electromagnetic_2018} transmission through the lens--screen system was analysed when there were two identical antennas at the two sides of the system.   Here we assume that on the left there is  a directive array antenna, while on the right there is a black body. Now, it may appear from the first sight that such a system  breaks reciprocity and allows transmission only from the left to the right.  Indeed, when the source is   the antenna, the  plane wave incident on the lens will be focused onto the hole on the slit and completely absorbed by the black body (see Fig.~\ref{hole_a}). However, in the ``reverse process'', the black body (effectively, it is an antenna)  will radiate  in all directions in  a reciprocal way as shown in Fig.~\ref{hole_b}. Thus, in the reciprocal scenario there will be exactly equal power which will pass over the hole via obliquely propagating waves and received by the antenna. In particular, this conclusion means that this ``asymmetrically transmitting" device cannot break the symmetry of thermal flux exchange between hot bodies on the opposite sides of the screen.   It is also important to understand that if we replace the positions of the \emph{antennas}, moving the directive antenna array to the right and the omnidirectional black body to the left, the power transmission will be very different from the original case.  There is no contradiction with the reciprocity theorem because in this thought experiment we change the \emph{system}, not only flip the positions of \emph{sources}.    Referring to Fig.~\ref{fig_antenna}, such  test would be equivalent to replacing the positions of the dipole and loop antennas in addition to the positions of external sources. 

It is also important to keep in mind fundamental restrictions on achievable nonreciprocal effects which come from the energy conservation principle and other fundamental physical laws. For instance, all resonators must to some extend allow coupling to outside (lossy) world just to allow excitation of fields inside the resonator. This imposes a fundamental limit on the quality factor defined by the internal resistance of the source. It would be most desirable to isolate a resonator from the sources inserting a nonreciprocal device (isolator) at the entry port~\cite{tsakmakidis_breaking_2017}. However, as discussed above, ideal isolators cannot reflect energy into the resonator. Instead, they absorb it inside the isolator, meaning that the loaded quality factor of the resonator does not change if one tries to isolate the resonator from the exterior using a nonreciprocal device.   

It is convenient to classify different systems providing asymmetric transmission into three basic types.  This classification can be made using the scattering matrix representation given in~\r{eh2}. Consider a two-port system where there are two possible modes in each port (alternatively, such system can be called a four-port system). These two modes can have different polarizations or different field distributions. The scattering matrix in the general form is given by
\e
\=S = 
\begin{pmatrix} \displaystyle %\vspace{2mm}
S_{11}^{x  x} & S_{11}^{x y} & S_{12}^{x x } & S_{12}^{x y}\vspace{1.5mm}\\  
S_{11}^{yx} & S_{11}^{yy} & S_{12}^{yx} & S_{12}^{yy}\vspace{1.5mm}\\ 
S_{21}^{xx} & S_{21}^{xy} & S_{22}^{xx} & S_{22}^{xy}\vspace{1.5mm}\\ 
S_{21}^{yx} & S_{21}^{yy} & S_{22}^{yx} &S_{22}^{yy}\\ 
\end{pmatrix},
\l{scatmat}
\f 
where subscripts $i,j$ denote the port number and superscripts $x,y$ denote the mode number.  According to~\r{eh4}, for reciprocal systems the scattering matrix is symmetry $\=S = \=S^T$. Nevertheless, even with symmetric matrix, we can achieve peculiar scenarios of asymmetric transmission of three different types.
\begin{enumerate}
    \item Asymmetric wave conversion characterized by \mbox{$S_{12}^{yx} \neq S_{21}^{yx}$}. See multiple examples of asymmetric conversion for linear and circular polarized waves~\cite{fedotov_asymmetric_2006,singh_terahertz_2009,menzel_asymmetric_2010,mutlu_diodelike_2012,wu_giant_2013,pfeiffer_high_2014}. The waveguide in Fig.~\ref{waveguide} also corresponds to this type (here  $S_{21}^{yx} =S_{12}^{xy} $, which is in agreement with the reciprocity condition $\=S = \=S^T$). 
    \item Asymmetric transmission characterized by \mbox{$S_{12}^{yx} \neq S_{21}^{xx}$}. See the characteristic example of asymmetric  propagation of waves with different field profiles~\cite{wang_-chip_2011,fan_comment_2012}.
    \item Unidirectional reflection characterized by \mbox{$S_{11}^{xx} \neq S_{22}^{xx}$}. The examples of this type include metasurfaces with asymmetric reflection properties~\cite{radi_tailoring_2014,alaee_magnetoelectric_2015,shevchenko_bifacial_2015}, asymmetric diffraction gratings~\cite{wang_extreme_2018}, and systems with parity-time symmetry~\cite{alaeian_parity-time-symmetric_2014}.
\end{enumerate}
We stress that all the works belonging to these three types correspond to reciprocal systems. 

Systems with parity-time (PT) symmetry deserve a separate discussion. In such  systems, the Hamiltonian is non-Hermitian, while it has an entirely real energy spectrum (see great reviews~\cite{longhi_parity-time_2017,huang_unidirectional_2017} for optical PT symmetry). In other words, waves propagating in such systems have real  propagation constants and oscillations have real frequencies. In the optical realm, a non-Hermitian Hamiltonian  corresponds to a complex   refractive index distribution, i.e. PT-symmetric system must have spatially alternating   regions of loss and gain $\freq{n}(x)=\freq{n}^\ast(-x)$, where $x$ is the coordinate. Thus, under parity (space) inversion, the lossy regions are interchanged by those with gain. Additional time reversal flips these regions again, returning to the original system. Therefore, a PT-symmetric system does not change under simultaneous space and time inversions.  
Importantly, the reciprocity theorem also applies  to PT-symmetric systems~\cite{fan_comment_2012,jalas_what_2013}. Although asymmetric reflection was demonstrated for the PT-symmetric structures ($S_{11}$ can be zero, while $S_{22}$ can have amplitude larger than unity due to the active regions), they all exhibit reciprocal transmission~\cite{alaeian_parity-time-symmetric_2014}. The only exceptions are PT-symmetric systems incorporating  nonlinear~\cite{Peng1,chang_paritytime_2014} or time-modulated components~\cite{song_direction-dependent_2019}.

% Some confusing terminology:  Acoustic wave rectifier, optical (acoustical) diode, one-way mode transmission, emulated nonreciprocity,\dots 

% As I mentioned in the previous email, a lossy medium turns into active under the proper time reversal. Active one will turn into lossy one. Thus, a PT-symmetric system incorporating gain and loss with refractive index profile n(x)=n*(-x) (x is coordinate) will satisfy TR too. See a good review “Longhi_2017….pdf” and also paper “nphys1515.pdf”. 
% The interesting feature of a PT-symmetric system is that it can work as an active laser and lossy absorber at the same real frequency (see “PhysRevA.82.031801.pdf”). 
% Also it should be mentioned that even though some people call PT-structures exhibiting unidirectionality, they are completely reciprocal (see “PhysRevA.89.033829.pdf”) due to the symmetric S-matrix (while S11 can be 0 and S22 can be even larger than 1 due to active regions), unless they are combined together with nonlinearity (see “nphys2927.pdf”). 

\section{Conclusions}
In this tutorial paper, we have attempted to provide an intuitive and concrete introduction to the concept of electromagnetic reciprocity. Since reciprocity is closely related to the  time-reversal operation, we first elaborated on the conventional definition and the essense of the latter. Based on the fundamental works by Casimir, Onsager, and Sachs, we demonstrated how   motion reversal with correct microscopic initial conditions results in the conventional mathematical form  $t\rightarrow -t$ of the time reversal operation.  Since, strictly speaking, almost all physical laws are time-reversal symmetric, it is justifiable to introduce an alternative notion of restricted time reversal. It was shown that  restricted time reversal holds only in pointwise reciprocal systems, which makes these two concepts  equivalent for electromagnetic systems. 

Based on the microscopic time reversibility (same as time reversal defined in physics literature) and three other assumptions, a comprehensive derivation of the Onsager reciprocal relations was given, and its application to several physical processes was demonstrated. We  showed that, using these fundamental  relations, one can derive the  reciprocity theorem proposed by Lorentz in 1896. Employment of this theorem was presented for different types of electromagnetic systems. Following Casimir's extension of the Onsager relations, we  expressed the universal restrictions imposed on nonreciprocal systems due to the microscopic reversibility. 
Role and meaning of reciprocity in systems characterized as  multi-port networks were covered.

Next, we indicated different routes to break reciprocity in electromagnetic systems, subsequently discussing the three most common ones in detail: Materials with an external field bias, nonlinear systems, and time-varying composites. Regarding the first route, properties of multiple  materials biased by magnetic field were compared for applications in different parts of the frequency spectrum. We explained the operation principles of several most important nonreciprocal components. Furthermore, space- and time-symmetry considerations were examined with the application to material tensors. A general classification of linear time-invariant media based on these considerations was presented. Building upon this classification, we were able to relate seemingly irrelevant electromagnetic effects and even envision novel ones in artificial   composites. 
Finally, we discussed about several systems which were erroneously referred to in the literature as nonreciprocal.

\section*{Acknowledgments}
This work was supported in part by the Finnish Foundation for Technology Promotion, the  Academy of Finland (project 287894), Nokia Foundation (project 
201920030), and by the U.S. Air Force Office of  Scientific Research MURI project (Grant No. FA9550-18-1-0379).
The authors thank Mr.~Cheng Guo, Dr.~Momchil Minkov, Prof.~Steven G. Johnson, and Dr.~Adi Pick   for 
useful comments and discussions about the manuscript.

\bibliography{bibliography}
\bibliographystyle{ieeetr}

% \bibliographystyle{IEEEtran}
% \bibliography{IEEEabrv,bibliography}

\end{document}